\documentclass[aps,prx,twocolumn,longbibliography,notitlepage]{revtex4-2}
\usepackage{graphicx}
\usepackage{amsfonts}
\usepackage{amssymb}
\usepackage{amsmath}
\usepackage{mathtools}
\usepackage{afterpage}
\usepackage[mathscr]{euscript}
\usepackage{braket}
\usepackage{subfigure}
\usepackage{multirow}
\usepackage{float}
\usepackage{qcircuit}

\usepackage[utf8]{inputenc} 
\usepackage{yfonts}
\usepackage{dsfont}
\usepackage{cancel}
\usepackage[scr=boondox]{mathalpha}

\begin{document}
\newcommand{\Z}{\mathbb Z}
\newcommand{\R}{\mathbb R}
	\renewcommand{\L}{\mathcal L}
	\newcommand{\A}{\mathcal A}
	\renewcommand{\H}{\mathcal H}

	
	\title{Crystalline gauge fields and quantized discrete geometric response for Abelian topological phases with lattice symmetry}
	\author{Naren Manjunath}
        \author{Maissam Barkeshli}
	\affiliation{Department of Physics, Condensed Matter Theory Center, and Joint Quantum Institute,
		University of Maryland, College Park, Maryland 20742, USA}
	
              \begin{abstract}
                Clean isotropic quantum Hall fluids in the continuum possess a host of symmetry-protected quantized invariants, such as the Hall conductivity, shift and Hall viscosity. Here we develop
                a theory of symmetry-protected quantized invariants for topological phases defined on a lattice, where quantized invariants with no continuum analog can arise. We develop topological field theories using discrete crystalline gauge fields to fully characterize quantized invariants of (2+1)D Abelian topological orders with symmetry group $G = U(1) \times G_{\text{space}}$, where $G_{\text{space}}$ consists of orientation-preserving space group symmetries on the lattice. We show how discrete rotational and translational symmetry fractionalization can be characterized by a discrete spin vector, a discrete torsion vector which has no analog in the continuum or in the absence of lattice rotation symmetry, and an area vector, which also has no analog in the continuum. The discrete torsion vector implies a type of crystal momentum fractionalization that is only non-trivial for $2$, $3$, and $4$-fold rotation symmetry. The quantized topological response theory includes a discrete version of the shift, which binds fractional charge to disclinations and corners, a fractionally quantized angular momentum of disclinations, rotationally symmetric fractional charge polarization and its angular momentum counterpart, constraints on charge and angular momentum per unit cell, and quantized momentum bound to dislocations and units of area. The fractionally quantized charge polarization, which is non-trivial only on a lattice with $2$, $3$, and $4$-fold rotation symmetry, implies a fractional charge bound to lattice dislocations and a fractional charge per unit length along the boundary.
                An important role is played by a finite group grading on Burgers vectors, which depends on the point group symmetry of the lattice. 
	\end{abstract}
	
	\maketitle

        \section{Introduction}

One of the most striking discoveries in physics is the quantized Hall conductivity of integer and fractional quantum Hall (FQH) systems \cite{girvin1999quantum,goerbig2009quantum}. The quantized Hall conductivity \cite{Laughlin1983}, which requires $U(1)$ charge conservation to define, is however only one of many symmetry-protected topological invariants of FQH systems. In the continuum, clean isotropic quantum Hall systems possess additional symmetry-protected invariants, such as a quantized Hall viscosity \cite{Avron1995hvisc,Tokatly2007,Read2009hvisc,Read2011hvisc,Bradlyn2015,Klevtsov2015}, the shift, and fractional orbital spin of quasiparticles \cite{Wen1992shift}. These invariants define quantized responses to deformations of the spatial geometry \cite{Wen1992shift,Cho2014,Gromov2014,Gromov2015,Bradlyn2015}.

The problem of interacting particles in the continuum is in many cases an approximation to interacting particles on a lattice. This approximation is typically only valid in a dilute limit where the lattice effects can be ignored. However topologically ordered phases can also occur when lattice effects are strong, such as in fractional Chern insulators or quantum spin liquids \cite{parameswaran2013,savary2017}. The crystalline symmetry can in principle allow for new topological invariants that are not possible in continuum systems, while also modifying the known invariants of continuum systems. It is therefore important to understand the possible topological invariants that are protected by the crystalline symmetry of the lattice, together with the on-site symmetry. 

In this paper we develop such an understanding in the case of (2+1)D Abelian topological phases with symmetry group $G = U(1) \times G_{\text{space}}$, where $G_{\text{space}} = \Z^2 \rtimes \Z_M$, for $M = 1,2,3,4,6$, is a discrete orientation-preserving space group symmetry of a lattice. To do this, we develop a theory of discrete ``crystalline gauge fields'' coupled to the emergent dynamical $U(1)$ gauge fields that describe the topological order. The crystalline gauge fields include gauge fields associated with the discrete translation and rotation symmetries, which keep track of certain geometric properties of the lattice, such as the presence of dislocations and disclinations, and areas and lengths of closed cycles in lattice units. As such, they form a discrete analog of the coframe field and spin connection used in continuum geometry. While crystalline gauge fields have been discussed before in the theory of elasticity \cite{Kleinert}, previous treatments in elasticity theory have not fully taken into account the non-Abelian nature of the space groups involved. We note that recently crystalline gauge fields have also been used in the study of quantum phases of matter, see e.g. Ref. \cite{Thorngren2018,song2019electric}, although effective actions involving both translation and rotation gauge fields have not to our knowledge been discussed previously.

         \begin{table*}[t]
        	\begin{tabular}{ l||l|l|l|l||l|l|l|l|l|l|l }
        		\hline
        		& \multicolumn{4}{c||}{Characterizing symmetry fractionalization}& \multicolumn{7}{|c}{SPT terms: integer contributions to response theory} \\ \hline
        		Parameter & $\vec{q}$& $\vec{s}$&$\vec{t}$&$\vec{m}$&$k_1$&$k_2$&$k_3$&$\vec{k}_4$&$\vec{k}_5$&$k_6$&$k_7$ \\ \hline
        		Allowed values& $\Z^D$ &$\Z^D$ &$\Z^D\times\Z^D$&$\Z^D$&$\Z$&$\Z$&$\Z$&$\Z^2$&$\Z^2$&$\Z$&$\Z$ \\
        		Trivial values & $K \vec{\Lambda}$& $K \vec{\Lambda}_1+M\vec{\Lambda}_2$&$\binom{K \vec{\Lambda}_1}{K \vec{\Lambda}_2}+(1-U(\frac{2\pi}{M}))\binom{\vec{\Gamma}_1}{\vec{\Gamma_2}}$&$K \vec{\Lambda}$&$0$&$M\Z$&$M\Z$&$(1-U(\frac{2\pi}{M}))\Z^2$&$(1-U(\frac{2\pi}{M}))\Z^2$&$0$&$M\Z$ \\
        		Classification & $\A$& $\A/M\A$&$K_M\otimes\A$&$\A$&$\Z$&$\Z_M$&$\Z_M$&$K_M$&$K_M$&$\Z$&$\Z_M$ \\        		
        		\hline    
        	\end{tabular}
        	\caption{ Summary of the parameters defining topologically non-trivial terms in the effective action (Eq. \ref{Eff_Action})
                  for $G=U(1)\times G_{\text{space}}$, and their classification.  $\vec{q}, \vec{s}, \vec{t}, \vec{m}$ characterize symmetry fractionalization,
                  while $k_i$ parameterize additional SPT (Dijkgraaf-Witten) terms in effective action. The topological order is characterized by a
                  $D\times D$ $K$ matrix, and the vectors $\vec{\Gamma}_i,\vec{\Lambda}_i$ are arbitrary $D\times 1$ integer vectors.
                  $\mathcal{A}$ is the Abelian group arising from fusion of the anyons. $K_M = \Z_1, \Z_2^2, \Z_3, \Z_2, \Z_1$, for $M = 1,2,3,4,6$ respectively.
                  Relabelling the dynamical gauge fields can give redundancies among different choices of the above parameters.}
        	\label{table:Summary}
        \end{table*} 

\begin{table*}[t]
	\centering
	\begin{tabular} {l|p{0.8\textwidth} }
		\hline
		\multicolumn{2}{c}{Fractional symmetry quantum numbers}\\ \hline
		Generalized charge vector & Associated quantum number \\ \hline
		$\frac{q_I}{2\pi} a^I \cup dA$& $Q_{\vec{l}} = \vec{l}^T K^{-1} \vec{q}$, fractional charge of $\vec{l}$\\
		$\frac{s_I}{2\pi} a^I \cup dA$& $ L_{\vec{l}} = \vec{l}^T K^{-1} \vec{s}$, fractional angular momentum of $\vec{l}$\\
		$\frac{\vec{t}_I}{2\pi} a^I \cup d\vec{\cancel{R}}$& $\vec{P}_{\vec{l}} = (1-U^T(\frac{2\pi}{M}))^{-1} (\vec{l}^T K^{-1} \vec{t}_x, \vec{l}^T K^{-1} \vec{t}_y)^T $, fractional linear momentum of $\vec{l}$ \\
		$\frac{m_I}{2\pi} a^I \cup A_{XY}$& $\tau_{\vec{l}} = \vec{l} K^{-1} \vec{m}$,  fractionalization of translation algebra: $T_{x,\vec{l}} T_{y,\vec{l}} = T_{y,\vec{l}} T_{x,\vec{l}} e^{i\tau_{\vec{l}}}$ \\ \hline
		\multicolumn{2}{c}{Quantized fractional response terms}\\ \hline
		Response theory term & Associated response property \\ \hline
		$\frac{\sigma_H}{2} A \cup dA$& $\sigma_H = $ Hall conductivity \\
          $\frac{\mathscr{S}}{2\pi} A \cup dC$& Defines discrete analog of shift. Charge of $2\pi/M$ disclination is          $\mathscr{S}/M$, angular momentum of $\phi$ flux given by $\mathscr{S}\phi/2\pi$. \\
		$\frac{\ell_s}{4\pi} C \cup dC$ &Angular momentum of elementary disclination equals $\ell_s/M$ (up to framing anomaly) \\
		$\frac{\check{\vec{P}}_c}{2\pi} \cdot A \cup d\vec{R}$ & Fractional quantized charge polarization: (i) Charge of dislocation with Burgers vector $\vec{b}$ equals $\check{\vec{P}}_c\cdot \vec{b}$; (ii) Charge per unit length on a boundary along $\hat{e}$ equals $\check{\vec{P}}_c\cdot\hat{e}$; (iii) A $U(1)$ flux of $\phi$ has linear momentum equal to $\check{\vec{P}}_c \phi/2\pi$ \\
          $\frac{\check{\vec{P}}_s}{2\pi} \cdot C \cup d\vec{R}$ & Fractional quantized angular momentum polarization: Angular momentum of dislocation with Burgers vector $\vec{b}$ equals $\check{\vec{P}}_s\cdot \vec{b}$ \\
          		$\frac{\Pi_{ij}}{4\pi} R_i \cup dR_j$ & Fractional quantized torsional response: Momentum of dislocation with Burgers vector $\vec{b}$ is $\vec{P}_{\text{disloc}, \vec{b}} = \Pi \vec{b}$ \\
		$\frac{\nu_c}{2\pi} A \cup A_{XY}$ & $\nu_c = $ charge per unit cell (filling)\\
		$\frac{\nu_s}{2\pi} C \cup A_{XY}$ & $\nu_s = $ angular momentum per unit cell \\
		$\frac{\vec{\nu}_p}{2\pi} \cdot \vec{R} \cup A_{XY}$ & $\vec{\nu}_p = \vec{P}_{\vec{m}}$ linear momentum per unit cell \\
		\hline
	\end{tabular}
	\caption{Summary of the quantized topological terms that arise in the effective action for a topological order coupled to a background crystalline gauge field $B=(A,\vec{R},C)$ for the symmetry $G=U(1)\times G_{\text{space}}$. $A,\vec{R}$ and $C$ refer to the $U(1)$, translation ($\Z^2$) and point group rotation ($\Z_M$) components of the gauge field, while $A_{XY}$ denotes the area element and can be written in terms of $(\vec{R},C)$. $d\cancel{R}$ is defined in Eq. \ref{cancelR}.
          Symmetry quantum numbers are associated to the coupling terms between $B$ and the Abelian topological order, specified by a $K$ matrix of internal gauge fields. $U(\frac{2\pi}{M})$ is the elementary point group rotation matrix. The response coefficients are obtained by integrating out the internal gauge fields. The classification of the parameters in the effective action is summarized in Table \ref{table:Summary}.}\label{table:SummaryofTerms}
\end{table*}

Recently a powerful algebraic theory using G-crossed braided tensor categories has been developed to comprehensively characterize and classify (2+1)D topologically ordered phases of matter with symmetry \cite{Barkeshli2019}. In the case of Abelian topological orders with symmetries whose action does not permute distinct quasiparticle types, an alternate approach using topological effective actions, which we develop here, is significantly simpler and yields insight into the physical response.

Our results may be of particular relevance in a number of physical systems. These include the experimentally realized fractional Chern insulators in van der Waals heterostructures \cite{spanton2018,parameswaran2013} and synthetic quantum Hall systems in photonics \cite{schine2019,ozawa2019} or ultracold atoms \cite{cooper2019,hafezi2007}. These platforms may in particular be able to directly measure the (fractionally) quantized charges bound to lattice dislocations and disclinations. Our results are also of relevance for the study of quantum Hall systems with crystalline symmetries on orbifolds \cite{Gromov2014,gromov2016,klevtsov2017}, polygons, and two-dimensional surfaces of polyhedra.

Our results are summarized in Tables \ref{table:Summary} and \ref{table:SummaryofTerms}. We find that in general symmetry fractionalization for $G = U(1) \times G_{\text{space}}$ is determined by four invariants, which are specified by a charge vector $\vec{q}$, a discrete spin vector $\vec{s}$, a discrete torsion vector $(\vec{t}_x, \vec{t}_y)$, and an area vector $\vec{m}$. The discrete spin vector $\vec{s}$ is a discrete version of the well-known spin vector used in continuum FQH states \cite{Wen1992shift}, which specifies a fractional orbital angular momentum for the anyons \cite{gromov2016}. The discrete torsion vector $\vec{t}$ has no analog in the continuum and can only be non-trivial for $M = 2,3,4$-fold lattice rotational symmetry; it specifies a fractional linear momentum for the anyons that does not appear to have been discussed in previous studies of topological phases of matter. Finally the area vector $\vec{m}$, which also has no analog in the continuum, specifies the anyon per unit cell \cite{Cheng2016} and determines how the anyons effectively fractionalize the translation algebra \cite{Jalabert1991,Sachdev_2018,sachdev1999translational,Essin2013SF,Essin2014spect,Cheng2016}. The discrete spin and torsion vectors $\vec{s}$ and $\vec{t}$ furthermore can only be non-trivial when there is some appropriate commensuration between $M$, the order of the point group symmetry, and the group structure of the fusion rules of the anyons. 

The quantized response theory, obtained by integrating out the dynamical $U(1)$ gauge fields, provides the response of the system to background gauge fields describing background electromagnetic fields and geometrical defects of the lattice (see Eq. \ref{responseAction}). We find, for example,
\begin{enumerate}
\item A discrete analog of the shift of FQH states. This binds a quantized fractional charge (modulo the charge of the anyons) to disclinations and angular momentum to magnetic flux.
\item Fractional quantized angular momentum for disclinations.
\item Fractional quantized charge polarization for $M = 2,3,4$-fold rotational symmetry. This implies a fractional charge bound to lattice dislocations (modulo the charge of the anyons), fractional charge per unit length along boundaries (modulo the charge of the anyons), and associates a quantized momentum to $U(1)$ flux.
\item An angular momentum analog of the fractional charge polarization, which associates a fractional angular momentum to dislocations (modulo the angular momentum of the anyons).
\item Fractional quantized charge $\nu_c$, angular momentum $\nu_s$, and linear momentum $\vec{\nu}_p$ per unit cell. The charge filling $\nu_c$ gives a generalized Lieb-Schulz-Mattis constraint that imposes constraints on the topological order, $\vec{m}$, and $\vec{q}$ given the charge per unit cell \cite{Cheng2016}.
\item Fractional quantized torsional response which associates momentum to dislocations. In particular, this addresses a long-standing issue raised by Ref. \cite{Hughes2011thv,Hughes2013thv}, where the coupling to continuum geometry gave an unquantized torsional Hall response; our work predicts that properly taking into account the discrete crystalline space group symmetry gives rise to a fractional quantized torsional response, but only for $M = 2,3,4$.
\end{enumerate}

Our effective field theory allows us to explicitly classify all distinct symmetry-enriched topological phases for a given Abelian topological order (for the case where symmetries do not permute the anyons). We find, for example, that there are 2304 distinct symmetry-enriched topological states with the intrinsic topological order of the $1/2$ Laughlin state on the square lattice, once the integer part of the filling and Hall conductivity are fixed. 

The outline of this paper is as follows. In Section \ref{Sec:CGF} we define the background crystalline gauge field for $G = U(1)\times G_{\text{space}}$ on a manifold $\mathcal{M}$ with a triangulation, and in Section \ref{Sec:SF} we study their gauge transformations and the symmetry fluxes associated to them. The effective action for SET phases with $G$ symmetry is discussed in Section \ref{Sec:EA+RT} by coupling the crystalline gauge field and background $U(1)$ gauge field to the dynamical gauge fields that specify the intrinsic topological order. In this Section we also study the effective response theory obtained by integrating out the internal gauge fields. Specific examples involving the 1/2 Laughlin topological order and the $\Z_2$ gauge theory are discussed in Section \ref{Sec:Examples}. In Section \ref{Sec:Classif} we obtain the SET classification from the effective action and discuss with examples how this classification is reduced when we account for relabellings of the gauge fields. In Section \ref{Sec:Continuum}, we compare our formulation of crystalline gauge theory on a discrete triangulation to the more standard continuum field theory approach and compare the crystalline gauge fields to the coframe fields and spin connection used in continuum geometry. We conclude with a discussion in Section \ref{Sec:Disc}. 
	
	\section{Crystalline gauge fields}
	\label{Sec:CGF}

        At a formal mathematical level, our theory of crystalline gauge fields is equivalent to treating the discrete space group symmetry $G_{\text{space}}$ as an internal symmetry of the topological effective field theory. The main difference with usual internal symmetries, which arise from on-site symmetries of a microscopic lattice model, is the physical interpretation of the crystalline gauge fields, which in turn requires certain gauge-invariant quantities to be determined by geometric properties of the underlying lattice, as we describe below. 

        Ultimately, the topological field theory that we develop in terms of the quantum Chern-Simons theory possesses an implicit dependence on a space-time metric, which is the framing anomaly associated with the chiral central charge \cite{witten1989,Gromov2015}. To be physically meaningful, this space-time metric must be determined by the crystalline gauge fields (see Section \ref{Sec:GravAnom}). Further discussion regarding the relation between the space group symmetry in lattice systems and internal symmetries of the topological effective field theory is presented in Sec. \ref{Sec:Disc}. 
        
	We consider a $(2+1)$D space-time manifold $\mathcal{M} = \Sigma^2 \times \mathbb{R}$, where $\Sigma^2$ is the space on which the clean lattice system is defined. We fix an arbitrary triangulation of $\mathcal{M}$ and we define on the links a gauge field valued in the symmetry group $G = U(1) \times G_{\text{space}}$. $G_{\text{space}} = \mathbb{Z}^2 \rtimes \Z_M$ contains translation symmetry and a discrete $M$-fold rotation symmetry for $M = 1,2,3,4,6$. Physical results will be independent of triangulation. We define a $U(1)$ gauge field $A_{ij}$ on the link $ij$ of the triangulation, with the link directed towards $j$ (with $A_{ij} = -A_{ji}$ and $A_{ij} \sim A_{ij} + 2\pi$). Next, we define the crystalline gauge field
\begin{align}
B_{ij} = (\vec{R}_{ij}, C_{ij}).
\end{align}
Here,
\begin{align}
\vec{R}_{ij}^T = (X_{ij},Y_{ij}) = ((R_{ij})_x, (R_{ij})_y) \in 2\pi \Z^2
\end{align}
is an integer gauge field corresponding to $\Z^2$ translations. The field $C$ corresponds to point group rotations, where we take
\begin{align}
C_{ij} \in \frac{2\pi}{M}\Z,
\end{align}
with $C_{ij} \sim C_{ij} + 2\pi$. Group multiplication is given by $(\vec{R}_1,C_1) (\vec{R}_2,C_2) = (\vec{R}_1+{U(C_1)}\vec{R}_2,C_1+C_2)$, where we use addition in place of multiplication when the group is abelian.  $U(C_1)$ is the $2\times 2$ rotation matrix corresponding to $C_1$. Formally $\frac{1}{2\pi} B$ is a lift of an element of $G_{\text{space}}$ to $\Z^2 \rtimes \frac{1}{M}\Z$, while $A$ is a lift from $U(1)$ to $\R$.

        The gauge freedom in $\vec{R}$ corresponds to the freedom to relabel lattice coordinates. It arises from the well-known ambiguity in elasticity theory that the displacement vector is
        only meaningful up to an integer lattice vector \cite{Kleinert}, which we discuss further in Appendix \ref{Sec:elast}. 
        The gauge freedom in $C$ corresponds to the freedom in locally orienting the $x$ and $y$ axes at every point in space and time. For example, if for $M = 4$
        we have $C_{ij} = \pi/2$ on some link $ij$, this means the local coordinate axes at $i$ and $j$ will be rotated relative to each other by an angle $\pi/2$.
        
        Under a gauge transformation which places the gauge variable $(\vec{r}_i, h_i)$ at the vertex $i$, we have:
	\begin{align}\label{GspaceMultLaw}
          &B_{ij} \rightarrow (\vec{r}_i,h_i)^{-1}  B_{ij}  (\vec{r}_j,h_j) \nonumber \\
	&= ({U(-h_i)}  (\vec{R}_{ij} + {U(C_{ij})}\vec{r}_j - \vec{r}_i), -h_i+ C_{ij}+ h_j)
	\end{align}

        The underlying lattice of the physical system specifies the gauge invariant quantities of the crystalline gauge field. Flux of $C$ corresponds to disclinations: $\oint_\gamma C$ gives the total angle of disclinations
        within the cycle $\gamma$. If $C$ vanishes everywhere, then $\oint_\gamma \vec{R}$ gives the total Burgers vector of dislocations contained in $\gamma$. If space is a torus and $C$ vanishes everywhere, then
        $\oint_x X $, $\oint_y Y$ give the lengths of the torus in the $x$ and $y$ directions, while $\oint_y X$ gives the shear in the $x$ direction upon traversing the $y$ cycle, and similarly for $\oint_x Y$.

        When $C$ is non-zero, one needs to take into account the local change of coordinate frame along $\gamma$. Consider the product $B_{01} B_{12} \dots B_{n-1,n}$, where $B_{ij} \in G_{\text{space}}$. The translation component of this product is given by 
        \begin{equation}
        \int \vec{\mathcal{R}}^{(0)} := \sum\limits_{k=0}^{n-1} U(C_{01}+ C_{12}+\dots +C_{k-1,k}) R_{k,k+1}
        \end{equation} 
        Motivated by this, we define a Burgers vector
        $\oint_\gamma \vec{\mathcal{R}}^{(0)}$, where
        \begin{align}
          \vec{\mathcal{R}}^{(0)}_{k,k+1} = {U(C_{01}+C_{12}+ \dots +C_{k-1,k})} \vec{R}_{k,k+1}
        \end{align}
        for some arbitrary choice of origin $0$ and path from $0$ to $k$.
        The extra $C$ factors play a role analogous to the covariant
        derivative allowing parallel transport of $\vec{R}$ on the lattice. Under a gauge transformation,
        \begin{align}
          \oint_\gamma \vec{\mathcal{R}}^{(0)} \rightarrow U(-h_0) \oint_\gamma \vec{\mathcal{R}}^{(0)},
          \end{align}
          corresponding to the fact that the Burgers vector rotates under rotation of the local coordinate
          system at the origin $0$. 
        The value of this Burgers vector is invariant under the $\vec{r}$-dependent part of the gauge transformation
        (i.e. the translation gauge transformations), but is only well-defined up to an overall rotation. In general the value of this integral around a closed loop $\gamma$ defines the total Burgers vector for any dislocations located inside $\gamma$. In the special case of a closed loop in a flat configuration, $\oint_C \vec{\mathcal{R}}^{(0)}=0$.
        To compare Burgers vectors in different regions, it is important that a common origin $0$ is chosen.

        $(\vec{R},C)$ thus play a role similar to the coframe field and spin connection used in continuum geometry
        (see Section \ref{framespin} for further discussion); it is useful to distinguish them because $(\vec{R},C)$
        have discrete gauge transformations, which plays a crucial role in the classification of topological terms.
        Note that we do not consider the continuous elastic response of the crystal due to stresses and strains,
        which does not receive any topological, quantized contributions \cite{barkeshli2012ph, rao2020}.
        
        \section{Symmetry fluxes} 
        \label{Sec:SF}

        In order to construct the effective topological field theory, we need to understand how to construct symmetry fluxes that can be used in the effective action. While symmetry fluxes for
        $A$ and $C$ are relatively straightforward, the symmetry fluxes for the translation gauge field are more complicated, particularly in the presence of the rotation gauge field $C$. Mathematically,
        when the gauge fields are flat the symmetry fluxes define representative $2$-cocycles associated with the second group cohomology $\mathcal{H}^2(G, \Z)$. 
        
        The $U(1)$ gauge flux
        \begin{align}
          d A [012] = A_{01} + A_{12} - A_{02}
         \end{align}
          defined on a 2-simplex [012] of the triangulation is gauge-invariant,
          with $dA \sim dA + 2\pi$. Note that mathematically $d$ corresponds to the coboundary operation on the triangulation.

          $C$ behaves mathematically like a discrete version of $A$; the flux $\int_D dC$ for any region $D$
          is gauge-invariant and gives the total angle of disclinations within $D$. Below
          we will discuss the fluxes associated to translation symmetry, which are less familiar.
        
        \subsection{The flux $d\cancel{R}$ and its relation to dislocation density}

        Naively one may think that $d \vec{\mathcal{R}}^{(0)}$ should be the gauge-invariant physical quantity corresponding to the dislocation density. However $d \vec{\mathcal{R}}^{(0)}$ depends on a choice of origin together with a choice of local coordinate frame at that origin. Therefore $d \vec{\mathcal{R}}^{(0)}$ is both non-local in general and also not gauge invariant. Moreover, in the presence of a disclination, the value of $\vec{\mathcal{R}}^{(0)}_{ij}$ depends on the precise path chosen between the origin and $i$, and is therefore ambiguous up to a rotation by the disclination angle.
        
        The solution is to instead use the $\vec{R}$ fields themselves, which are local. But there is considerable ambiguity in $\vec{R}$ under gauge transformations. In particular, we now show that gauge transformations preserve the value of $d\vec{R}$ only
        up to terms of the form $(1-U\left(\frac{2\pi}{M}\right)) d\vec{\Gamma}$ where $\frac{1}{2\pi}\vec{\Gamma} \in \Z^2$.
        
        We argue as follows. From the definition of $\mathcal{\vec{R}}^{(0)}$ we have
        \begin{align}
        \vec{R}_{ij}= \vec{\mathcal{R}}_{ij}^{(0)} + (1-{U(C_{0\rightarrow i})}) \vec{R}_{ij}.
        \end{align}
        Here we have defined $C_{0\rightarrow i} = \int_{\gamma} C$ for some given path $\gamma$ from the origin 0 to the point $i$. The last term is of the form $(1-U\left(\frac{2\pi k}{M}\right)) \vec{R}_{ij}$, for some integer $k$. Let $C_{0\rightarrow i} = \frac{2\pi k_{0, i}}{M}$. Using the fact that $1-U^k = (1-U)(1 + U + \dots + U^{k-1})$, we conclude that
        \begin{widetext}
        	\begin{align}
        	\vec{R}_{ij} &= \vec{\mathcal{R}}_{ij}^{(0)} + (1-U\left(2\pi/M \right))(1 + U\left(2\pi/M \right) + \dots + U^{(k_{0, i}-1)}\left(2\pi/M \right)) \vec{R}_{ij} \\
        	&:= \vec{\mathcal{R}}_{ij}^{(0)}+ (1-U\left(2\pi/M \right)) \vec{\Gamma}_{ij}
        	\end{align}
        \end{widetext}
        The last line defines the vector field $\vec{\Gamma}$ in terms of $\vec{R}$, with $\frac{1}{2\pi}\vec{\Gamma} \in \Z^2$. 
        
        A general gauge transformation sends $\vec{R}_{ij} \rightarrow U(-h_i)(\vec{R}_{ij} + U(C_{ij}) \vec{r}_j - \vec{r}_i)$. But the above relation will still hold with $\vec{\Gamma}$ replaced by some $\vec{\Gamma}'$ where $\frac{1}{2\pi}\vec{\Gamma}' \in \Z^2$. Now under gauge transformations, assume that the coordinate axes at the origin are rotated by the angle $2\pi m/M$. Then $d\vec{R}$ transforms as  
        
        \begin{align}
        d\vec{R} &= d\vec{\mathcal{R}}^{(0)}+ (1-U\left(2\pi/M \right)) d\vec{\Gamma} \\
        & \rightarrow U(2\pi m/M)d\vec{\mathcal{R}}^{(0)}+ (1-U(2\pi/M)) d\vec{\Gamma}' \\
        &= d\vec{\mathcal{R}}^{(0)}+ (1-U(2\pi/M)) (d\vec{\Lambda} + d\vec{\Gamma}')
        \end{align}
        where $\vec{\Lambda} = (1+U(2\pi/M)+ \dots + U^{m-1}(2\pi/M))\vec{\mathcal{R}}^{(0)}$, and also satisfies $\frac{1}{2\pi}\vec{\Lambda} \in \Z^2$. 
        
        Therefore gauge transformations preserve the value of $d\vec{R}$ only
        up to terms of the form $(1-U\left(\frac{2\pi}{M}\right)) d\vec{\Gamma}$.

        To summarize, the correct definition of a Burgers vector, given in terms of $\vec{\mathcal{R}}^{(0)}$, is nonlocal due to the choice of origin $0$, and so we are forced to use the field $\vec{R}$
        instead in the effective action. $d\vec{R}$ is not gauge-invariant: it is determined only up to terms of the form $(1-U\left(\frac{2\pi}{M}\right)) d\vec{\Gamma}$.
        However, $d\vec{R}$ and $d\vec{\mathcal{R}^{(0)}}$ are gauge-equivalent up to such terms. Therefore the fractional part of $(1-U\left(\frac{2\pi}{M}\right))^{-1} d\vec{R}$
        is (i) local, (ii) gauge-invariant, and (iii) equal to the physically meaningful quantity $\frac{1}{2\pi} (1-U\left(\frac{2\pi}{M}\right))^{-1} d\vec{\mathcal{R}}^{(0)} \text{ mod } 1$.
        This motivates us to define the local quantity
        \begin{align}
          \label{cancelR}
          d\vec{\cancel{R}}=\left(1 - U\left(\frac{2\pi}{M}\right)\right)^{-1} d\vec{R},
          \end{align}
        which captures the local, gauge-invariant part of a Burgers vector.

 \begin{table}[t]
	\begin{tabular}{ |l|l|l|l| }
		\hline
		$M$ & ${{(1-U\left(\frac{2\pi}{M}\right))^{-1}}}(a,b)^T$ & Gauge invariants $\mod 1$ & $K_M$ \\ \hline
		2 &$\frac{1}{2}(a,b)^T$ &  $\frac{1}{2}\{(0,0),(1,0),(0,1),(1,1)\}$ & $\Z_2^2$\\
		3 &$\frac{1}{3}(2a+b,b-a)^T$ &$\{(0,0),(1/3,1/3),(2/3,2/3)\}$ & $\Z_3$ \\
		4 &$\frac{1}{2}(a+b,b-a)^T$ &$\{(0,0),(1/2,1/2)\}$ & $\Z_2$ \\
		6 &$(a-b,a)^T$ &${(0,0)}$ & $\Z_1$\\
		
		\hline    
	\end{tabular}
	\caption{ Gauge invariant, locally well-defined part of the Burgers vectors for different rotation point groups, with $\frac{1}{2\pi} \oint \vec{\mathcal{R}}^{(0)} = (a,b)^T \in \Z^2$. We use a lattice basis where $U(2\pi/M)$ takes $\hat{x} \rightarrow \hat{y}$ for $M \ne 2$ (see Appendix \ref{Appendix:PtGrpMatrices}).}
	\label{table:KM}
\end{table}     

The possible holonomies thus fall into different classes based on the distinct values taken by
\begin{align}
  \label{KMdef}
  \frac{1}{2\pi}\oint_{\partial D} \vec{\cancel{R}} \text{ mod } 1.
\end{align}
Eq. \ref{KMdef} defines a finite group grading on Burgers vectors, where we denote the finite group as $K_M$, and which is formally defined as
\begin{align}
K_M = \Z^2/(1 - U(2\pi/M) \Z^2 .
\end{align}
To understand this physically, note that to each region $D$ we can assign a local Burgers vector with the choice of origin $0 \in D$. Without picking a common origin, the Burgers vector for a region containing two subregions $D$ and $D'$ is thus ambiguous up to separate local rotations of the coordinate axes for the origins $0 \in D$ and $0' \in D'$. This is explained below in more detail. The part of the Burgers vector that is gauge invariant and can be defined locally defines a finite group grading on Burgers vectors, where we denote $K_M$ as the finite group. The results for various $M$ are given in Table \ref{table:KM}. 

Note that a nontrivial Burgers vector is associated to dislocation defects as well as disclination defects, which additionally have a nonzero holonomy of $C$. A disclination dipole is a composite of two defects in which the individual $C$ holonomies are equal and opposite; however, the net $\vec{R}$ holonomy may still be nonzero. This is the gauge-theoretic formulation of the well-known fact that a disclination dipole is physically equivalent to a dislocation.
  
  	\begin{figure*}
  	\centering
  	\includegraphics[width=15cm,height=10cm]{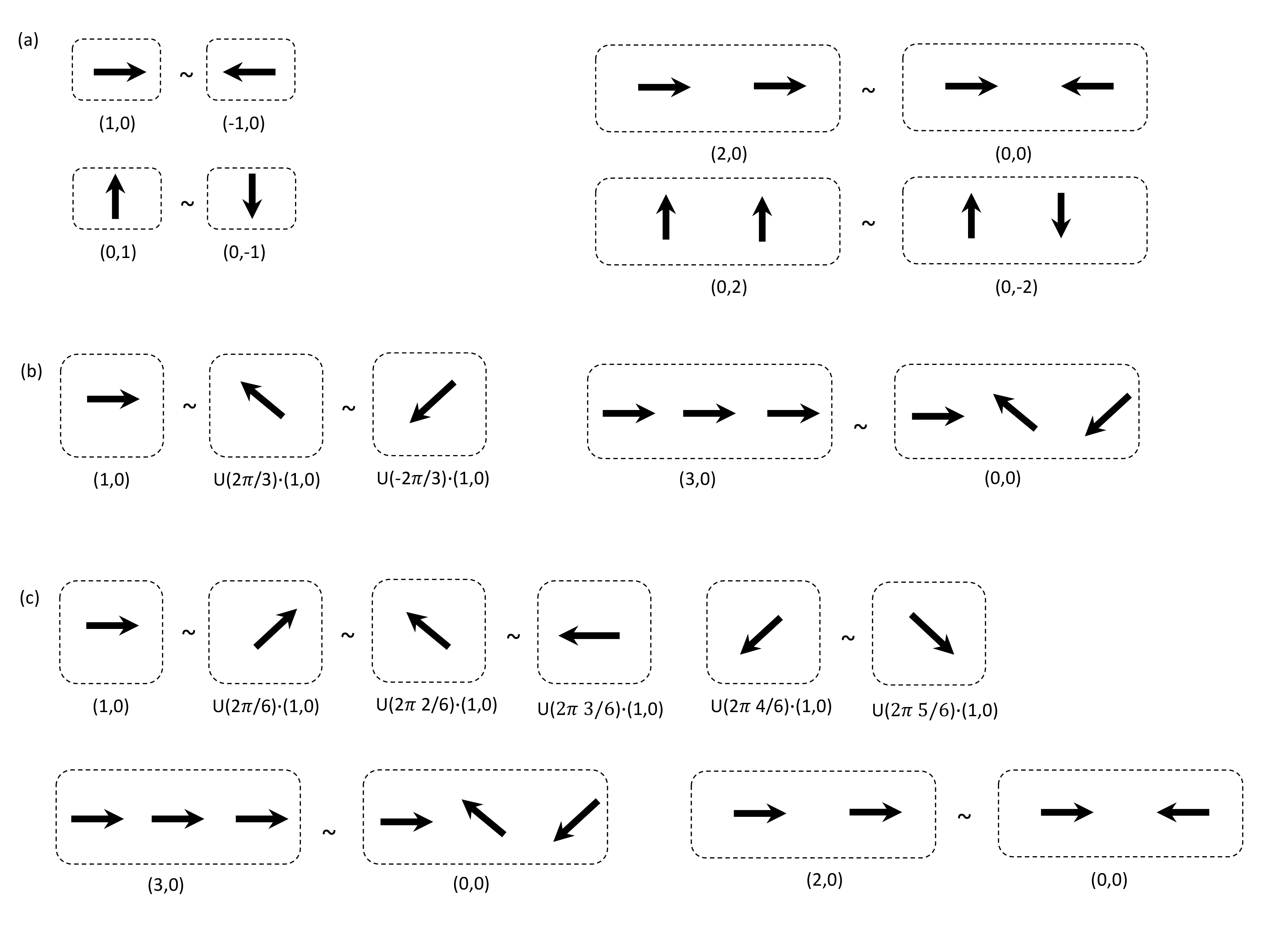}
  	\caption{Visual representation of how the groups $K_M$ classify dislocation Burgers vectors. (a) For $M=2$, the vectors $(a,b)$ and  $(-a,-b)$ are in the same equivalence class. Moreover, the sum of two neighbouring Burgers vectors can be viewed as either $(a,b) + (a',b')$ or $(a,b) - (a',b')$; this gives the relations $(0,0) \sim (2,0) \sim (0,2)$, which reduce the classification to a group $\Z_2\times\Z_2$. (b) For $M=3$, we see that $(3,0) \sim (1 + U(2\pi/3) + U(4\pi/3)) (1,0)^T = (0,0)$; in general $(2a+b,b-a) \sim (0,0)$, so the classification is given by $K_3 \cong\Z_3$. (c) For $M=6$, we can combine the $M=2$ and $M=3$ results to show that $(0,0) \sim (2,0) \sim (3,0)$; thus $(0,0) \sim (1,0)$, and similarly $(0,0) \sim (0,1)$. Therefore every Burgers vector can be trivialized. Similar reasoning applied to the $M=4$ case gives $K_4 \cong \Z_2$.}
  	\label{Fig:KM}
  \end{figure*}
  
  \subsubsection{Understanding the group $K_M$}
  \label{Sec:KM}
  
  There are a number of ways to understand $K_M$ more intuitively and physically. Let us consider the most direct way following the mathematical derivation above. A second derivation based on rotationally symmetric configurations of boundary charge is discused in Section \ref{Sec:RT}).

  Let us first consider the case $M = 2$, and start by considering a small region with a locally defined Burgers vector $(a,b)$ (see Fig. \ref{Fig:KM}). Under a local rotation of the space, this
  Burgers vector transforms to $(a,b) \rightarrow (-a,-b)$. Thus the Burgers vector $(1,0) \sim (-1,0)$ and $(0,1) \sim (0,-1)$. Now consider two regions, each with a
  locally defined Burgers vector $(a,b)$ and $(a',b')$. The combined Burgers vector thus would be $(a+a', b + b')$.  Upon a $\pi$ rotation of the second region however,
  $(a',b') \rightarrow (-a',-b')$, so $(a+a',b+b') \rightarrow (a-a', b-b')$. Therefore, when considering the Burgers vector of a large region containing Burgers vectors in smaller regions,
  $(2,0) \sim (0,0)$ and similarly $(0,2)\sim (0,0)$. We see that the Burgers vectors form the group $\Z_2 \times \Z_2$, due to the fact that the Burgers vector of a region, when
  including these local rotations, is only partially well-defined. An equivalent analysis for $M = 3,4,6$ gives the groups $\Z_3$, $\Z_2$, and the trivial group (see Fig. \ref{Fig:KM}).

  In general, dislocations whose Burgers vectors are of the form $(1-U(2\pi/M)) \vec{b}$ are equivalent to zero. If we have two neighbouring dislocations with $\vec{b}$ and $-\vec{b}$, the total Burgers vector associated to a loop containing the dislocations is zero. However a local rotation of $-\vec{b}$ by the angle $2\pi/M$ will give a net holonomy equal to $\vec{b}-U(2\pi/M)\vec{b}$ around the same loop. These values of Burgers vectors are therefore considered to be in the trivial equivalence class. This is what we mean by the statement that rotation gauge symmetry induces a finite group grading on Burgers vectors. They are thus classified by elements of $\Z^2$ modulo $(1-U(2\pi/M))\Z^2$, which can be taken as the mathematical definition of $K_M$.

  Mathematically, if we consider a generic group element in $G_{\text{space}}$, we can define the $K_M$ grading of the translation component of the group element. One can show that this $K_M$ grading is invariant under conjugation; therefore this $K_M$ grading can be viewed as an invariant of conjugacy classes of $G_{\text{space}}$. The same idea can be expressed intuitively as follows. Suppose we have two well-separated symmetry defects $p$ and $q$ which are defined by the holonomies of $B$ as follows: $\oint_p B = B_p = (\vec{R}_p,C_p)$ and $\oint_q B = B_q = (\vec{R}_q,C_q)$. Now the holonomy of $B$ around a loop encircling both $p$ and $q$ can be measured equally by the group element $B_p B_q = (\vec{R}_p + U(C_p) \vec{R}_q,C_p + C_q)$ or by the group element $B_q B_p = (\vec{R}_q + U(C_q) \vec{R}_p,C_q + C_p)$. These values of the holonomy should therefore be treated as physically equivalent. They are in fact gauge-equivalent: the difference in the translation component of the two holonomies equals 
  \begin{align}
  &\vec{R}_p + U(C_p) \vec{R}_q - (\vec{R}_q + U(C_q) \vec{R}_p) \nonumber \\&= (1-U(C_q)) \vec{R}_p -(1-U(C_p))\vec{R}_q.
  \end{align} 
  In the most general case, the rhs is a multiple of the matrix $(1-U(\frac{2\pi}{M}))$ by an integer vector. Therefore, a dislocation Burgers vector which takes such values should be regarded as trivial. Indeed, we can always find a gauge transformation which sets these values of Burgers vectors to zero. 
  
  In the same way, we can consider three well-separated defects $p,q,r$, whose holonomies are given by the group elements $B_p, B_q$ and $B_r = B_q^{-1}$. Now, the holonomy of the gauge field can be written either as $B_p B_q B_{q}^{-1} = B_p$ or as $B_q B_p B_q^{-1}$. Therefore two defects in the same conjugacy class must be regarded as physically equivalent; the corresponding translation components will be gauge equivalent and thus have the same $K_M$ grading.
  
  The groups $K_M$ arise naturally while classifying the allowed fractional $U(1)$ charges associated to the $\vec{R}$ holonomy of a disclination. This was shown for topological crystalline insulators in free fermion systems in Ref. \cite{Li2020disc}; we will carry out a similar analysis for bosonic SET phases in Section \ref{Sec:RT}. 

\subsection{Area flux}
\label{Sec:Area}	
In terms of the translation gauge fields we can also construct a flux $A_{XY}$, which is quadratic in $\vec{R}$ and corresponds to an area element. We will see that $A_{XY}$ is not by itself gauge-invariant, which is analogous to the fact that area elements are not invariant under general diffeomorphisms in continuum geometry. Nevertheless, we will see that under a gauge transformation, when $B = (\vec{R}, C)$ is flat, $A_{XY}$ changes by an integer-valued coboundary, so that it gives a well-defined area on closed manifolds. Physically this area corresponds to the number of unit cells of the clean (defect free) lattice. With some minor modifications, we will see that $A_{XY}$ can also provide a well-defined area for spaces with boundary.

We define
\begin{align}
A_{XY}[ijk] = \frac{1}{4\pi}\vec{R}_{ij} \times(U(C_{ij})\vec{R}_{jk}),
\end{align}
where $\times$ is the cross product of vectors. When $C =0$ everywhere, this gives the usual area element as expected, and it is easy to verify that on a torus $T^2$ whose side lengths are $L_x$ and $L_y$,
\begin{align}
  \frac{1}{2\pi} \int_{T^2} A_{XY} = L_xL_y.
  \end{align}
The factor $U(C_{ij})$ keeps track of the relative orientation of the coordinate axes at $i$ and $j$ when $C \neq 0$. In the absence of dislocations, $A_{XY}$ is gauge-invariant up to a boundary term, so that $A_{XY}$ integrated over a closed surface is gauge invariant. To obtain a well-defined area on spaces with boundary, we require the translation gauge transformations to reduce from $\Z^2$ to the subgroup of translations preserved by the boundary. Here we study the behavior of the area flux $A_{XY}$ under a gauge transformation and discuss its properties in the presence of dislocations and boundaries. 

The area flux on a 2-simplex $[ijk]$ can be written as
\begin{align}
A_{XY}[ijk] &= \frac{1}{4\pi} \vec{R}_{ij} \times U(C_{ij}) \vec{R}_{jk} \\
&= \frac{1}{4\pi} U(C_{0\rightarrow i})\vec{R}_{ij} \times U(C_{0\rightarrow i})U(C_{ij}) \vec{R}_{jk} \\
&= \frac{1}{4\pi}\mathcal{\vec{R}}_{ij}^{(0)} \times \mathcal{\vec{R}}_{jk}^{(0)} 
\end{align}
where $\times$ refers to the cross product: $\vec{v} \times \vec{u} = v_x u_y - v_y u_x$. The second line uses the fact that the cross product is invariant under an equal rotation of both arguments;
the symbol $C_{0\rightarrow i}$ refers to the sum of $C$'s on any given path from the origin 0 to the point $i$. The last line uses the definition of $\mathcal{\vec{R}}^{(0)}$. Note that the
cross product of two  $\mathcal{\vec{R}}^{(0)}$ fields is thus local even though a single such field is not. Since $A_{XY}$ is independent of the choice of origin $0$, we drop this
superscript and simply write
\begin{align}
A_{XY} = \frac{1}{4\pi} \vec{\mathcal{R}}_{ij} \times \vec{\mathcal{R}}_{jk} ,
\end{align}
with the understanding that $\vec{\mathcal{R}}_{ij}$ is defined with respect to an arbitrary choice of origin $0$. 

Under a gauge transformation, this equality implies that

\begin{align}
4\pi A_{XY}[ijk]&=  \mathcal{\vec{R}}_{ij} \times \mathcal{\vec{R}}_{jk}\\
&\rightarrow (\mathcal{\vec{R}}_{ij}+d\tilde{\vec{r}}_{ij})\times (\mathcal{\vec{R}}_{jk}+d\tilde{\vec{r}}_{jk})
\end{align}
Here we have defined $\tilde{\vec{r}}_{i} = {U(C_{0\rightarrow i})\vec{r}_i}$ (for the same arbitrary choice of origin $0$ used to define $\mathcal{R}$).
The difference $\delta A_{XY}$ can be written as    
\begin{align}
4\pi \delta A_{XY}[ijk] = (\mathcal{\vec{R}}_{ij})\times d\tilde{\vec{r}}_{jk} + d\tilde{\vec{r}}_{ij} \times (\mathcal{\vec{R}}_{jk} + d\tilde{\vec{r}}_{jk})
\end{align}
Defining
\begin{align}
f_{ij} =  \mathcal{\vec{R}}_{ij} \times \tilde{\vec{r}}_{j}+ \tilde{\vec{r}}_{i} \times (\mathcal{\vec{R}}_{ij} +d\tilde{\vec{r}}_{ij})  ,
\end{align}
we see that $\delta A_{XY}$ is a coboundary whenever $d \vec{\mathcal{R}} = 0$:
\begin{align}
4\pi \delta A_{XY}[ijk] = df [ijk]
\end{align}
Therefore when $A_{XY}$ is integrated over the entire manifold, this property implies that a gauge transformation will only contribute boundary terms to the integral
(assuming $\vec{\mathcal{R}}$ is flat). Therefore $\frac{1}{2\pi}\int_{\Sigma^2} A_{XY}$ over a closed 2-manifold $\Sigma^2$ is gauge-invariant (when $\vec{\mathcal{R}}$ is flat),
which we physically interpret as the area of the space $\Sigma^2$. Note that since the cross-product gives the area of a parallelogram, the integration over the whole space covers
the manifold twice, such that $A_{XY}$ is quantized to be an integer multiple of $2\pi$ when integrated over a 2-cycle. 

Although we have defined a gauge-invariant area only for closed manifolds, we can also define a gauge-invariant area for manifolds with boundary by restricting the gauge transformations
on the boundary. Specifically, we require that the quantity $f$ defined above must vanish for every boundary 1-simplex. For this to occur, it is sufficient that the boundary fields
$\mathcal{\vec{R}}_{ij}$ and the boundary gauge transformation variables $\tilde{\vec{r}}_j$ be parallel to each other. This requirement can also be viewed as a consequence of the fact
that a boundary can be chosen to break one of the two $\Z$ translation symmetries, so that the $\vec{R}$ field essentially reduces to a $\Z$ gauge field on the boundary.

For example, suppose the space is a square formed by the region $0 \le x,y \le a$ with origin (0,0). For simplicity let $C=0$ everywhere on the boundary except on links associated
with the corners, which have $C = \pi/2$. Let the fields $\vec{R}_{ij}$ on the $y=0$ line have zero $Y$-component. Now as we meet the corner $(a,0)$, we meet a 1-simplex with $C=\pi/2$.
The above condition on $\mathcal{\vec{R}}$ now means that on the $x=a$ line, $\vec{R}$ has zero $X$-component. In fact, one component of $\vec{R}$ is always constrained to vanish on the
boundary. 

The discussion above has so far required that $\vec{\mathcal{R}}$ be flat. In particular, when $\vec{\mathcal{R}}$ is flat, $\mathcal{\vec{R}}_{ij}\times \mathcal{\vec{R}}_{jk} =
\mathcal{\vec{R}}_{jk}\times\mathcal{\vec{R}}_{ki}$, so the definition of $A_{XY}$ does not depend on the ordering of the vertices. But if we assume that the simplex $[ijk]$ contains a dislocation, this equality no longer holds. we instead
have $\mathcal{\vec{R}}^{(0)}_{ij}+\mathcal{\vec{R}}^{(0)}_{jk}+\mathcal{\vec{R}}^{(0)}_{ki} = \vec{b}^{(0)}$ (where we have reinstituted the explicit dependence on the origin $0$), so 
\begin{align}
\mathcal{\vec{R}}^{(0)}_{ij}\times \mathcal{\vec{R}}^{(0)}_{jk} - \mathcal{\vec{R}}^{(0)}_{jk}\times\mathcal{\vec{R}}^{(0)}_{ki} =\vec{b}^{(0)}\times \mathcal{\vec{R}}^{(0)}_{jk} \ne 0
\end{align}
This means that the area of a simplex with nonvanishing holonomy of $\vec{\mathcal{R}}$ is not well defined. This is physically expected: on a lattice with a dislocation, the number of unit cells within a region
containing a dislocation cannot be obtained purely from the dimensions of the boundary. In fact, the number of unit cells in a small region containing a dislocation is not well-defined.
Moreover, as the dislocation moves, additional unit cells are added or removed. Therefore extensive observables such as the total charge or angular momentum will no longer be
gauge-invariant. However, intensive quantities such as the filling or angular momentum per unit cell will still be well-defined, because they are a ratio of two extensive quantities
computed with the same triangulation.

A well-defined area can be defined for a given fixed configuration of dislocations by cutting out the regions containing the dislocations. Then the system is viewed as a
manifold with boundary, and a gauge-invariant area can be defined as discussed above by restricting the gauge transformations on the boundary. Effectively this approach treats the dislocation as a hole in the simplicial formulation. In principle we can consider alternatively treating it as a puncture (for example, a sphere $S^2$ with a puncture would correspond to the plane $\R^2$), but then we cannot describe the open set near the puncture in terms of a finite triangulation.

\section{Effective action and response theory} 
\label{Sec:EA+RT}

With this understanding of the local gauge-invariant fluxes of the crystalline gauge fields, we are now ready to study the effective action. 

\subsection{Effective action}

To derive the effective action, we rely heavily on group cohomology, which classifies the distinct, inequivalent topological terms that can appear. The derivation of these terms from group cohomology is detailed in Appendix \ref{Appendix:GrpCoho}. 

The topological effective Lagrangian is
        \begin{widetext}
          \begin{align} \label{Eff_Action}
            \mathcal{L} &=- \frac{1}{4\pi} a^I \cup K_{IJ} da^J +\mathcal{L}_{frac}+\mathcal{L}_{SPT} \nonumber \\
	\mathcal{L}_{frac} &= \frac{1}{2\pi} a^I \cup (q_I dA + s_I dC + \vec{t}_I \cdot d \vec{\cancel{R}} + m_I A_{XY} ) \nonumber \\
	\mathcal{L}_{SPT} &= \frac{k_1}{2\pi} A \cup dA + \frac{k_2}{2\pi} A \cup dC + \frac{k_3}{2\pi} C \cup dC 
	+ \frac{1}{2\pi} A \cup (\vec{k}_4 \cdot  d \vec{\cancel{R}})+\frac{1}{2\pi} C \cup (\vec{k}_5 \cdot d \vec{\cancel{R}}) 
	+ \left(\frac{k_6}{2\pi} A + \frac{k_7}{2\pi} C\right) \cup A_{XY} .
	\end{align}
      \end{widetext}
      We have used the cup product from cohomology: $(A \cup d A)[ijkl] = A_{ij} dA[jkl]$ for a 3-simplex $[ijkl]$. 
      
      The non-degenerate $D \times D$ symmetric integer matrix $K$ \cite{Wenbook}, which couples the dynamical $U(1)$ gauge field $a^I$, characterizes the intrinsic topological
      order \footnote{Formally $a^I$ here are the lifts from $U(1)$ to $\R$.}. Topologically distinct quasiparticles correspond to integer vectors
      $\vec{l} \sim \vec{l} + K \vec{\Lambda}$, where $\vec{l}, \vec{\Lambda} \in \Z^D$. The quasiparticles form an Abelian group
      $\A = \Z_{n_1}\times\dots\times\Z_{n_D}$ under fusion, where the $n_i$ are the diagonal entries in the Smith normal form of $K$.

      This simplicial formulation of the Abelian CS theory was recently used in Ref. \cite{demarco2019} to develop a local bosonic model for chiral topological phases. 
      
      $\mathcal{L}_{frac}$, which contains the coupling between the background gauge fields and the $a^I$, specifies symmetry fractionalization, i.e. how the anyons carry fractional symmetry quantum numbers. Mathematically this is classified by the second group cohomology $\mathcal{H}^2(G, \mathcal{A})$ \cite{Essin2013SF,Barkeshli2019} \footnote{Note that the symmetry fractionalization anomaly \cite{Barkeshli2019,Barkeshli2020Anomaly,Bulmash2020} always vanishes here because $\mathcal{H}^4(U(1)\times G_{space}, U(1))$ is trivial.}. The distinct terms in $\mathcal{L}_{frac}$ are consistent with, and in fact can be derived from, the group cohomology classification (see Appendix \ref{Appendix:GrpCoho})
      \begin{align}
\mathcal{H}^2(G, \mathcal{A}) = \A\times(\A/M\A)\times(K_M\otimes\A)\times\A,
      \end{align}
      for $G = U(1) \times [\Z^2 \rtimes \Z_M]$. Here $\otimes$ denotes the tensor product of groups, defined in
      Appendix \ref{Appendix:GrpCoho}; for example, $\Z_q \otimes \Z_p = \Z_{\text{gcd}(p,q)}$. 

      The terms in $\mathcal{L}_{SPT}$ correspond to Dijkgraaf-Witten (DW) terms, classified by $\mathcal{H}^3(G, U(1)) \cong \H^4(G,\Z)$ \cite{Dijkgraaf1989pz}.
      In our case, we have
      \begin{align}
\mathcal{H}^3(U(1) \times [\mathbb{Z}^2 \rtimes \Z_M]), U(1)) =  \Z^2\times\Z_M^3\times K_M^2
      \end{align}
      The terms in $\mathcal{L}_{SPT}$ correspond explicitly to representative cocycles in $\mathcal{H}^3(G, U(1))$, as discussed in detail in Appendix \ref{Appendix:GrpCoho}. Physically the DW terms can be understood in terms of stacking symmetry-protected topological (SPT) states \cite{Chen2013,Barkeshli2019}.

      While we have defined our topological field theory using the framework of discrete gauge theory, we can equivalently use integral, real-valued differential forms, as discussed in Section \ref{Sec:Diffforms}.

      The terms we have written above are complete for bosonic systems. For fermionic systems, a partial understanding can be achieved by
      changing the quantization of the integers $k_i$, to allow them to be half-integer; a complete understanding of this should be determined by
      group supercohomology \cite{Gu2014Supercoh,Wang2017Supercoh}. In the fermionic case, there may also be symmetry-enriched topological phases
      beyond group supercohomology, which cannot be fully described by the above effective action. We leave a comprehensive understanding of the fermionic
      case for future work. 
  
      Note that the above action is only uniquely defined when the gauge fields are flat: $da^I, dA, dC, d\vec{\cancel{R}}_i \in 2\pi \Z$.
      When the gauge fields are not flat, the action is not invariant under the shift of $a$, $A$, or $C$ by $2\pi$ on a single 1-simplex.
      More generally, for non-flat gauge fields, one can add additional terms to the action which depend on the field strength and which are
      not uniquely specified \cite{Kapustin2014}. Non-trivial fluxes of $a$, $A$, $C$, and $\vec{\cancel{R}}$ can be included by treating them as
      punctures or holes in the spatial manifold around which the gauge fields have non-trivial holonomy, such that the gauge fields remain flat.        
      The above also implies the action is invariant under changes of lift $a_{ij} \rightarrow a_{ij} + 2\pi$ as long as
      $\vec{q},\vec{s},\vec{t},\vec{m}$ are integer vectors.

        In what follows, to read off physical properties, we use the fact that objects charged under $A$, $C$, $R$ correspond to
        $U(1)$ charge, angular momentum, and linear momentum. The generalized charges can be defined physically through the Berry phase
        obtained by adiabatically braiding charges around the associated fluxes. 
        
        \subsubsection{Charge vector $\vec{q}$}
        
        The charge vector $\vec{q} \in \Z^D$ assigns fractional electric charge
        \begin{align}
          Q_{\vec{l}} = \vec{q}^T K^{-1} \vec{l}
        \end{align}
          to the anyon $\vec{l}$.
          Alternatively, this term induces an anyon $\vec{q}$ under insertion of $2\pi$ flux. As such,
          $\vec{q}$ is also sometimes referred to as a vison or fluxon. 

          Two charge vectors $\vec{q}, \vec{q'}$ describe the same anyon if $\vec{q'} = \vec{q} + K \vec{\Lambda}$ for some $\vec{\Lambda} \in \Z^D$. Therefore the group of inequivalent choices for $\vec{q}$ is $\A$. Note that for a fixed state, this equivalence is realized in the effective action by relabelling $\vec{a} \rightarrow \vec{a} - \vec{\Lambda} A$. Shifting $\vec{q}$ thus also changes the values of $k_1,k_2,k_4$ and $k_6$, which couple $A$. The full equivalence relation is
          \begin{align}
            (\vec{q};k_1,k_2,k_{4,i},k_6) \sim (&\vec{q} + K \vec{\Lambda}; k_1 - \vec{q} \cdot \vec{\Lambda} - \vec{\Lambda}^T K \vec{\Lambda}/2,
            \nonumber \\
            & k_2-\vec{s} \cdot \vec{\Lambda}, k_{4,i}-\vec{t}_i \cdot \vec{\Lambda}, k_6 - \vec{m} \cdot \vec{\Lambda}).
          \end{align}
          
          \subsubsection{Discrete spin vector $\vec{s}$}
          
          The discrete spin vector $\vec{s} \in \Z^D$ is the analog for discrete rotational symmetry of the spin
          vector defined previously for continuum FQH systems \cite{Wen1992shift}. However, as we discuss below, this term is
          only non-trivial when there is a compatibility between the intrinsic topological order and the order $M$ of the point group symmetry. 

          This term induces an anyon $\vec{s}$ under the insertion (fusion) of $M$ elementary disclinations. In particular,
          this term contributes a phase $e^{i2\pi\vec{s}^TK^{-1} \vec{l}}$ to the adiabatic transport of an anyon $\vec{l}$ around
          $M$ elementary disclinations. Alternatively, this term associates a fractional orbital angular momentum
        \begin{align}
         L_{\vec{l}} = \vec{s}^T K^{-1} \vec{l}
          \end{align}
          to the quasiparticle $\vec{l}$, which contributes a braiding phase $e^{2\pi i L_{\vec{l}} /M }$ to an anyon $\vec{l}$ encircling a $2\pi/M$ disclination.

          Consider a continuum FQH state where we adiabatically transport an anyon $\vec{l}$ around a region $\Sigma$ of a manifold
          with curvature. The resulting Aharonov-Bohm phase $\gamma_{AB} = \gamma_{AB,1} + \gamma_{AB,2}$ receives two
          contributions \cite{gromov2016}. The first contribution $\gamma_{AB,1}$ is associated to the fractional $U(1)$ charge of $\vec{l}$ and equals 
          \begin{align}
            \gamma_{AB,1} &= Q_a \Phi(\Sigma),
          \end{align}
          where $\Phi(\Sigma)$ is the total magnetic flux through $\Sigma$.
          The second contribution is due to coupling to the spatial curvature:
           \begin{align}
            \gamma_{AB,2} &= \left(\frac{\vec{l}^TK^{-1}\vec{l}}{2}+\vec{l}^TK^{-1}\vec{s}\right) N_R(\Sigma)
          \end{align}
         
          Here $N_R(\Sigma)$ is the integrated curvature flux through $\Sigma$. The quantity in parantheses defines the total spin of $\vec{l}$,
          \begin{align}
            S_{\vec{l}} = L_{\vec{l}} + \frac{\vec{l}^TK^{-1}\vec{l}}{2} .
          \end{align}
          The first contribution is the orbital angular momentum, which comes from the symmetry fractionalization, and
          can be understood as the braiding of $\vec{l}$ with the anyon $\vec{s}$ associated to a $2\pi$ curvature flux.
          The second contribution arises because of self-interaction effects that result in the anyon $\vec{l}$ braiding
          around itself as it is transported around a closed loop.  For an explicit calculation of the full A-B phase in a
          continuum geometry the reader is referred to Refs. \cite{gromov2016,EINARSSON1995}.

          In the discrete case that we are considering in this paper, the same equations are expected to hold, with the modification that
          the curvature $N(\Sigma)$ arises only due to point sources of $2\pi/M$ curvature flux arising from disclinations and corners. 
          
          Note that taking $\vec{s} = M \vec{\Lambda}$ for $\vec{\Lambda} \in \Z^D$ is trivial, since it can be completely accounted for
          by binding an anyon $\vec{\Lambda}$ to an elementary disclination, which can in turn always be done by adjusting the
          local energetics at disclinations. The non-trivial case cannot be captured simply by
          associating an anyon to an elementary disclination. Therefore we
          have two equivalence relations:
          \begin{align}
            (\vec{s},\{k_i\}) \sim (\vec{s} + K \vec{\Lambda},\{k_i'\})
          \end{align}
          (by relabelling $a \rightarrow a - \vec{\Lambda} C$), and
          \begin{align}
              (\vec{s},k_3) \sim (\vec{s} + M \vec{\Lambda}',k_3).
          \end{align}
          The choices of $\vec{s}$ inequivalent under both relations constitute the group $\A / M \A$. For
          $\A = \Z_{n_1}\times\dots\times\Z_{n_D}$, $\A/M\A = \Z_{(n_1,M)}\times\dots\times\Z_{(n_D,M)}$, where $(n,M) = \gcd (n,M)$.
          We see, therefore, that the order of the group $\A$ must be compatible with $M$ to obtain a non-trivial fractionalization class. 

          The equivalence on $\vec{s}$ implies that the theory predicts the angular momentum of an anyon
          $\vec{l}$ modulo $M (\vec{\Lambda}^T K^{-1} \vec{l})$.
		
          \subsubsection{Discrete torsion vector $(\vec{t}_x,\vec{t}_y)$}
          The integer vector $(\vec{t}_{x},\vec{t}_{y})$, with $\vec{t}_i \in \Z^D$, which we refer to as the discrete torsion vector,
          does not have an analog in the continuum because torsion (i.e. the gauge-invariant part of the dislocation density)
          is not quantized in continuum geometry. Furthermore this term is
          non-trivial only in the presence of rotational symmetry, with $M = 2,3,4$, because, as summarized in Table \ref{table:KM}, the
          gauge-invariant part of the dislocation density (defined by the group grading $K_M$) is nontrivial only when $M = 2,3, 4$.

          $\vec{t}$ associates an anyon $(\vec{t}_x,\vec{t}_y) \cdot (a,b)$ to a region with Burgers vector $(1 - U(2\pi/M)) \cdot (a,b)$. Note that
          an anyon is attached only for Burgers vectors in the trivial class in $K_M$. Values of $\vec{t}$ which can be accounted for by
          attaching an anyon to an elementary dislocation are topologically trivial, as they can be accounted for by adjusting the
          local energetics of a dislocation. It follows that the topologically distinct values
          of $(\vec{t}_x, \vec{t}_y)$ are classified by the group $K_M \otimes \A$, which for $M=2,3,4$ equals $\A/2\A\times\A/2\A, \A/3\A$
          and $\A/2\A$ respectively (see Appendix \ref{Appendix:GrpCoho} for a definition of the symbol $\otimes$).

          The term defining the torsion vector can be written in full as $\frac{1}{2\pi} a^I \cup (t_{I,i} (1-U(2\pi/M))^{-1}dR_i)$.
          From this we obtain that the discrete torsion vector furthermore associates a fractional (linear) momentum
        \begin{align}
          \vec{P}_{\vec{l}} &=  (1 - U(2\pi/M)^T)^{-1}\vec{p}_{\vec{l}}
          \nonumber \\
          (\vec{p}_{\vec{l}})_i &= \vec{l}^T K^{-1} \vec{t}_i
          \end{align}
          to the anyon $\vec{l}$, which is well-defined (i.e. topologically robust) modulo the equivalence on $\vec{t}$:
          \begin{align}
            \vec{t}_I &\sim \vec{t}_I + (1-U(2\pi/M)) \Z^2 , \;\;\text{ for } I = 1,\cdots, D
            \nonumber \\
            \vec{t}_{i} &\sim \vec{t}_i + K \Z^D , \;\; \text{ for } i = x,y).
          \end{align}          
          The momentum $\vec{P}_{\vec l}$ of an anyon can be defined by the Berry phase
          $e^{2\pi i \vec{P}_{\vec{l}} \cdot \vec{b}}$ obtained upon adiabatically braiding the anyon
          $\vec{l}$ around a dislocation with Burgers vector $\vec{b}$.  

          Under a $2\pi/M$ rotation, the momentum transforms as $\vec{P}^T_{\vec{l}} \rightarrow  P^T_{\vec{l}}U(2\pi/M) $;
          in other words, under a $2\pi/M$ rotation, the change in momentum is
          $ (U^T(2\pi/M) - 1) \vec{P}_{\vec{l}} = - \vec{p}_{\vec{l}}$.
          However this is precisely the first ambiguity in $\vec{t}_I$: shifting $\vec{t}_I \rightarrow \vec{t}_I +
          (1-U(2\pi/M))(-\vec{t}_I)$ changes the momentum $\vec{P}_{\vec{l}}$ by $-\vec{p}_{\vec{l}}$.
          Therefore, the topologically robust part of the fractional momentum $\vec{P}_{\vec{l}}$ is consistent with rotational invariance. 
          
          We emphasize that this ``crystal momentum fractionalization,'' which is only non-trivial for $M = 2,3,4$,
          is fundamentally distinct from the more
        familiar notion usually discussed in the context of quantum spin liquids (see e.g. \cite{Essin2013SF,Essin2014spect}). The latter case is
        associated with non-commutativity of the translation operator restricted to a given anyon and arises from the existence of an anyon
        per unit cell (discussed below), which can be non-trivial even in the case $M =1$.

        \subsubsection{Area vector $\vec{m}$}
        
 	Finally, $\vec{m} \in \Z^D$, which we refer to as the area vector, also has no analog in the continuum.
        This associates an anyon $\vec{m}$ per unit cell, as has been discussed algebraically in previous work \cite{Cheng2016,Lu2017fillingenforced} and
        gives rise to certain notions of ``crystal momentum fractionalization'' discussed previously \cite{Jalabert1991,Sachdev_2018,sachdev1999translational,Essin2013SF,Essin2014spect}.
        This means that if a quasiparticle $\vec{l}$ is taken around a region $S$ containing $\text{Num}(S)$ unit cells,
        the wave function acquires a braiding phase $e^{2\pi i \vec{l}^T K^{-1} \vec{m} \text{Num(S)}}$. Algebraically, this means that the
        translation operators satisfy a
        magnetic translation algebra when its action is restricted to the anyon $\vec{l}$:
        \begin{align}
          T_{x, \vec{l}} T_{y, \vec{l}} = e^{2\pi i \vec{l}^T K^{-1} \vec{m}  } T_{y, \vec{l}} T_{x, \vec{l}} ,
        \end{align}
        where $T_{x, \vec{l}}$ and $T_{y,\vec{l}}$ are the translation operators in the $x$ and $y$ direction, restricted to the anyon $\vec{l}$.
        See Ref. \cite{Barkeshli2019} for a
        precise formulation of symmetry operations restricted to anyons. 

  \subsection{Response theory}
  \label{Sec:RT}

  Given the topological effective action, we can integrate out the dynamical $a$ gauge fields to obtain an effective response theory:
\begin{widetext}
  \begin{align}
    \label{responseAction}
\L_{\text{eff}} &= \frac{\sigma_H}{2} A \cup dA + \frac{\mathscr{S}}{2\pi} A \cup dC + \frac{\ell_s}{4\pi} C \cup dC  
+ \frac{\check{\vec{P}}_c }{2\pi} \cdot (A \cup d\vec{R}) +\frac{\check{\vec{P}}_s}{2\pi} \cdot (C \cup d\vec{R}) 
+ \frac{1}{2\pi}\left(\nu_c A + \nu_s C\right) \cup A_{XY} \nonumber \\
    & + \frac{\vec{\nu}_p}{2\pi} \cdot \vec{R} \cup A_{XY} + \frac{\Pi_{ij}}{4\pi} R_i \cup d R_j + \frac{\alpha}{4\pi} A_{XY} \cup d^{-1} A_{XY} + \mathcal{L}_{\text{anom}} ,
\end{align}
\end{widetext}
where
\begin{align}
\mathcal{L}_{\text{anom}} = - \frac{\text{sgn}(K)}{48\pi} C \cup dC .
\end{align}
Note that as usual, the effective response theory is not well-defined on compact manifolds due to the fractional values of the coefficients; nevertheless, the response theory can be used to read off the fractionally quantized responses of the system on an open patch of space.

The first term is the well-known Hall conductivity, which is given by
\begin{align}
\sigma_H = (2k_1 + \vec{q}^T K^{-1} \vec{q})/2\pi
\end{align}

The second and third terms are discrete analogs of the known continuum geometric response of FQH states \cite{Wen1992shift, Avron1995hvisc, Tokatly2007, Cho2014,Gromov2014,Gromov2015,Bradlyn2015,Klevtsov2015}.

The remaining terms in $\L_{\text{eff}}$ are intrinsic to the lattice and have no analog in continuum FQH states. In what follows, we discuss them individually in detail.

We note that the term formally written as $A_{XY} \cup d^{-1} A_{XY}$, with $\alpha = \vec{m}^T K^{-1} \vec{m}$, corresponds to $A_{XY} \cup c$, where $d c = A_{XY}$. This
term arises from the fact that an anyon is associated with each unit cell. However it is not clear how or whether this term can be physically measured as a
quantized geometric response. We thus do not discuss this term further below. 

\subsubsection{Discrete shift $\mathscr{S}$ and fractional charge of disclinations}

The second term gives a discrete analog of the shift \cite{Wen1992shift, Biswas2016,Han2019,Liu2019ShiftIns} $\mathscr{S}$, where
\begin{align}
  \mathscr{S} &= k_2 + \vec{q}^T K^{-1} \vec{s}.
\end{align}
In particular, this term implies that lattice corners and disclinations carry fractional $U(1)$ charge. Both an elementary $2\pi/M$ disclination and a corner of angle $2\pi/M$ carry a fractional $U(1)$ charge of
\begin{align}
Q_{\text{disclin}, 2\pi/M} = \mathscr{S}/M = \frac{k_2 + \vec{q}^T K^{-1} \vec{s}}{M}.
\end{align}  
For example, if $M = 4$ and the system is defined at the surface of a 3D cube, there are effectively $8$ disclinations, each one carrying a fractional charge $\mathscr{S}/4$. If the system is defined on a square, each corner also has a fractional charge $\mathscr{S}/4$. This term therefore implies the system is a fractional ``higher order'' topological state \cite{You2018HOTI,Benalcazar2019HOTI,Rasmussen2020HOSPT}. Note that when the edge of the system is gapped, the corner charge is clearly well-defined; however when the corner lies along a chiral gapless boundary of the system, it is not clear whether any remnant of the corner charge persists. 

Since the $A \cup dC$ term defining the shift can also be written as $C \cup dA$, this term also associates an angular momentum to a $\phi$ flux given by
\begin{align}
 L_{A, \phi} = \frac{\phi}{2\pi} \mathscr{S} = \frac{\phi}{2\pi}(k_2 + \vec{q}^T K^{-1} \vec{s}).
\end{align}
Note that the fractional part of the angular momentum of a $2\pi$ flux equals $L_{\vec{q}} = \vec{q}^T K^{-1} \vec{s} \; (\text{mod } 1)$, which is the angular
momentum of the anyon $\vec{q}$ associated to a $2\pi$ flux.

Note that the response theory only predicts the fractional charge, angular momentum, and linear momentum of the dislocations and disclinations up to those of the elementary anyons, as anyons can always be bound to these defects by adjusting the local energetics. Therefore in this case, the fractional charge $\mathscr{S}/M$ is determined only modulo the charges $Q_{\vec{l}} = \vec{q}^T K^{-1} \vec{l}$, for any integer vector $\vec{l}$.

\subsubsection{Disclination angular momentum}

The third term contributes to a fractionally quantized contribution $L_{2\pi/M}$ to the  angular momentum of the elementary $2\pi/M$ disclination,
\begin{align}
  \label{discAM}
  L_{2\pi/M} &= \frac{\ell_s}{M} - \frac{1}{M}\frac{c}{12}
  \nonumber \\
  \ell_s &= (2k_3 + \vec{s}^T K^{-1} \vec{s})
\end{align}
The contribution proportional to the chiral central charge $c = \text{sgn}(K)$, where $\text{sgn}(K)$ is the signature of $K$, arises from the framing anomaly, $\mathcal{L}_{\text{anom}}$, which we discuss further in Sec. \ref{Sec:GravAnom}.

\subsubsection{Fractional quantized charge polarization $\check{\vec{P}}_c$ and fractional charge of dislocations}

The term with
\begin{align}
                     \check{\vec{P}}_c &= (1 - U^T(2\pi/M))^{-1} (\vec{k}_4 + \vec{p}_{\vec{q}})
                     \nonumber\\
  &= \vec{P}_{\vec{q}} + (1 - U^T(2\pi/M))^{-1} \vec{k}_4 
\end{align}
is referred to as a fractionally quantized charge polarization. As we discuss, this leads to three basic properties that are predicted by the topological response theory:
\begin{enumerate}
\item Fractionally quantized charge of dislocations (modulo charge of anyons)
\item Fractionally quantized momentum of $U(1)$ flux
\item Fractionally quantized charge per unit length along boundaries (modulo charge of anyons) 
\end{enumerate}
The quantization arises due to the rotational symmetry of the lattice. Without rotational symmetry ($M = 1$), the polarization is a non-quantized topological response \cite{song2019electric}.  Furthermore, $\check{P}_{c,j}$ is only well-defined modulo $\Z$.

This term associates a fractional charge
\begin{align}
Q_{\text{disloc}; \vec{b}} = \check{\vec{P}}_c \cdot \vec{b}
  \end{align}
  to a dislocation with Burgers vector $\vec{b}$. Note that, as in the case of the disclination charge, the topological response theory only predicts the dislocation charge modulo the charges of the anyons.

  Whether the fractional charge of a dislocation is non-trivial because of non-trivial values of the discrete torsion vector is a somewhat subtle issue. In principle, the dislocation charge can be fractions of the minimal anyon charge even when the discrete torsion vector is trivial, due to the interplay between the SPT term $\vec{k}_4$ and the minimal anyon charge. Observe that the fractional charge receives two contributions: one from the intrinsic topological order and symmetry fractionalization, which arises from $\vec{P}_{\vec{q}}$, and one from the
  SPT term $k_4$. The SPT term
  can contribute a fractional charge in multiples of $1/2$ (for $M = 2, 4$) or $1/3$ (for $M = 3$). Together with the charge of the anyons $Q_{\vec{l}}$ which can be trivially bound to
  dislocations due to local energetics, this implies that in principle one can obtain fractional charges at dislocations that may be fractions of the anyon charge, but which arise from a
  trivial value of $\vec{t}$. For example, consider the case of the $1/2$ Laughlin topological order on a honeycomb lattice ($M = 3$) and $k_4 = 1$. There, all choices of discrete
  torsion vector $\vec{t}$ are trivial, because $\Z_3 \otimes \Z_2 = \Z_1$; nevertheless, even a trivial value of $\vec{t}$ can give rise to a dislocation charge $1/3 - 1/2 = 1/6$. On the
  other hand, on the square lattice ($M =4$), a dislocation charge of $1/4$ can only occur for the non-trivial choice of discrete torsion vector $\vec{t} \in \Z_2 \otimes \Z_2 = \Z_2$, while
  the trivial choice can only give multiples of $1/2$.   

Let us compare the charge of a dislocation with Burgers vector $\vec{b}$ and its rotated counterpart $U(2\pi/M) \vec{b}$. The difference is given by
\begin{align}
  \Delta Q &= Q_{\text{disloc}; \vec{b}} - Q_{\text{disloc}; U(2\pi/M) \vec{b}}
             \nonumber \\
           &= \check{\vec{P}}_c^T (1 - U(2\pi/M)) \vec{b}
             \nonumber \\
  &= (k_{4;i} + \vec{q}^T K^{-1} \vec{t}_i) b_i
  \nonumber \\
           &= (k_{4;i} + Q_{\vec{t}_i}) b_i.
\end{align}
In other words, the difference is given in integer multiples of the fractional charge of $\vec{t}_x$ and $\vec{t}_y$. Thus the contribution to the dislocation charge from the topological response theory, which is only well-defined modulo the charges of the anyons, is rotationally invariant.

The charge polarization term also contributes to the charge of a disclination, if it has nontrivial $\vec{R}$ holonomy. Interestingly, the classification of free fermion SPT phases based on their disclination charges is shown to equal $\Z_M \times K_M$ in Ref. \cite{Li2020disc}. This agrees with the bosonic crystalline gauge theory, which predicts that the disclination charge is classified by the terms $\frac{1}{2\pi} A \cup (k_2 dC + \vec{k}_4 \cdot d\cancel{\vec{R}})$, where $k_2 \in \Z_M$ and $\vec{k}_4 \in K_M$. (The coefficients in a crystalline gauge theory of fermions can in principle have different quantization conditions than in the bosonic case, but we will not discuss the fermionic case in detail here.) The classification approach in Ref. \cite{Li2020disc} based on Wannier orbitals centred at high-symmetry points is an example of a defect network construction. The problem of establishing a correspondence between the topological responses in the defect network picture and the group cohomology picture is briefly alluded to in Section \ref{Sec:Disc}.

If the dislocation described by $\vec{b}$ is connected to an edge of the system, the holonomy at the edge is changed by the amount $-\vec{b}$. Hence there must be a compensating fractional charge at the edge. However since the dislocation line is not by itself well-defined, this boundary fractional charge can be delocalized along the boundary. 

  This term also associates a momentum
  \begin{align}
\vec{P}_{A, \phi} = \check{\vec{P}}_c \phi/2\pi
    \end{align}
    to a $U(1)$ flux of $\phi$ spread uniformly throughout the system. The momentum of $2\pi$ flux has been discussed previously in the context of Dirac spin liquids in Refs. \cite{Song_2019monopole,Song_2020monopole}; our results are consistent with these works for systems with orientation-preserving symmetries. Note that the contribution to $\check{\vec{P}}_c$ from the intrinsic topological order is equal to the momentum $\vec{P}_{\vec{q}}$ of the anyon $\vec{q}$, which is the anyon associated to a $2\pi$ flux.  

    Finally, this term associates a fractional charge per unit length $\check{\vec{P}}_c \cdot \hat{e}$ to a boundary along the direction $\hat{e}$. This corresponds to a fractional charge polarization $\vec{P}_c = \check{\vec{P}}_c \times \hat{z}$ for a system defined on a space with boundary. As above, this fractional charge per unit length is only topologically robust modulo the charge of the anyons. Under a rotation, the charge per unit length along the boundary stays invariant up to the charge of the elementary anyons. Therefore the contribution of the topological response theory to the boundary charge per unit length is rotationally invariant.

 We note that because the boundary charge per unit length is only topologically protected modulo the charge of the anyons, the system does not necessarily have a non-zero polarization on a space with boundary; one can arrange the local energetics along the boundary so that the boundary charge per unit length is the same on all boundaries. Nevertheless, the three physical effects described above are all intimately related to the quantum theory of polarization in higher dimensions \cite{song2019electric}, which is why we refer to this term as the fractional charge polarization. 

\begin{figure*}
	\centering
	\includegraphics[width=18cm,height=10cm]{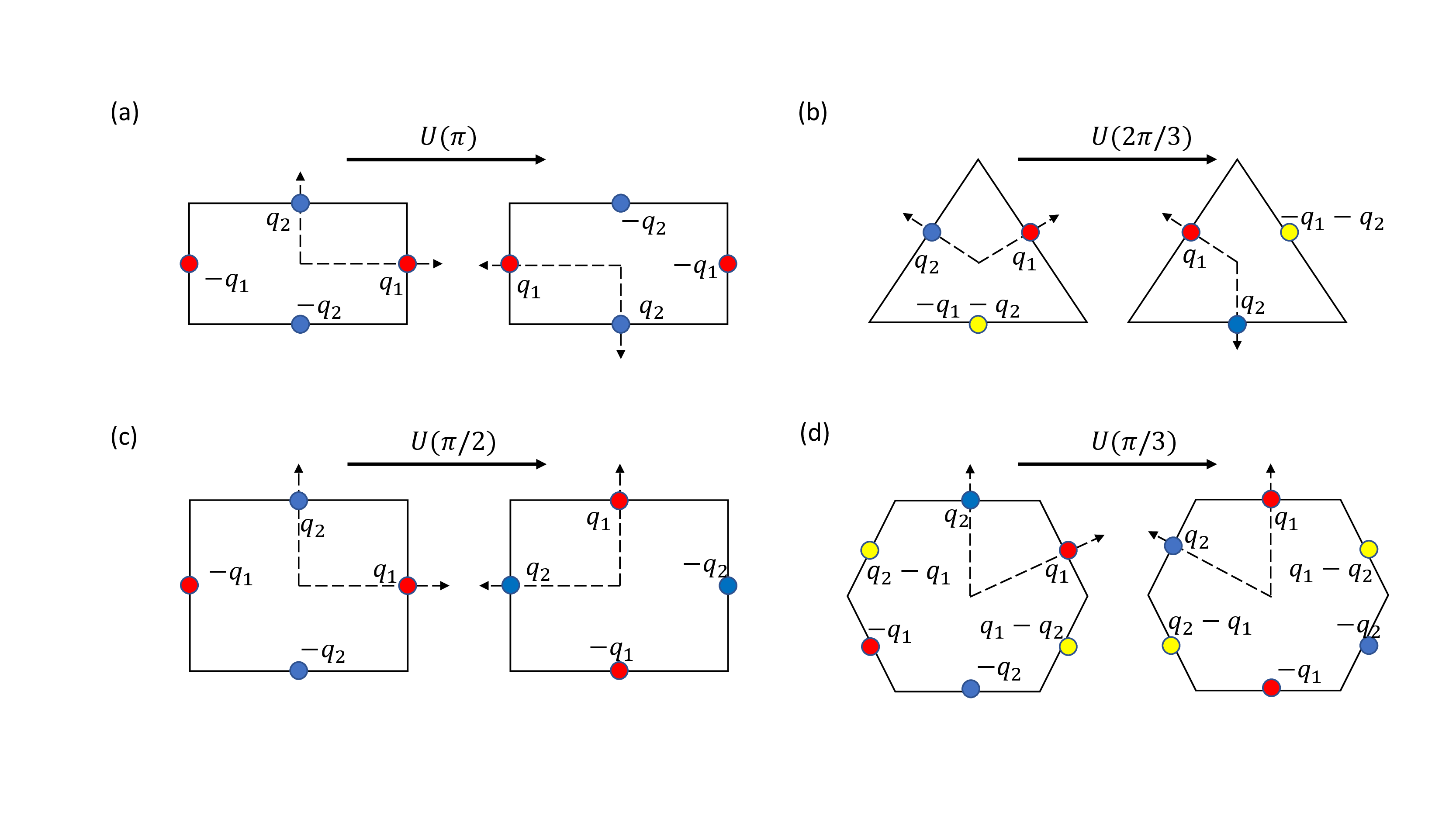}
	\caption{The $K_M$ classification of rotationally symmetric configurations of boundary charge for (a) $M=2$, (b) $M=3$, (c) $M=4$, and (d) $M=6$. We choose our coordinate axes to be normal to the boundaries, and place a charge per unit length equal to $(q_1,q_2)^T \vec{n}$ on the boundary with normal vector $\vec{n}$. Thus in (b), for $M=3$ we have the arrangement $A = (q_1,q_2,-q_1-q_2)$ as we proceed anticlockwise around the boundary segments. Now under a $2\pi/3$ rotation of axes, the charge per unit length at the same three segments gets redefined as $A' = (-q_1-q_2,q_1,q_2)$. Since the fractional charge per unit length on each boundary segment remains the same if we only rotate the coordinate axes, we should have $A = A' \mod 1$. This implies that $q_1 = q_2$ and $3q_1 \in \Z$; the three distinct choices of $q_1$ now determine the group $K_3$. We can follow similar reasoning in (a),(c) and (d).}     
	\label{Fig:BdryCharge}
\end{figure*}

The polarization response can be used to obtain another simple way to understand the group $K_M$. The group $K_M$ corresponds to the group of allowed fractional charges per unit length along the boundary when the bulk has no intrinsic topological order, as we explain below. 

Consider a system with fractional charge per unit length along its boundary given by $\vec{P} \cdot \hat{n}$, where $\vec{P}$ is the polarization vector and $\hat{n}$ is the normal to the boundary. An integer value of $\vec{P}$ corresponds to placing an integer charge per unit length on the boundary, which can always be done locally. This is shown pictorially in Fig. \ref{Fig:BdryCharge}, where we assign fractional charge per unit length to each boundary segment under one choice of coordinate axes. For example, if we consider a system with $M=4$, the charge per unit length on the boundaries normal to $\hat{x},\hat{y},-\hat{x},-\hat{y}$ are $(q_1,q_2,-q_1,-q_2)$ respectively. Now we can perform rotations of the axes by $2\pi/M$, which will relabel the charge on each segment since the normal vectors $\vec{n}$ get redefined. In this case, the coordinate axes are rotated by an angle $\pi/2$, and the charges on the same boundary segments will now be labelled as $(-q_2,q_1,q_2,-q_1)$ (see Fig. \ref{Fig:BdryCharge}).  However, the fractional charge on each edge should be the same from either calculation. Therefore we must have $(q_1,q_2,-q_1,-q_2) = (-q_2,q_1,q_2,-q_1) \mod 1$. We can see that the only solutions are $(q_1,q_2) = (0,0)$ or $(1/2,1/2)$. Therefore the group of distinct assignments of charge at the boundary is $K_4 \cong \Z_2$. One can work out the other cases similarly. 

We note that in our initial discussion of Section \ref{Sec:KM}, the $K_M$ classification arose from general properties of the dislocations that do not depend on a particular Lagrangian, while in the second derivation given here, it arose from demanding rotational invariance of a physical response related to the term $\frac{\check{\vec{P}}}{2\pi} A \cup dR$ in the Lagrangian. 

Finally, we look at the case with $M=1$, corresponding to the absence of rotation symmetry. We cannot directly apply the previous reasoning in this case to obtain a useful classification. In a system without rotation symmetry, the Burgers vector of any dislocation is well-defined: the value of $d\vec{R}$ is gauge-invariant. Since there is no gauge transformation relating them, there is no grading of Burgers vectors. In the example of boundary charge, one can now have any assignment of fractional charges per unit length on the boundary of such a system. In either case, the group classifying inequivalent dislocations or fractional boundary charge configurations is not a finite group. However, if we define $K_1$ so that it classifies the \textit{quantized} fractional charges per unit length that can be assigned to a boundary, the group is trivial. The quantization was a direct result of discrete rotation symmetry, which is broken when $M=1$.

\subsubsection{Quantized angular momentum polarization $\check{\vec{P}}_s$}

The term with $\check{\vec{P}}_s$ is the rotational analog of $\check{\vec{P}}_c$, where
\begin{align}
                      \check{\vec{P}}_s &= (1 - U^T(2\pi/M))^{-1} (\vec{k}_5 + \vec{p}_{\vec{s}})
                     \nonumber\\
  &= \vec{P}_{\vec{s}} + (1 - U^T(2\pi/M))^{-1} \vec{k}_5 
\end{align}
It associates a fractionally quantized angular momentum
\begin{align}
L_{\text{disloc}; \vec{b}} =\check{\vec{P}}_s \cdot \vec{b}
\end{align}
to a dislocation with Burgers vector $\vec{b}$. Note that the contribution to $\check{\vec{P}}_s$ coming from the symmetry fractionalization is simply
$\vec{P}_{\vec{s}}$, the linear momentum of $\vec{s}$.

However it is not clear whether dual response, which is the analog of attaching momentum to a $U(1)$ flux, which here would formally correspond to a momentum of a disclination, is well-defined. It is also unclear whether the analog of the boundary charge per unit length has any meaning in this context, because the boundary is not fixed by a rotation. 

\subsubsection{Charge, linear momentum, and angular momentum filling: $\nu_c,\nu_s, \vec{\nu}_p$}

The term proportional to $A \cup A_{XY}$ corresponds to a charge of
\begin{align}
\nu_c = k_6 + \vec{q}^T K^{-1} \vec{m}
\end{align}
per unit area. This gives a generalized Lieb-Schulz-Mattis constraint \cite{Cheng2016} which imposes constraints on $\vec{q}$, $K$, and $\vec{m}$ in terms of the filling $\nu_c$. Likewise, the term proportional to $C \cup A_{XY}$ associates a fractional angular momentum of
\begin{align}
\nu_s = (k_7 + \vec{s}^T K^{-1} \vec{m})
\end{align}
to each unit area.

The term $\vec{R} \cup A_{XY}$ associates a momentum of
\begin{align}
 (\vec{\nu}_{p})_j = \sum_i \vec{t}_i^T K^{-1} \vec{m} (1- U(2\pi/M)^{-1}_{ij}
\end{align}
per unit area of the system. It arises from the fact that there is an anyon $\vec{m}$ per unit cell, which in turn carries a momentum as specified by the coupling $\vec{t}$. Indeed, observe that
\begin{align}
\vec{\nu}_p = \vec{P}_{\vec{m}} . 
\end{align}
Remarkably, this implies that the ground state may carry momentum, depending on the area of the system; only for certain commensurate areas is the ground state momentum trivial. Further, this term is also only non-trivial for $M = 2,3,4$-fold rotational symmetry. 

\subsubsection{Fractionally quantized torsional response}

The term with $\Pi_{ij}$ associates a fractionally quantized momentum of
\begin{align}
\vec{P}_{\text{disloc}, \vec{b}} = \Pi \vec{b}
\end{align}
to a dislocation with Burgers vector $\vec{b}$. Here
\begin{align}
  \Pi_{ij} &= e_i^T K^{-1} e_j,
  \nonumber \\
  e_i &= (1- U(2\pi/M))^{-1}_{ii'}\vec{t}_{i'}
\end{align}
This is closely related to the torsional Hall response that has been discussed for continuum Dirac theories \cite{Hughes2011thv,Hughes2013thv}, although there the corresponding term is not quantized and is sensitive to the ultraviolet cutoff. The non-trivial quantization only occurs for lattice systems with $M = 2,3,4$-fold rotational symmetry.

We note that here we read off momentum as being defined by the charge of the translation gauge field. It is not clear how to define the momentum of a dislocation microscopically. For example, naively one would define the charge in terms of the Aharonov-Bohm phase obtained by braiding with a flux; in this case this naively corresponds to the phase obtained by braiding dislocations around each other. However to define this microscopically, the restricted mobility of the dislocations on a lattice with a conserved number of atoms must be taken into account. 

\subsubsection{Framing anomaly}
\label{Sec:GravAnom}

We note that the topological field theory itself does possess a continuous space-time symmetry corresponding to diffeomorphism invariance, which corresponds to the retriangulation invariance of the path integral for a given fixed configuration of flat gauge fields. For chiral topological phases, a gravitational CS term, proportional to the chiral central charge $c$, for the full $SO(2,1)$ spin connection $\Omega$ also arises upon evaluating the path integral. This arises from the implicit metric dependence in the path integral measure required for gauge fixing and quantizing the CS theory, and is referred to as the framing anomaly \cite{witten1989,Gromov2015}. In a continuum formulation, this is written as:
\begin{align}
\mathcal{L}_{\text{anom}} = -\frac{c}{96\pi} \text{Tr } \left( \Omega d \Omega + \frac{2}{3} \Omega^3 \right) .
\end{align}
This term may also be viewed as the gravitational anomaly of the (1+1)D boundary of the system, which hosts a chiral CFT with central charge $c$.

We note that the quantization of the CS theory also gives rise to another contribution to the effective theory, given by the Ray-Singer analytic torsion \cite{witten1989}. This term is a topological invariant of the underlying space-time manifold, and is unimportant for our discussion. 

Mathematically we may consider $\Omega$ to be a separate quantity depending on an underlying space-time metric, and to be distinct from $C$ and $\vec{R}$. However to be physically meaningful, the space-time manifold $\mathcal{M}$ should split into space and time separately as assumed in this work, with the time-components of $\Omega$ vanishing:
\begin{align}
  \Omega^a_{0,\mu} &= \Omega^0_{b,\mu} = 0
  \nonumber \\
  \omega_\mu &\equiv \Omega^{1}_{2,\mu} 
\end{align}
Furthermore, the spatial component of $\Omega$, which we have denoted $\omega_\mu$, is an $SO(2)$ gauge field whose field strength corresponds to the curvature of the system. The physical origin of this curvature in a lattice system arises from lattice disclinations, so we require that $\omega$ should be determined by the lattice rotation gauge field $C$. We can relate the continuum definition of $\omega$ to the definition of $C$ on the triangulation by integrating over a $1$-simplex $[ij]$ of the triangulation:
\begin{align}
  \label{Comega}
  \int_{ij} \omega_{\mu} &= C_{ij}
\end{align}

We see therefore that the framing anomaly contributes the following term to the effective response theory:
\begin{align}
\mathcal{L}_{\text{anom}} = -\frac{c}{48\pi} C \cup dC 
  \end{align}
This term will then contribute an additional angular momentum to disclinations proportional to $c$ (see Eq. \ref{discAM}). 

\subsubsection{Additional coboundary terms in response theory}

When we consider a system with only $U(1)$ charge conservation and $\Z^2$ translation symmetries (i.e. $M = 1$), the charge polarization corresponds to a non-quantized topological term \cite{song2019electric}.
In our notation, this term has the form $\frac{\check{\vec{P}}}{2\pi} \cdot \vec{R} \cup dA$ in $2+1$ dimensions, where $\check{\vec{P}}$ is a pair of real numbers defined modulo 1. This non-quantized term is not associated to symmetry fractionalization or to SPTs; rather than corresponding to a non-trivial 3-cocycle, the above term can be understood as a 3-coboundary of the group $U(1) \times \Z^2$. Nevertheless, such a term can be physically meaningful. This means that for a complete understanding of the topological terms, we should also study response terms that are not associated to SPT responses but which correspond to group 3-coboundaries. In this section we consider these possibilities when $G = U(1)\times G_{\text{space}}$, and the rotation symmetry is nontrivial.

We first note that in the presence of rotation symmetry, we do not find any non-quantized topological terms (i.e. terms that are retriangulation invariant in our simplicial formulation). For example, the nonquantized polarization term mentioned above becomes quantized as a result of the rotation symmetry. However, we do find that we can add certain additional quantized topological terms beyond the SPT terms in the effective action, Eq. \eqref{Eff_Action}. Although we have not explicitly found a coboundary representation for these cocycles, these terms correspond to coboundaries because the SPT terms already present in the effective action form a complete set of cocycle representatives of $\H^3(G,U(1))$ (see Appendix \ref{Appendix:GrpCoho}).

First consider the response term $\frac{\Pi_{ij}}{4\pi} R_i \cup dR_j$, where $\Pi_{ij} = (1-U^T(2\pi/M))^{-1} (t_i^T K^{-1} t_j) (1-U^T(2\pi/M))^{-1}$. This coefficient can be modified in a manner that preserves gauge-invariance, as follows: we can define
\begin{equation}
\tilde{\Pi}_{ij} \equiv (1-U^T(2\pi/M))^{-1} ((\vec{t}_i)^T K^{-1} \vec{t}_j + k_{ij}) (1-U^T(2\pi/M))^{-1}
\end{equation}
where $k_{ij} \in \Z$. In some cases, this shift in the momentum of a dislocation due to $k_{ij}$ can be considered to be trivial, and part of the equivalence in the definition of $\vec{t}_i$. However, in general this contribution may not be completely accounted for by the equivalences on $\vec{t}_i$. 

Similarly, consider the response term $\frac{\vec{\nu}_{p}}{2\pi} \cdot \vec{R} \cup A_{XY}$. We can modify the coefficient of this term as follows:
\begin{equation}
 (\tilde{\vec{\nu}}_{p})_j = \sum_i (\vec{t}_i^T K^{-1}\vec{m} + k_i) (1- U(2\pi/M))^{-1}_{ij} 
\end{equation}
where $k_i \in \Z$.

Finally, in principle we can have terms which are not related to the response terms already present in Eq. \eqref{responseAction}. For example, we can consider terms proportional to $R \cup R \cup R$, or terms composed of various powers of $A,R$ and $C$. Most terms of this kind will not be topological, i.e. will not satisfy the requirement of retriangulation invariance. Those terms that are retriangulation invariant will be coboundaries or equivalent to one of the existing SPT terms, since we already have a complete set of SPT cocycles. To our knowledge, none of these terms are associated to non-quantized topological responses. However, we have not checked all the possibilities systematically.

\section{Examples}
\label{Sec:Examples}
\subsection{$1/2$ Laughlin topological order}

Consider the 1/2 Laughlin topological order on a square lattice ($M = 4$), with symmetry $U(1)\times [\Z^2\rtimes\Z_4]$. We have $K = 2$ and $\A = \Z_2$, with the anyons given by $I = 0 \; (\text{mod } 2)$ and $S = 1 \; (\text{mod } 2)$. The symmetry fractionalization classification is $\H^2(G,\A) \cong \Z_2^4$, with $\A/4\A \cong K_4\otimes\A \cong \Z_2$. Thus there are two inequivalent symmetry fractionalization classes associated to each of $q,s,\vec{t}$ and $m$. Throughout this discussion we will define the elementary rotation matrix as $U(\pi/2) = \begin{pmatrix}
0 & 1 \\ -1 & 0
\end{pmatrix}$. At various points, we will comment on the differences in the analysis when we consider different values of $M$.

The charge, spin and area vectors are each determined by choosing $q,s,m \in \{I,S\}$. The fractional charge and angular momentum of the anyon $a$ are thus given by
\begin{align}
 Q_{a} = \frac{q a}{2} \mod 1
\end{align}
and
\begin{align}
L_a = \frac{s a}{2} \mod 1
\end{align}
respectively.
The charge filling gives a LSM constraint on $m$ and $q$:
\begin{align}
  \nu_c = \frac{q m}{2} \mod 1 
\end{align}
Therefore half-filling (i.e. half charge per unit cell) necessarily fixes $q = m = S$, while integer filling requires at least one of $q$ or $m$ to be trivial. 

There are two inequivalent choices of discrete torsion vector, corresponding to $(t_x , t_y) = (I,I)$ and $(S,I)$, with $(I,I) \sim (S,S)$ and $(S,I) \sim (I,S)$. To see this, note that naively
the possible discrete torsion vectors are $(t_x,t_y) \in \{(I,I),(I,S),(S,I),(S,S)\}$. The equivalence condition satisfied by them is 
\begin{equation}
\binom{t_x}{t_y} \sim \binom{t_x}{t_y}+(1-U(\pi/2))\binom{t'_x}{t'_y} = \binom{t_x}{t_y}+\binom{t'_x-t'_y}{t'_x+t'_y}.
\end{equation}
This condition implies that the symmetry fractionalization class is completely determined by the value of $(t_x+t_y) \mod 2$. Therefore the assignments $(I,S),(S,I)$ are equivalent and correspond to nontrivial symmetry fractionalization, while the assignment $(S,S)$ is in fact trivial. The latter assignment is seen to be trivial because we have $\binom{1}{1} = (1-U(\pi/2)) \binom{1}{0}$, and thus $(S,S)$ corresponds to attaching the anyon $S^{b_x}I^{b_y}$ to a dislocation with Burgers vector $(b_x,b_y)$.

The momentum associated to each anyon $a$ is 
\begin{equation}
\vec{P}_{a} = (1-U(\pi/2))^{-1} \vec{t}^T \frac{1}{2} a = \frac{a}{4} \binom{t_x+t_y}{t_y-t_x}.
\end{equation}
Recall this momentum is only well-defined (i.e. topologically robust) up to the equivalences on $\vec{t}$ (and the representative $a$ of the anyon). Therefore for trivial choice of $(t_x, t_y)= (I,I) \sim (S,S)$, we have
$2 P_{a,i} \in \Z$ ; that is, we have $\vec{P}_I, \vec{P}_S = \binom{0}{0} \sim \binom{1/2}{0} \sim \binom{0}{1/2}$. Therefore half-integer momentum components should be regarded as trivial. Physically this can be understood from the fact that the change in the braiding phase between $a$ and an elementary dislocation can be compensated for by attaching a semion to the elementary dislocation. For the non-trivial choice of $(t_x,t_y) = (S,I) \sim (I,S)$, we have $\vec{P}_{S} = \binom{1/4}{-1/4} \sim \binom{1/4}{1/4}$. Observe that under a rotation, $2P_S$ is invariant modulo $1$. 

The above analysis shows that for the $1/2$ Laughlin state, a momentum whose components are integer or half-integer is an indication of trivial symmetry fractionalization: it corresponds to attaching the anyon $S$ to an elementary dislocation in some fixed direction. Thus, while considering some other $M$, we will continue to associate the momentum $1/2$ with trivial symmetry fractionalization. For the $1/N$ Laughlin state with $N$ even, an analogous argument would imply that a momentum of $1/N$ units corresponds to trivial symmetry fractionalization, and therefore it is enough to check whether $N P_{a,x}$ and $N P_{a,y}$ are nontrivial.

Let us consider the classification of spin vectors in more generality. For the 1/2 Laughlin state with $M$ even, the spin vector is always nontrivial if it equals $1 \mod 2$. However, for $M=3$, we have $s = 3s \mod 2$. This means that every spin vector can be thought of as associating the anyon $s$ to an elementary $2\pi/3$ disclination. Therefore in this case, the choice $s=1$ is in fact trivial. If we generalize to $1/N$ Laughlin states with $\A=\Z_N$, the number of distinct spin vectors equals $\Z_N/M\Z_N = \Z_{(M,N)}$.   

Next we discuss the fractionally quantized responses. The Hall conductivity is given by $\sigma_H = \frac{1}{2\pi}(q^2/2+2k_1)$, as usual.

The discrete shift is defined as $\mathscr{S} = \frac{q s}{2} + k_2$. Therefore the fractional charge associated to a $\pi/2$ disclination is
\begin{align}
Q_{\text{disclin},\pi/2} = \frac{\mathscr{S}}{4} = \frac{qs}{8} + \frac{k_2}{4}. 
\end{align}
Thus we see that shifting $k_2$ by an integer changes the fractional charge by $1/4$; shifting $k_4 \rightarrow k_4+4$ adds a trivial integer charge to the elementary disclination. Furthermore, when
$q = s = S$, we obtain a $1/8$ charge at the elementary disclination. 

The angular momentum of a $\pi/2$ disclination is 
\begin{equation}
  L_{\pi/2} = \ell_s/4 -c/48 = \frac{s^2}{8} + \frac{k_3}{2} - \frac{1}{48}
\end{equation}
where we have included the contribution $c/48$ from the central charge $c = 1$ which arises due to the framing anomaly. Note that the fractional part of the angular momentum remains the same when we shift $k_3 \rightarrow k_3+2$, even though $k_3$ has a $\Z_4$ classification.

The charge of a dislocation with Burgers vector $\vec{b}$ is $Q_{disloc, \vec{b}} = \check{\vec{P}}_c \cdot \vec{b}$, where
\begin{equation}
\begin{pmatrix} \check{P}_{c,x}\\ \check{P}_{c,y}\end{pmatrix} = \frac{1}{2}\begin{pmatrix} k_{4,x}+k_{4,y}\\ k_{4,x}-k_{4,y}\end{pmatrix} + \frac{q}{4}\begin{pmatrix} t_x+t_y\\ t_x-t_y\end{pmatrix}.
\end{equation} 
Observe that the SPT contribution from $\vec{k}_4$ can only take two inequivalent values: $(0,0)^T$ or $(1/2,1/2)^T$. This follows from demanding rotational invariance of the polarization up to integers, i.e. of $\check{\vec{P}}_c$ modulo integers. The non-trivial symmetry fractionalization ($q = S$ and $(\vec{t}_x, \vec{t}_y) = (S, I) \sim (I,S)$, then contributes $(1/4, 1/4)^T \sim (1/4, -1/4)^T$. Therefore dislocations can carry charge of $\pm 1/4$, even though the minimal anyon charge is $1/2$. A similar calculation can be performed for the angular momentum polarization. 

On a space with boundary, the non-trivial symmetry fractionalization class ($q = S$ and $(\vec{t}_x, \vec{t}_y) = (S, I) \sim (I,S)$) therefore contributes a charge of $1/4$ (mod $1/2$) per unit length along the boundary. The other symmetry fractionalization classes contribute a $0$ charge (mod $1/2$) per unit length along the boundary.

The momentum per unit cell is given by the momentum of the anyon per unit cell, $\vec{\nu}_p = \vec{P}_{\vec{m}} = \frac{m}{4}\begin{pmatrix} t_x+t_y \\ t_x-t_y \end{pmatrix}$. For trivial fractionalization (either $m$ or $\vec{t}$ trivial), $2\vec{\nu}_p$ is an integer vector. The non-trivial fractionalization gives rise to $\vec{\nu}_p = (1/4, 1/4) \sim (1/4, -1/4)$. 

Finally, we compute the momentum of a dislocation with Burgers vector $\vec{b}$. The $i$ component of the momentum equals $\Pi_{ij} b_j$, where $\Pi_{ij} = (1-U^T(\pi/2))^{-1}t_i^T K^{-1}t_j(1-U(\pi/2))^-1$. Thus we obtain
\begin{equation}
\Pi_{ij} = \frac{1}{8}\begin{pmatrix} (t_x+t_y)^2& t_x^2-t_y^2\\t_y^2-t_x^2 & (t_x-t_y)^2\end{pmatrix}.
\end{equation}
For nontrivial symmetry fractionalization , where $t_x \pm t_y$ is odd, we see that $\Pi_{ij}$ has diagonal components equal to $1/8$. On the other hand, if we have trivial symmetry fractionalization, the only possible values of the components are $0$ and $1/2$, which correspond to trivial values of crystal momentum, as discussed above.

Finally, we note that different choices of the parameters $k_i$ do not necessarily give different SET phases. This is because of redundancies that arise when we consider gauge field relabellings, as we disuss for the $1/2$ Laughlin state in Section \ref{Sec:Classif} and in Appendix \ref{Appendix_RelabellingExample}. 

\subsection{$\Z_2$ gauge theory}

In this example we consider the case where the intrinsic topological order is given by $\Z_2$ gauge theory (i.e. that of the $\Z_2$ toric code), which has $\A = \Z_2\times \Z_2$. This is the case relevant for gapped $\Z_2$ quantum spin liquids. 

The system is described by a $K$-matrix $K = \begin{pmatrix}0 & 2 \\2&0\end{pmatrix}$. The anyons are given by $I = (0,0)^T,e=(1,0)^T,m=(0,1)^T$ and $\psi = (1,1)^T$, and we
have $\vec{a}^T K^{-1}\vec{b} = \frac{a_1b_2+a_2b_1}{2}$. 

The symmetry fractionalization classes are specified by inequivalent choices of $\{\vec{q},\vec{s},\vec{t}_i,\vec{m}\} \in \A\times(\A/M\A)\times (K_M\otimes\A)\times\A$. Thus the classification of $\vec{q}$ and $\vec{m}$ is individually $\Z_2\times\Z_2$, irrespective of the value of $M$. Suppose we wish to determine $\vec{q}$. To do so, we first compute the fractional charge $Q_{\vec{l}} = \vec{q}^T K^{-1}\vec{l} = \frac{q_1l_2+q_2l_1}{2}$, for each anyon $\vec{l}$. From this data, we can uniquely determine the integers $q_I \mod 2$, which fix the charge fractionalization. A similar method allows us to determine the anyon $\vec{m}$. Together these determine the charge filling, which places a LSM-type constraint on the fractionalization data:
\begin{align}
\nu_c \text{ mod 1} = \vec{q}^T K^{-1} \vec{m} = \frac{q_1 m_2 + q_2 m_1}{2} . 
\end{align}

Next we turn to the spin vector. We have $\A/M\A = \Z_2\times\Z_2$ if $M$ is even, and is $\Z_1$ otherwise. As in the previous example, we see that the anyons $\vec{s}$ and $M\vec{s}$ are equivalent when $M=3$, so any choice of $\vec{s}$ can be understood in terms of attaching anyons to elementary disclinations in this case. Now suppose $M$ is even. When $\vec{s} = e$, the fractional angular momentum of each anyon (modulo 1), given by $\vec{s}^T K^{-1} \vec{l}$, equals
\begin{equation}
L_I = 0, L_e = 0, L_m = 1/2, L_{\psi} = 1/2.
\end{equation}
If we choose $\vec{s}=m$, a similar calculation yields
\begin{equation}
L_I = 0, L_e = 1/2, L_m = 0, L_{\psi} = 1/2;
\end{equation}
and choosing $\vec{s}=\psi$ gives
\begin{equation}
L_I = 0, L_e = 1/2, L_m = 1/2, L_{\psi} = 0.
\end{equation}
Note that measuring the angular momentum for any one anyon does not uniquely fix the value of $\vec{s}$. This result emphasizes that in general we need to know the angular momentum of \textit{every} anyon in order to fix the symmetry fractionalization class. All these calculations could formally be done in the same manner for $M$ odd; however, each set of angular momentum values thus calculated would correspond to trivial symmetry fractionalization.

For $M=2,3,4$, the distinct torsion vectors are classified by
\begin{align}
K_2\otimes(\Z_2\times\Z_2) &= \Z_2^4 \\
K_3\otimes(\Z_2\times\Z_2) &= \Z_1 \\
K_4\otimes(\Z_2\times\Z_2) &= \Z_2\times\Z_2
\end{align}
We can understand the $\Z_2^4$ classification as follows: when $M=2$, the anyons $\vec{t}_x$ and $\vec{t}_y$ are fixed independently, and each can be equal to $I,e,m$ or $\psi$. The equivalence relation on $\vec{t}_x,\vec{t}_y$ does not provide any additional constraint. 

For $M=3$, the equivalence relation is 
\begin{equation}
\binom{\vec{t}_x}{\vec{t}_y} \sim \binom{\vec{t}_x}{\vec{t}_y}+(1-U(2\pi/3))\binom{t'_x}{t'_y} = \binom{\vec{t}_x}{\vec{t}_y}+\binom{t'_x-t'_y}{t'_x+2t'_y}
\end{equation}
(see Appendix \ref{Appendix:PtGrpMatrices} for the explicit forms of the rotation point group matrices). Notice that every anyon can be written in the form $\binom{t'_x-t'_y}{t'_x+2t'_y} \mod 2$. Therefore every assignment is trivial. In this case, although we can certainly adjust $\vec{t}_x,\vec{t}_y$ so as to obtain nontrivial values of momentum for the anyons, the symmetry fractionalization class is still trivial.

For $M=4$, we find, as in the previous example, that $\vec{t}_x$ and $\vec{t}_y$ are not independent: we can only fix $\vec{t}_x+\vec{t}_y \in \{I,e,m,\psi\}$. This leads to the $\Z_2\times\Z_2$ classification. Finally, for $M=6$, the torsion vectors are always trivial, irrespective of the structure of $\A$.

Let us consider below the fractional $U(1)$ charges of the defects, for the special case of the standard gapped $\Z_2$ spin liquid at half-filling and on the square lattice, where $M=4$, $\vec{q} = m = (0,1)^T$, $\vec{m} = e = (1,0)^T$ and all $k_i = 0$. In this case, the Hall conductivity vanishes, $\sigma_H = 0$, and depending on the value of $\vec{s}$ the $U(1)$ charge of a pure $\pi/2$ disclination is calculated as $\mathscr{S}/4= \vec{q}^T K^{-1} \vec{s}/4$. To find the $U(1)$ charge of a dislocation, we compute
\begin{align}
\begin{pmatrix} \check{P}_{c,1}\\ \check{P}_{c,2}\end{pmatrix} &=  (1-U(\pi/2))^{-1}\begin{pmatrix} \vec{q}^T K^{-1} \vec{t}_x\\ \vec{q}^T K^{-1}\vec{t}_y\end{pmatrix}\\
&=  \frac{1}{4}\begin{pmatrix} \vec{t}_{x,1}+\vec{t}_{y,1}\\ \vec{t}_{x,1}-\vec{t}_{y,1}\end{pmatrix}
\end{align} 
The four fractionalization classes related to the torsion vector are specified by choosing $\vec{t}_x + \vec{t}_y \sim I,e,m$ or $\psi$. Note that $\vec{t}_{i,1} = 1$ if $\vec{t}_i = e,\psi$, while $\vec{t}_{i,1}=0$ if $\vec{t}_i = I,m$. Thus, if the momentum fractionalization class is specified by $I$ or $m$ (i.e. it is trivial), the polarization will take values of the form $(0,0)^T, (0,1/2)^T, (1/2,0)^T$. The charge of a dislocation computed using these values will be a multiple of $1/2$ and can be understood as the charge of some anyon associated to that dislocation. If the momentum fractionalization is specified by $e$ or $\psi$, the polarization will take values of the form $(1/4,\pm 1/4)$. Then the charge of a dislocation can take the values 1/4 or 3/4, which cannot be understood through the attachment of anyons to each dislocation. This is a feature of nontrivial momentum fractionalization. Note that the seeming asymmetry between $e$ and $m$ in this example is due to our choice of $\vec{q} = m$.

The rest of the responses are straightforward to compute in this example given our general theory and we leave them for more detailed studies of $\Z_2$ spin liquids. 

\section{Classification of SETs and reduction of $\mathcal{H}^3(G,U(1))$}
\label{Sec:Classif}

\subsection{Recovering the $\H^2(G,\A)$ and $\H^3(G,U(1))$ classification}

The four generalized charge vectors $\vec{q}, \vec{s}, \vec{t}, \vec{m}$ described above can all be included independently in the effective action for $G = U(1)\times
G_{\text{space}}$. Therefore the group classification of the generalized charge vectors is $\A\times(\A/M\A)\times(K_M\otimes\A)\times\A$, which equals $\mathcal{H}^2(G, \mathcal{A})$ as expected. For 
$M = 1$ the correct $\A\times\A$ classification is produced by taking $K_1$ to be trivial. When the magnetic flux
per unit cell is not an integer, the group structure becomes a non-trivial central extension of $G_{\text{space}}$ by $U(1)$ due to the magnetic translation algebra. This case is left for future work. 

The full classification of the allowed SPT terms is given by $ \Z^2\times\Z_M^3\times K_M^2$, which indeed equals $\H^3(G,U(1))$, as we derive in Appendix \ref{Appendix:GrpCoho}. The classification based on $\H^2(G,\A) $ and $\H^3(G,U(1))$ is summarized in Table \ref{table:Summary}. However not all of these choices give topologically distinct
phases of matter \cite{Barkeshli2019,Lu2016} as some of them can be trivialized by field redefinitions. The particular redundancies that appear depend sensitively on the choice of $K$-matrix and
the generalized charge vectors \cite{Lu2016} as we will describe below. For example, we find that the $1/2$ Laughlin topological order on a square lattice ($M = 4$) posseses
2304 distinct symmetry-enriched topological states when the integer part of the charge filling per unit area $(k_6)$ and the Hall conductivity $(k_1)$ are fixed.

\subsection{Reduction of $\H^3(G,U(1))$ due to relabellings} 
\label{Appendix_RelabellingExample} 

\begin{table*}[t]
	\centering
	\begin{tabular} {l |l |l}
		\hline
		$N$ & Generalized charge vectors & Relabelled SPT parameters $ (k_2',k'_3,k'_{5,i},k'_7)$\\
		\hline
		$4N'$	&$(q \mod N,s \mod 4, \vec{t} , m \mod N)$ &$(k_2-q,k_3-\frac{N}{2}-s,k_{5,i}-t_i,k_7-m)$  \\
		$4N'+2$ & $(q \mod N,s \mod 2, \vec{t}, m \mod N)$ &$(k_2-2q,k_3-2s,k_{5,i},k_7 - 2m)$ \\
		\hline
	\end{tabular}
	\caption{The effect of relabellings on the SET classification for the $1/N$ Laughlin state (for $N$ even) with $U(1)$ charge conservation and $p4=\Z^2\rtimes\Z_4$ space group symmetry. In the left column, $N'$ is an integer. The charge, spin and area vectors are now integers, while the torsion vector $\vec{t}_i$ is valued in $\Z^2$, and $t_x+t_y \mod 2$ specifies the symmetry fractionalization. The internal gauge fields can be relabelled so as to leave the partition function invariant but transform the SPT parameters $(k_2,k_3,k_5,k_7)$ to $(k_2',k_3',k_5',k_7')$ (shown in the last column). Note that the precise transformation of these coefficients is different for different symmetry fractionalization classes and for different values of $N$. }\label{table:Z4Relabellings}
\end{table*}
As discussed above, $\mathcal{L}_{frac}$ specifies the symmetry fractionalization class through the choice of the generalized charge vectors,
which corresponds to the classification $\mathcal{H}^2(G, \mathcal{A})$. $\mathcal{L}_{SPT}$ contains additional terms depending only on the background gauge fields, and is classified by $\mathcal{H}^3(G, U(1))$. The choice of $\mathcal{H}^3(G, U(1))$, which corresponds to
changing the coefficients $k_i$ in $\mathcal{L}_{SPT}$, can be understood as stacking (2+1)D SPT states. Physically, the effect of changing the action by a choice of $\mathcal{H}^3(G, U(1))$ is to change the braiding and fusion properties of the symmetry defects \cite{Barkeshli2019}.

\begin{table*}[t]
	\centering
	\begin{tabular} {l|l|l|l|l}
		\hline 
		\multicolumn{5}{c}{Count of SETs for 1/2 Laughlin topological order ($\A=\Z_2$) with $G=U(1)\times[\Z^2\rtimes\Z_M]$} \\ \hline
		$M$ & $\H^2(G,\Z_2)$ & $\H^3(G,U(1))$ &Naive SET count ($k_1,k_6$ fixed)& Reduced SET count ($k_1,k_6$ fixed)\\
		\hline
		2 &$\Z_2^5$ &$\Z^2\times\Z_2^7$ &4096 &800 \\
		3 &$\Z_2^2$ &$\Z^2\times\Z_3^5$ &972 &972 \\
		4 &$\Z_2^4$ &$\Z^2\times\Z_2^2\times\Z_4^3$ &4096 &2304 \\
		6 &$\Z_2^3$ &$\Z^2\times\Z_6^3$ &1728 &972 \\  
		\hline
	\end{tabular}
	\caption{Count of SETs for the 1/2 Laughlin topological order with $G=U(1)\times[\Z^2\rtimes \Z_M]$.
		We have fixed $k_1$ and $k_6$, which are the integer parts of the Hall conductivity and the charge filling. See Appendix \ref{Appendix:LaughlinRelab} for the derivation. 
	}\label{Table:SETcount}
\end{table*}

Depending on the choice of symmetry fractionalization class and the precise topological order involved, it is possible that changing the action by a non-trivial choice of $\mathcal{H}^3(G, U(1))$ does not yield a distinct phase of matter. Therefore, keeping the symmetry fractionalization choice fixed,
the true classification of distinct symmetry-enriched topological states (SETs) is reduced from $\mathcal{H}^3(G, U(1))$ to a subgroup. In the G-crossed braided tensor category formulation \cite{Barkeshli2019}, this reduction corresponds to cases where changing the algebraic theory of defects
by an element of $\mathcal{H}^3(G, U(1))$ can be completely accounted for by a relabeling of the symmetry defects. 

We can also see this reduction from $\mathcal{H}^3(G, U(1))$ in the context of our topological effective action. In this context, we see that field redefinitions can be made to absorb the effect of changing the couplings in $\mathcal{L}_{SPT}$ by certain amounts. Since this
analysis is heavily dependent on the precise topological order (precise choice of $K$ matrix) involved, here we will focus on some simple examples.

To illustrate the main idea, let us begin by considering the case where $G = \Z_M$, with the symmetry fractionalization class specified by the spin vector $\vec{s}$, and the associated defect class given by $k \in \Z_M$. The $\Z_M$ gauge field $C$ couples to $a$ as follows:
\begin{align}
\mathcal{L} = -\frac{1}{4\pi} K_{IJ} a^I \cup d a^J + \frac{s_I}{2\pi} a^I \cup dC + \frac{k}{2\pi} C \cup dC
\end{align}

In this case, there are naively $M$ distinct choices of $k$, $k = 0,\cdots, M-1$, corresponding to $\mathcal{H}^3(\Z_M, U(1)) = \Z_M$. 
First, we note that the choice of couplings $(\vec{s}, k)$ has the following redundancies:
\begin{align}
\label{redund1}
(\vec{s} + M \vec{\Lambda}, k) \sim (\vec{s}, k) \sim (\vec{s}, k+M)
\end{align}
The first equivalence is because $M\frac{\vec{\Lambda}}{2\pi} \cdot \vec{a} \cup d C$ is trivial, as explained in the main text. The second equivalence follows from $\mathcal{H}^3(\Z_M, U(1)) = \Z_M$. 

Next, observe that we can rewrite the Lagrangian as
\begin{widetext}
	\begin{align}
	\mathcal{L} &= -\frac{1}{4\pi} K_{IJ} (a^I + u^I C)  \cup d (a^J + u^J C) 
	+ \frac{s_I + K_{IJ} u_J}{2\pi} (a^I + u^I C) \cup dC +\frac{2k - \vec{u}^T K \vec{u}-2\vec{s} \cdot \vec{u}}{4\pi} C \cup dC ,
	\end{align}
\end{widetext}
where $\vec{u} \in \Z^D$. Since $a^I$ is dynamical, the shift $a^I \rightarrow a^I + u^I C$ can be trivially absorbed by redefining the integration variables. Note that $a^I + u^I C$ still obeys the flux quantization condition since $dC$ integrates to $2\pi \Z$ over any 2-cycle. 

Therefore, we have the additional equivalence
\begin{align}
\label{redund2}
(\vec{s}, k) \sim ( \vec{s} + K \vec{u}, k - \frac{\vec{u}^T K \vec{u}}{2} - \vec{s}^T \vec{u}) . 
\end{align}
Combining the equivalences in \eqref{redund2} and \eqref{redund1}, we see that whenever $K \vec{u} = M \vec{\Lambda}$ , we get
\begin{align}
(\vec{s}, k) \sim ( \vec{s}, k - \frac{\vec{u}^T K u }{2} - \vec{s}^T \vec{u}) . 
\end{align}
For a fixed choice of $\vec{s}$, this corresponds in general to a reduction of $\mathcal{H}^3(\Z_M, U(1)$. 

Now we can work out some specific examples. Consider the $1/N$ Laughlin state with $N$ even, for which $K = N$, and take $M = 2$. 
Since $s \sim s + M$, there are two possible spin vector classes, given by $s$ odd or $s$ even. Suppose we choose $u = 1$ and $\Lambda = N/2$. Then we have
\begin{align}
(s, k) \sim (s, (k - N/2 - s) \text{ mod } 2)
\end{align}

For $s=1$, then this relabelling will take $k \rightarrow (k-1 - N/2) \mod 2$. Hence, if $N$ is a multiple of 4, the SET classes corresponding to $(s,k) = (1,0)$ and $(1,1)$ are the same, while the two classes $(s,k) = (0,0)$ and $(1,1)$ are distinct. 
The above result was previously also obtained using the edge physics of Chern-Simons theories in Ref \cite{Lu2016}; here we have reproduced their result with the field theory in the bulk.

One can use similar reasoning to obtain the SET classification for general $K$-matrix states and for general symmetries $G = U(1) \times G_{\text{space}}$, as explained in Appendix \ref{Appendix:LaughlinRelab}. The summary of the relabeling analysis for $1/N$ bosonic Laughlin topological orders and for the case where $G_{\text{space}} = \Z^2 \rtimes \Z_4$ is summarized in Table \ref{table:Z4Relabellings}. In Appendix \ref{Appendix:LaughlinRelab} we further do an explicit counting of SET states for the $1/2$ Laughlin topological order, with $G = U(1) \times G_{\text{space}}$ and considering all 5 orientation-preserving 2d space group symmetries, $G_{\text{space}} = \Z^2 \rtimes \Z_M$ for $M = 1,2,3,4,6$; these results are shown in Table \ref{Table:SETcount}. 

\section{Crystalline gauge theory: continuum approach}
\label{Sec:Continuum}

In this section we discuss two aspects of crystalline gauge theory. The first is that our method of defining discrete crystalline gauge fields on simplices and using simplicial calculus to evaluate the action was a practical choice to make direct the relation with the group cohomology classifications of symmetry-enriched topological states (SETs) \cite{Barkeshli2019}. However we expect that the same results can also be obtained by working with real-valued differential forms.

The second aspect is that the discrete translation and rotation gauge fields defined in this work are directly related to the coframe field and the spin connection that arise in continuum geometry and are known to be closely related to elasticity theory. (In Appendix \ref{Sec:elast}, we provide some background on the origin of crystalline gauge fields in terms of the gauge theory of elasticity as discussed in Ref. \cite{Kleinert}.) 

\subsection{Crystalline gauge fields as differential forms}
\label{Sec:Diffforms}

In order to construct actions from discrete gauge fields, it was convenient to work in terms of simplicial cohomology and simplicial calculus (see Appendix A of Ref \cite{Kapustin2014} for a review). There, our translation gauge fields could be viewed as $\Z^2$-valued $1$-cochains defined on the triangulated space-time manifold $\mathcal{M}$; that is, $X,Y \in C^1(\mathcal{M}, \Z)$. Similarly, the rotation gauge field can be viewed as a $\Z_M$ valued $1$-cochain, $C \in C^1(\mathcal{M}, \Z_M)$ (strictly speaking, in the main text $C$ corresponded to a lift of the $\Z_M$ gauge field to $\frac{2\pi}{M} \Z$. The action is then invariant under changes of lift, e.g. shifting $C_{ij} \rightarrow C_{ij} + 2\pi$ for a single 1-simplex $ij$).

We can consider instead a formulation where we take the gauge fields to be real-valued differential $1$-forms. We thus can define 
\begin{align}
a^I, A, X, Y, C \in \Omega^1(\mathcal{M}, \R),
\end{align}
where $\Omega^k(\mathcal{M},\R)$ denotes the space of real-valued differential $k$-forms. $a^I$, $A$, $\vec{R} = (X,Y)$, and $C$ are the internal, $U(1)$, translation, and rotation gauge fields, respectively, now defined as differential $1$-forms.

The discreteness of the gauge fields enters through constraints on the holonomies of these gauge fields. Given a cycle $\gamma$, we require
\begin{align}
\oint_\gamma \vec{R} \in 2\pi\Z^2, \;\; \oint_\gamma C \in \frac{2\pi}{M} \Z ,
\end{align}
with the equivalence
\begin{align}
\oint_\gamma C &\sim \oint_\gamma C + 2\pi, \;\; \oint_\gamma A \sim \oint_\gamma A + 2\pi
\nonumber \\
\oint_\gamma a^I &\sim \oint_\gamma a^I + 2\pi
\end{align}
Dislocations and disclinations must therefore correspond to singular sources of flux for $X$, $Y$, $C$. Differential forms which are required to integrate to
discrete values along cycles are referred to as integral differential forms.

The gauge transformations are also real-valued. In particular large gauge transformations for $a^I$, $A$ and $C$ must be quantized in units of $2\pi$.

We then write the effective action using the wedge product:
\begin{widetext}
	\begin{align}
	\mathcal{L} &= -\frac{1}{4\pi} K_{IJ} a^I \wedge d a^J + \mathcal{L}_{frac} + \mathcal{L}_{SPT}
	\nonumber \\
	\mathcal{L}_{frac} &= \frac{1}{2\pi} a^I \wedge (q_I dA + s_I dC + \vec{t}_I \cdot d \vec{\cancel{R}} + m_I A_{XY} ) \nonumber \\
	\mathcal{L}_{SPT} &= \frac{k_1}{2\pi} A \wedge dA + \frac{k_2}{2\pi} A \wedge dC + \frac{k_3}{2\pi} C \wedge dC 
	+ \frac{1}{2\pi} A \wedge (\vec{k}_4 \cdot  d \vec{\cancel{R}})+\frac{1}{2\pi} C \wedge (\vec{k}_5 \cdot d \vec{\cancel{R}}) 
	+ \left(\frac{k_6}{2\pi} A + \frac{k_7}{2\pi} C\right) \wedge A_{XY}
	\end{align}
\end{widetext}
Here $A_{XY}$ is the continuum analog of the area element we defined in the simplicial formulation. For example, when $C = 0$, $A_{XY} = \frac{1}{4\pi} (X \wedge Y - Y \wedge X)$. 
Note that the terms in $\mathcal{L}$ aside from those involving $A_{XY}$ and $d \vec{\cancel{R}}$ are standard. To ensure that the terms involving $d \vec{\cancel{R}}$ are invariant under large gauge transformations of $a^I$, $A$, and $C$, we require $\frac{1}{2\pi} \int_{W} d \vec{\cancel{R}} \in \mathbb{Z}^2$
over any closed 2-cycle $W$. When $W$ is the space, for example, this physically corresponds to the fact that the total Burgers vector of the whole closed space is trivial. 

While we do not pursue a formal proof here, we expect that the effective action defined using this continuum formulation yields identical physical results as compared with the lattice gauge theory formulation used in previous sections.

Given a triangulation of the space-time manifold $\mathcal{M}$, we can understand the relation between the discrete formulation and the continuum formulation as follows. Given a link (1-simplex) $ij$ with vertices $i$ and $j$, the discrete gauge fields, $A_{ij}$, $C_{ij}$, $X_{ij}$ and $Y_{ij}$ are taken to be the integral from $i$ to $j$ along the 1-simplex $ij$ of their continuum counterparts. Note that only those continuum gauge field configurations can be used that give rise to the appropriate discrete values of $C$, $X$, and $Y$. Since the only gauge invariant quantities for $\vec{R}$ and $C$ are associated with disclinations and dislocations, we expect that such gauge configurations can always be found. 

We can see how to specify the action of $C$ on $\vec{R}$ by noting that in the continuum setting, $\vec{R}$ and $C$ correspond exactly to the continuum coframe fields $e$ and spin connection $\omega$. In the following section we discuss this correspondence in more detail. 

\subsection{Gauge fields for continuous spacetime symmetries: coframe field and spin connection}
\label{framespin}

The Euclidean group $\mathbb{E}^2 = \R^2 \rtimes SO(2)$ is a semidirect product of the group of continuous rotations in 2D, $SO(2) = U(1)$ and the group of continuous translations, $\mathbb{R}^2$.
In this case we can consider background gauge fields associated with the continuous translation and rotation symmetries.

The translation gauge fields in the continuum setting now correspond to the 1-form coframe fields $e^{a}_{\mu}, a = 1,2$ associated with the space $\Sigma^2$. For physically realistic space-time manifolds of the form $\mathcal{M} = \Sigma^2 \times \R$, where $\Sigma^2$ is space, we choose $e^1,e^2$ to be of the form $e^a_{i} dx^{i} = e^a_x dx + e^a_y dy$. There is also a fixed time-component of the coframe field, $e^0 = dt$. Below we will assume the space $\Sigma^2$ can be curved, but time is separate, as is appropriate for directly describing a condensed matter system. That is, the metric tensor $g = g_{ij}dx^i dx^j + g_{tt} dt^2$. 

The coframe fields diagonalize the metric tensor
\begin{align}
g_{ij} = e^a_i e^b_j \delta_{ab} ,
\end{align}
where $\delta_{ij}$ (the Kronecker delta) is the flat space metric. In the linearized approximation where $e^a_i = \delta^a_i + \tilde{e}^a_i$, we have
\begin{align}
g_{ij} = \delta_{ij} + \tilde{e}^{i}_j + \tilde{e}^j_i  ,
\end{align}
where $\delta^a_i = \delta_{ai}$ is the Kronecker delta.

A translation gauge transformation can be identified as an infinitesimal diffeomorphism:
\begin{align}
x^i \rightarrow f^i(x) =  x^i + \epsilon^i(x) ,
\end{align}
under which
\begin{align}
e^a_i &\rightarrow \partial_i f^j e^a_j
= (\delta_{i}^j + \partial_i \epsilon^j ) e^a_j
\nonumber \\
&= (\delta_{i}^j + \partial_i \epsilon^j ) (\delta^a_j + \tilde{e}^a_j)
= \delta^a_i + \tilde{e}^a_i + \partial_i \epsilon^a + \cdots
\nonumber \\
&= e^a_i + \partial_i \epsilon^a + \cdots,
\end{align}
where the $\cdots$ indicate the subleading term which we ignore in the linearized approximation. We see therefore that in the linearized approximation, the gauge transformations of $e_i^a$ are the continuous analog of the discrete translation gauge transformations
$\vec{R}_{ij} \rightarrow \vec{R}_{ij} + \vec{r}_{j} - \vec{r}_{i}$ on the lattice. Note that as in the discrete case, the continuous translation gauge transformations should preserve the gauge-invariant holonomies associated with $e^a$. In particular, the gauge transformations therefore correspond to diffeomorphisms that preserve the lengths along non-contractible cycles. 

Physically, the continuous translation gauge fields $e_i^a$ correspond to the plastic distortion tensor discussed in Ref. \cite{Kleinert}. The full strain tensor $u$ is the sum of the elastic strain tensor $u^{el}$ and the plastic strain tensor $u^p$: $u = u^{el} + u^p$. The gauge-invariant combination is $u^{el} = u - u^p$.

In addition to the translation gauge transformations, there are also rotation gauge transformations. These correspond to locally rotating the coordinate axes by an element of $SO(2)$, at every point. The gauge field associated with these gauge transformations is the spin connection, which is a 1-form gauge field $\omega$ that corresponds to the continuous spatial rotation symmetry. The spin connection specifies how the frame fields at nearby points are rotated relative to each other. In terms of the full 3D space-time spin connection  $\Omega_{b,\mu}^a$, the spin connection associated with spatial rotations corresponds to $\omega_{\mu} = \Omega_{2,\mu}^1$. In this language, we can explicitly write the correspondence between the continuum and discrete gauge fields as $\vec{R} \sim (e^1,e^2)$ and $C \sim \omega$. We emphasize that when the continuous $\mathbb{E}^2$ symmetry is broken down to a discrete space group symmetry, there is no distinction between $(\vec{R},C)$ and $(\vec{e},\omega)$. The gauge-invariant properties associated to $(\vec{R},C)$ can equally be calculated using $e^a,\omega$. 

To further clarify the correspondence between the discrete translation and rotation gauge fields and the continuum coframe fields and spin connection, we calculate the contribution of $\omega$ to the covariant derivative of $e^a$ using our discrete formulation with certain limiting arguments. At a point $\vec{r} + \delta \vec{r}$, the coframe field (written here using the translation gauge field notation of the main text) is
\begin{align}
R_j(\vec{r} + \delta \vec{r}) \approx R_j(\vec{r}) + \partial_i R_j \cdot \delta r_i.
\end{align}
As stated in the main text, the vector $\vec{R}(\vec{r})$ parallel transported to $\vec{r} + \delta \vec{r}$ is $U(C_{\vec{r},\vec{r}+\delta\vec{r}})\vec{R}(\vec{r})$, where we have chosen $\vec{r}$ as the origin. In the continuum, we can write $U(C_{\vec{r},\vec{r}+\delta\vec{r}} = \theta)$ as a rotation matrix $e^{i \theta \sigma_y}$ (this would not be appropriate on a lattice, where we need to use $GL(2,\Z)$ matrices in a lattice basis, but it is not a problem in the continuum). The total rotation applied between $\vec{r'}$ and $\vec{r'}+d\vec{r'}$ is written in terms of $\omega$ as $e^{-i \int\limits_{\vec{r}}^{\vec{r}+\delta\vec{r}} \vec{\omega}(\vec{r}') \cdot d\vec{r}' \sigma_y}$. Here we have written the spin connection as a vector with components $\omega_\mu$. This representation of $C$ shows that it directly corresponds to $\omega$ in the continuum.

To first order in $\delta \vec{r}$, we can approximate
\begin{align}
U(C_{\vec{r},\vec{r}+\delta\vec{r}}) &= e^{-i \int\limits_{\vec{r}}^{\vec{r}+\delta\vec{r}} \vec{\omega}(\vec{r}') \cdot d\vec{r}' \sigma_y} \\
&\approx 1 - i \sigma_y \vec{\omega}(\vec{r}) \cdot \delta\vec{r}  
\end{align} 
The covariant derivative of $\vec{R}$ in the direction $x^i$ can then be written as 
\begin{align}
D_i R_j (\vec{r}) &= \lim\limits_{x^i \rightarrow 0} \frac{1}{x^i}(R_j(\vec{r}+x^i) - (U(C_{\vec{r},\vec{r}+x^i})\vec{R} (\vec{r}))_j) \\
&= \lim\limits_{x^i \rightarrow 0} \frac{1}{x^i}(\partial_i R_j  x^i  + \omega_i(\vec{r}) x^i \times (i\sigma_y R)_j)  \\
&= \partial_i R_j +  \omega_i(\vec{r}) \epsilon_{jk} R_k 
\end{align} 
This is precisely the formula for the covariant derivative $D$ of $e^{a}$ in terms of $\omega$, which is written in the usual notation as 
\begin{equation}
T^a \equiv D e^a = d e^a + \epsilon^a_b \omega e^b = d e^a + \Omega^a_b \wedge e^b
\end{equation}  
Here $\Omega^a_{b,\mu}$ is the full spin connection. We have proved this formula using the fact that $\Omega^a_b = - \Omega^b_a$ is anti-symmetric, so that  $\Omega_{1}^1=\Omega_{2}^2 = 0$ and $\Omega_{2}^1 = -\Omega_{1}^2 = \omega$.

$T^a$ is the torsion 2-form, which characterizes how the frame field is rotated along the path traced by a curve in spacetime. The torsion as defined above can be directly related to the dislocation density, i.e. to the holonomy of translation gauge fields after accounting for parallel transport, similar to the quantity $d\mathcal{\vec{R}}$ used in our work. Furthermore, the flux associated to rotational symmetry alone ($d C$ in the lattice formulation, or $d \omega$ in the continuum) gives the curvature of the manifold, which is directly related to the disclination density. Therefore couplings involving $d C$ or $d \omega$ are essentially coupling the system to curvature. Given that torsion is not quantized in the continuum, there cannot be any quantized topological terms formed by coupling anyons or symmetry charges to the torsion (although nonquantized terms which are topological in the sense of being independent of changes in the underlying metric are well-known). 

The classification of SET phases with $U(1)\times\mathbb{E}^2$ symmetry is identical to the classification for $U(1)\times U(1)$ symmetry (this can be proved, for example, by computing the relevant cohomology groups) \cite{Manjunath2020Gx}. So while the translation group $\mathbb{R}^2$ has associated gauge fields $X$ and $Y$, the Lagrangian does not have any contribution from $X$ and $Y$; the only relevant terms for Euclidean group symmetry fractionalization and for the associated SPT states are given by $\frac{s_I}{2\pi} a^I \wedge d\omega$ and $\frac{k}{2\pi}\omega \wedge d\omega$ respectively.

\section{Discussion} 
\label{Sec:Disc}

\subsection{Spatial vs. internal symmetries}

As we have discussed, at a formal level our mathematical treatment of crystalline gauge fields is equivalent to treating the symmetry as an internal symmetry of the low energy quantum field theory. The main difference is (1) the physical interpretation of the fluxes in terms of geometrical properties of the lattice, with certain holonomies being restricted by the lattice area and lengths, and (2) the fact that we ultimately tie the space-time metric of the low energy topological quantum field theory, which arises from the framing anomaly, to the crystalline gauge fields. Here we will begin by discussing this issue in some more detail.

We have two levels of description of the system. The first is the microscopic lattice model, which has a global symmetry $G = U(1) \times G_{\text{space}}$, for some spatial symmetry group $G_{\text{space}}$. The second is the effective field theory description, which in our case is a topological field theory. The symmetry of the topological field theory is
  $G_{\text{IR}} \times \text{Diff}(M)$, where $G_{\text{IR}}$ is the internal symmetry of the field theory and $\text{Diff}(M)$ is the group of diffeomorphisms of the space-time manifold $M$. Here the internal symmetry $G_{\text{IR}}$ allows us to couple the field theory to background principal $G_{\text{IR}}$ bundles. The action of the microscopic $G$ symmetry in the low energy field theory is described by a group homomorphism:
  \begin{align}
\alpha: G \rightarrow G_{\text{IR}} \times \text{Diff}(M) . 
    \end{align}
    When ${\bf g} \in G$ is a purely on-site symmetry of the microscopic lattice model, then $\alpha({\bf g}) = ( \alpha({\bf g})|_{G_{IR}}, {\bf 1})$, where
    $\alpha({\bf g})|_{G_{IR}}$ denotes the restriction of $\alpha$ to the first factor and ${\bf 1}$ refers to the identity element of $\text{Diff}(M)$.
    That is, an on-site symmetry ${\bf g}$ in the microscopic lattice model is
    mapped to an internal symmetry in the field theory. On the other hand, if ${\bf g} \in G$ is a purely spatial symmetry of the microscopic lattice model, then
    $\alpha({\bf g})$ maps ${\bf g}$ to a combination of an internal symmetry and an element of $\text{Diff}(M)$. For example, a $\Z_M$ spatial rotation in the
    microscopic lattice model will be mapped in general to a $\Z_M$ internal symmetry combined with a $\Z_M$ rotation of space in the field theory. The
    distinct ways that a microscopic lattice symmetry can act in the field theory is taken into account by the different ways of coupling the effective field
    theory to background $G_{\text{IR}}$ gauge fields. 

    Another way to state the above is that given any spatial symmetry ${\bf g}$, one can always consider the combination $\alpha({\bf g})$
    followed by an appropriate element of $\text{Diff}(M)$, to obtain a symmetry action in the field theory that has trivial component in $\text{Diff}(M)$.
    Therefore, given any spatial symmetry in the microscopic lattice model, the effective field theory description can also in general contain a corresponding
    internal symmetry. To fully describe all possible SETs, we thus take $G_{\text{IR}} = G$, and we classify all the ways that the effective field theory
    can be coupled to $G$ bundles.

    Observe that in the above description, the symmetry defects associated with spatial symmetries in the micrscopic lattice model, such as
    dislocations and disclinations, should therefore be described in the field theory by symmetry fluxes of the internal symmetry of the field theory
    and, simultaneously, torsion and curvature defects in the space-time metric of the effective field theory. This reflects the fact that $\alpha$ maps to
    both $G_{\text{IR}}$ and $\text{Diff}(M)$. This explains why we equate the spin connection of the space-time metric to the rotation gauge field in
  Eq. \ref{Comega}.

  The above explanation is not a proof that spatial symmetries in lattice models can always be treated as internal symmetries in the effective field theory
  description. Nevertheless, all known examples of effective field theories of quantum many-body systems can be understood via the above paradigm.
  As a simple example, consider the action of translation symmetries in spin chains and their description in the low energy Luttinger liquid theory \cite{giamarchi2003}.

  The conjecture that spatial symmetries can always be treated as internal symmetries in the field theory has recently been formalized in Ref. \onlinecite{Thorngren2018} as the ``crystalline equivalence principle,'' where additional arguments have also been given in support of it. This principle has also received significant support from the theory of crystalline SPTs, where the SPT classifications obtained by treating spatial symmetries as internal symmetries can be compared with other more direct methods \cite{Song2017,Huang2017,Else2019,Song2020}, and the results agree with each other.   
   
  \subsection{Connection between points on the triangulation and points on the lattice}

  In our formulation of the crystalline gauge field, the underlying lattice model does not feature explicitly in the formulation, although one can give an interpretation to the crystalline gauge fields in terms of the microscopic lattice sites as done in the gauge theory of elasticity \cite{Kleinert}. The 0-cells of the triangulation of $\mathcal{M}$ need not be assumed to belong to any microscopic or coarse-grained lattice. The motivation for the gauge field itself is the assumption that the topological response is completely determined by the gauge-invariant data of the underlying lattice, defined as the lengths around non-contractible cycles, the area, the Burgers vectors of dislocations, and the angle of disclinations in the lattice. Now these quantities can all be specified by constructing loops which encircle all the defects, and which span the nontrivial cycles of the manifold, and then keeping track of the change in coordinate labels and the local orientation of coordinate axes as we go around each loop. This can all be achieved using a triangulation. Therefore it does not matter whether or not the vertices of the triangulation actually correspond to points or coarse-grained regions of the original lattice. As such, the precise locations of the lattice defects is unimportant for the analysis of the topological, quantized response properties. 
   
   Introducing a triangulation moreover has significant additional value: the condition that the effective action is indeed topological can be reformulated as a condition that the partition function is invariant under retriangulations. This in turn means that the action satisfies a group cocycle condition, which provides the link to the group cohomology classification of SETs, as we discuss in Appendx \ref{Appendix:GrpCoho}.

   \subsection{Relation to defect network constructions}

   Ref. \cite{Else2019} gives a general construction of crystalline SET phases in terms of defect networks; a similar approach has been studied for
   invertible phases in Ref. \cite{Song2020}. Here the manifold $\mathcal{M}$ is decomposed by means of a cellulation, and the defects in the theory, which include anyons as well as symmetry defects, are assumed to live on the 0-cells (vertices) of the cellulation. 
   
   The authors of Ref. \cite{Else2019} show that the defect network picture is equivalent to the crystalline equivalence principle. Our formalism is equivalent to assuming the crystalline equivalence principle and proceeding with the G-crossed braided tensor category \cite{Barkeshli2019} and associated group cohomology classifications of SET phases. In this sense, we expect that our approach formally yields the same classification results as the defect network picture. 
   
   However the two approaches differ in details of physical interpretation. Let us restrict to the SPT case for concreteness. In this special case the defect network picture is mathematically related to an equivariant cohomology theory, in which one considers the high-symmetry points of a space group unit cell and places symmetry charges on these high symmetry points. Two configurations of symmetry charge are in different SPT phases if they cannot be deformed into one another by local, symmetry-preserving unitaries. (This procedure is essentially the "block state" construction of SPT phases developed in Refs \cite{Huang2017,Song2017}.) It is not fully clear how this approach is equivalent to the topological response theory that we have described in our work. We can also express this distinction as follows: the equivariant cohomology approach has symmetry charges, but in this picture it is not apparent how these arrangements of charge give rise to different responses upon introducing symmetry fluxes. Reconciling the two pictures properly is an interesting direction, but beyond the scope of the present work.

   \subsection{Outlook}

   We have predicted a type of momentum fractionalization, characterized by the discrete torsion vector, which can only be non-trivial for
   $M = 2,3,4$-fold rotation symmetry together with translation symmetry. This term leads to a number of fractionally quantized response properties with no analog in the continuum.
   Perhaps most notably this includes a fractionally quantized charge polarization, which can assign non-trivial fractional charges to dislocations and fractional charges per unit length to
   boundaries (modulo the anyon charge). In addition to this, the theory predicts fractionally quantized linear and angular momenta for disclinations, dislocations, and units of area.
   It is important to verify the predictions of this crystalline gauge theory through microscopic studies of model Hamiltonians and wave functions. While the fractional charges of
   dislocations and disclinations can in principle also be probed by experiments on fractional Chern insulators with sufficiently weak disorder, it is an interesting theoretical
   question to understand the extent to which the fractionally quantized linear and angular momenta of anyons, lattice defects, and units of area can be experimentally measured.

   Our theory is expected to be complete for topological phases of bosons, where symmetries do not permute anyon types. For fermionic states, which are most
   relevant for experimental studies of fractional Chern insulators in solid state systems, our theory will still apply, although we expect some modifications in terms
   of different quantizations of certain coupling constants (e.g. some $k_i$ can be half-integer). There may also be additional fermionic SET phases and physical phenomena that cannot
   be fully captured with these effective actions, corresponding to ``beyond group supercohomology'' phases.

   When the space group symmetries do permute anyon types \cite{barkeshli2012a,barkeshli2013genon}, lattice defects can be non-Abelian and the classification of SETs is different.
   Furthermore, certain values of the coefficients of the response theory may be constrained by the symmetry permutation. A detailed study of this is left for future work. 

   The crystalline gauge theory we have developed treats the lattice defects as a fixed background configuration that is described in terms of a fixed background gauge field. Such a gauge theory apparently does not take
   into account the restricted mobility of dislocations and disclinations in a crystalline environment. The restricted mobility of these lattice defects can be described using higher rank tensor gauge fields, which are
   known to be dual to fracton theories (see e.g. Refs \cite{Pretko2017,Pretko2018, Radzihovsky2020}). It would be interesting to understand the relation between the topological field theory developed here and a formulation
   including higher rank tensor gauge fields which explicitly takes into account the restricted mobility of the lattice defects. 
   
   Finally, we note that in general, given a symmetry $G$ of a condensed matter system, the effective field theory must include coupling to background gauge fields of the symmetry in
   order to be fully specified. It would be interesting to revisit the large family of effective field theories used throughout condensed matter physics, including gapless
   theories, and to properly understand the coupling to background crystalline gauge fields. 
	 
\section{Acknowledgements} We thank Andrey Gromov, Su-Kuan Chu, and Max Metlitski for helpful discussions and comments. This work is supported by NSF CAREER (DMR- 1753240), an Alfred P. Sloan Research Fellowship, UMD startup funds, and the NSF Physics Frontier Center at the Joint Quantum Institute at UMD.

         
	\appendix

\section{2D Point group rotation matrices}
\label{Appendix:PtGrpMatrices}

An important role in the main text was played by the $2 \times 2$ rotation matrix $U\left(\frac{2\pi}{M}\right)$, associated with the generator of point group rotations. Due to the
presence of a lattice, there is a natural basis in which point group rotation matrices $U\left(\frac{2\pi}{M}\right)$ can be defined. We define the $x$ and $y$ axes to be the
lattice vectors, such that for $M$-fold point group rotations, the $x$ and $y$ axes subtend an angle $\frac{2\pi}{M}$. For $M=2$, $U\left(\frac{2\pi}{M}\right) = -1$, where
$1$ here denotes the $2 \times 2$ identity matrix.  For $M=3,4,6$ an elementary $\frac{2\pi}{M}$ rotation can always be defined to take $x \rightarrow y$.
In turn, the existence of a lattice ensures the rotated position of $y$ can be expressed as a linear combination of the original $x$ and $y$. The result
for $U\left(\frac{2\pi}{M}\right)$ is given in Table \ref{table:RotnMatrices}, along with the matrices $(1-U\left(\frac{2\pi}{M}\right))^{-1}$ that also arise frequently.

In our calculations we have assumed that the lengths $L_x,L_y$ are defined along these possibly nonorthogonal axes. Moreover, integrals $\int f(x,y) dx dy$ should be carried out with $x$ and $y$ defined by this lattice-specific coordinate system. The advantage of using these coordinates is that we always work with integer vectors and matrices, so the coefficients of the theory are always integers or fractions of integers.
\begin{table}[t]
	\centering
	\begin{tabular} {|l|l |l |l|l|}
		\hline
		$M$&2& 3& 4& 6 \\ \hline
		$U\left(\frac{2\pi}{M}\right)$
		&$\begin{pmatrix}
		-1 & 0 \\ 0 &-1
		\end{pmatrix}$
		& $\begin{pmatrix}
		0 & 1 \\ -1 &-1
		\end{pmatrix}$ 
		&$\begin{pmatrix}
		0 & 1 \\ -1 & 0
		\end{pmatrix}$ & 
		$\begin{pmatrix}
		0 & 1 \\ -1 &1
		\end{pmatrix}$ \\
		\hline
		$\left(1-U\left(\frac{2\pi}{M}\right)\right)^{-1}$
		&$\dfrac{1}{2}\begin{pmatrix}
		1 & 0 \\ 0 &1
		\end{pmatrix}$
		& $\dfrac{1}{3}\begin{pmatrix}
		2 & 1 \\ -1 &1
		\end{pmatrix}$ 
		&$\dfrac{1}{2}\begin{pmatrix}
		1 & 1 \\ -1 & 1
		\end{pmatrix}$ & 
		$\begin{pmatrix}
		0 & 1 \\ -1 &1
		\end{pmatrix}$ \\
		\hline
	\end{tabular}
	\caption{Elementary rotation matrices $U\left(\frac{2\pi}{M}\right)$ for different $M$. }\label{table:RotnMatrices}
      \end{table}

\section{Crystalline gauge theory and relation to gauge theories of elasticity}
\label{Sec:elast}
The discrete translation gauge field $\vec{R}$ that we use has previously been discussed in elasticity theory \cite{Kleinert}. Here we provide a brief review of how the discrete crystalline gauge fields arise in elasticity theory, following Ch. 9 of  Ref. \cite{Kleinert}. 

In elasticity theory, the basic variables are the displacements $u_i(\vec{r})$ of a particle on a lattice whose mean position is $\vec{r}$, along each direction $i$. The elastic energy is a function of the strain tensor components $\partial_i u_j$ and to lowest order has the form
\begin{align}
E &= \frac{1}{2}\sum\limits_{\vec{r}} \lambda_{ijkl} \partial_i u_j \partial_k u_l ,
\end{align}
where the operator $\partial$ is now interpreted as a discrete gradient. The corresponding classical partition function is given by 
\begin{equation}
Z = \prod\limits_{\vec{r},i} \left(\int\limits_{-\infty}^{\infty} \frac{du_i(\vec{r})}{a}\right) e^{-\beta E}
\end{equation}
Demanding that the energy is invariant under rigid rotations leads to the conditions $\lambda_{ijkl} = \lambda_{klij} = \lambda_{jikl}$ among the elastic moduli \cite{barkeshli2012ph}. This is the most general translation-invariant Lagrangian that can be written at lowest order in derivatives of $u_i$.

At low temperatures and in a classical theory, the displacements $u_i(\vec{r})$ are generally much smaller than the lattice spacing $a$. However, it is possible for thermal or quantum fluctuations to result in particles exchanging their positions over long times. Indeed, the diffusion of particles within the lattice means that it is appropriate to think of $u_i$ as being defined only up to a lattice constant; therefore, our partition function must be invariant under a transformation
\begin{align}
\label{uGauge}
u_i(\vec{r}) \rightarrow u_i(\vec{r}) + a N_i(\vec{r}),
\end{align}
where $a$ is the lattice spacing and $N_i$ is an integer vector field defined at the discrete positions $\vec{r}$. The transformation (\ref{uGauge}) is a gauge transformation which reflects the physical reality that the coordinates can be relabelled up to integers. To ensure gauge invariance under this transformation, we introduce new integer-valued gauge fields $\frac{1}{2\pi} R_{ij} \in \Z$ and replace
\begin{equation}
\partial_i u_j(\vec{r}) \rightarrow \partial_i u_j(\vec{r}) - \frac{a}{2\pi} R_{ij}(\vec{r}).
\end{equation}
(Here $i,j \in \{x,y\}$ and $R_{ij}(\vec{r})$ is a function defined on a lattice; this notation should not be confused with the notation $\vec{R}_{ij}$ in a simplicial formulation, where $ij$ is a 1-simplex on a triangulation.)

Note that to model the particles precisely, we should make sure that the gauge transformation induces a permutation of the location of all lattice sites. This requires that the integers $N_i(\vec{r})$ must in principle be correlated with each other, so that we do not allow multiple atoms to occupy the same lattice site, leaving other lattice sites completely empty. The assumption of the crystalline gauge theory \cite{Kleinert} is that the highly non-trivial interdependency of $N_i(\vec{r})$ can be ignored, and the $N_i(\vec{r})$ can be treated as independent integers. 

The partition function then includes a sum over all possible values of $R_{ij}$:
\begin{align}
Z &= \sum\limits_{\{R_{ij}(\vec{r})\}}\prod\limits_{\vec{r},i} \left(\int\limits_{-\infty}^{\infty} \frac{du_i(\vec{r})}{a}\right) e^{-\beta \tilde{E}}, \\
\tilde{E} &= \frac{1}{2}\sum\limits_{\vec{r}} \left( \lambda_{ijkl} (\partial_i u_j - \frac{a}{2\pi}  R_{ij}) (\partial_k u_l - \frac{a}{2\pi}  R_{kl})
\right)
\end{align}
The change of variables and subsequent sum over $R_{ij}$ encode the fact that the quantities $\partial_i u_j$ can change by any integer values at every lattice point, and that the different particle configurations are all treated equally. As originally desired, $Z$ is now invariant under the gauge transformation
\begin{align}
u_i(\vec{r}) &\rightarrow u_i(\vec{r}) + a N_i(\vec{r}) \\
R_{ij}(\vec{r}) & \rightarrow R_{ij}(\vec{r}) + 2\pi \partial_i N_j(\vec{r})
\end{align}

The $R_{ij}$ are precisely the discrete translation gauge fields suitably defined on a lattice: $R_{xi} = X_i$, $R_{yi} = Y_i$. Integrating out the displacements $u_i$ will result in a pure gauge theory in terms of the gauge fields $R_{ij}$. 

To further understand the fields $R_{ij}$, we next look at how this gauge theory treats dislocations. A lattice dislocation corresponds to a missing or extra line of atoms such that the number of nearest neighbours at the dislocation point changes. The fields $R_{ij}$ allow for such configurations, which are deviations from an ideal lattice configuration. These configurations would not be included in the partition function if we restricted ourselves to a change of variable $u_i(\vec{r}) \rightarrow u_i(\vec{r}) + aN_i(\vec{r})$, as this transformation amounts to a relabelling of coordinates but keeps the particles in an ideal lattice configuration. Another way to say this is that the integral $\oint_{\gamma} \partial_i N_j dl^i$, where $d\vec{l}$ is the infinitesimal line element along the loop $\gamma$, will always be zero in an ideal lattice and cannot represent a dislocation. A dislocation Burgers vector is obtained from the holonomy $\frac{1}{2\pi} \oint_\gamma \vec{R}$. The symmetrized quantity $\frac{1}{2}(R_{ij}+R_{ji})$ is the discontinuous part of the symmetrized strain tensor. A similar procedure can be followed for a continuous elastic medium, where the analog of $\frac{1}{2\pi} R_{ij}$ is referred to as the \textit{plastic} strain tensor $u_{ij}^{(p)}$ and is directly related to the coframe field used in differential geometry, as discussed in Sec. \ref{framespin}.  

We can also introduce disclinations in elasticity theory via a rotation symmetry gauge field. Disclinations, the fluxes of this rotation symmetry field, are related to the antisymmetric component of the strain tensor, which does not enter the action at the usual quadratic order. These effects can be included by adding higher derivative terms to the usual Lagrangian. Conventional elasticity theory does not, however, include translation as well as rotation symmetry via a nonabelian gauge field, as we have done. Instead, it makes certain approximations that allow rotations to be incorporated without dealing with the full space group symmetry. This does not affect the calculations greatly for thermodynamic purposes, but in dealing with topological properties we saw that the nonabelian gauge field led to a situation where only certain properties of dislocations are gauge-invariant. This feature cannot be reproduced by an approximate calculation.  

\section{Count of SETs for the Laughlin state with $G=U(1)\times G_{\text{space}}$}
\label{Appendix:LaughlinRelab}	

In Section \ref{Sec:Classif}, we discussed a general procedure to account for redundancies in the $\H^3(G,U(1))$ classification of SET phases using relabellings of the gauge field, when $G=\Z_M$. Here we will generalize that procedure to $G=U(1)\times G_{\text{space}}$. 

We first recall the effective action written in Eq. \eqref{Eff_Action}:

\begin{widetext}
	\begin{align}
	\mathcal{L} &=- \frac{1}{4\pi} a^I \cup K_{IJ} da^J +\mathcal{L}_{frac}+\mathcal{L}_{SPT} \nonumber \\
	\mathcal{L}_{frac} &= \frac{1}{2\pi} a^I \cup (q_I dA + s_I dC + \vec{t}_I \cdot d \vec{\cancel{R}} + m_I A_{XY} ) \nonumber \\
	\mathcal{L}_{SPT} &= \frac{k_1}{2\pi} A \cup dA + \frac{k_2}{2\pi} A \cup dC + \frac{k_3}{2\pi} C \cup dC 
	+ \frac{1}{2\pi} A \cup (\vec{k}_4 \cdot  d \vec{\cancel{R}})+\frac{1}{2\pi} C \cup (\vec{k}_5 \cdot d \vec{\cancel{R}}) 
	+ \left(\frac{k_6}{2\pi} A + \frac{k_7}{2\pi} C\right) \cup A_{XY} .
	\end{align}
\end{widetext}

The integer coefficients $k_1$ through $k_7$ have the following independent redundancies arising from the group structure of $\H^3(G,U(1))$ ($k_1$ and $k_6$ have no redundancy):
\begin{align}
k_2 &\sim k_2 + M \lambda_2\nonumber \\
k_3 &\sim k_3 + M \lambda_3\nonumber \\
\vec{k}_4 &\sim \vec{k}_4 + (1-U\left(\frac{2\pi}{M}\right)) \vec{\lambda}_4 \nonumber \\
\vec{k}_5 &\sim \vec{k}_5 + (1-U\left(\frac{2\pi}{M}\right)) \vec{\lambda}_5 \nonumber \\
k_7 &\sim k_7 + M \lambda_7\nonumber \\
\end{align}
where $\lambda_i \in \Z$. We also have the following independent equivalence for the anyon $\vec{s}$, introduced in Eq. \eqref{redund1}:
\begin{align}
s_I \sim s_I + M \Lambda_{s,I},
\end{align}
where $\Lambda_{s,I}$ are integers.

Now there is an additional independent equivalence involving the anyons $\vec{t}_i$. Note that the following term is trivial, and can therefore be added to the effective action with no change to the partition function:

\begin{equation}
\frac{(1-U\left(\frac{2\pi}{M}\right))}{2\pi}\vec{\Lambda}_{t,I} da^I \cup \vec{\cancel{R}} = \frac{1}{2\pi}\vec{\Lambda}_{t,I} da^I \cup \vec{R},
\end{equation}
where $\vec{\Lambda}_{t,I}$ are integer vectors. The second expression is always a multiple of $2\pi$ for a flat gauge field configuration of $a^I$, and is therefore trivial. This implies the following equivalence relation, which we have discussed previously:

\begin{align}
 \vec{t}_I \sim \vec{t}_I + (1-U\left(\frac{2\pi}{M}\right)) \vec{\Lambda}_{t,I}.
\end{align}

Now, the most general relabelling of the gauge fields $a^I$ which preserves the flux quantization condition $\int da^I \in 2\pi\Z$, includes the gauge field $\vec{\cancel{R}}$ as well as the gauge field $C$:

\begin{equation}
a^I \rightarrow a^I + u^I C + \vec{v}^I \cdot \vec{\cancel{R}},
\end{equation}
where $u^I,\vec{v}^I$ are all integers. For a flat background gauge field configuration, both $dC$ and $d\vec{\cancel{R}}$ are multiples of $2\pi$, and therefore this relabelling does not affect the flux quantization condition.

We can now repeat the procedure adopted in Section \ref{Sec:Classif}. First we relabel the fields $a^I$ as indicated above. Then we find the constraints on $u^I,\vec{v}^I$ such that $s_I$ and $\vec{t}_I$ can be shifted back to their original values by the addition of trivial terms. Finally, we compute the change in the coefficients $k_1$ through $k_7$ that is required in order to leave the effective action invariant after this relabelling.

The result is the following: $s_I$ and $\vec{t}_I$ can be shifted back to their original values when
\begin{align}\label{Eq:relabel_constraint}
K u^I &= M \Lambda_{s,I} \nonumber \\
K \vec{v}^I &= (1-U\left(\frac{2\pi}{M}\right)) \vec{\Lambda}_{t,I}.
\end{align}
For such relabellings, the SPT coefficients change simultaneously, in the following way:

\begin{align}\label{relabel_SPT}
k_2 &\rightarrow k_2 - q^I u_I \nonumber \\
k_3 &\rightarrow k_3 - \frac{u^I K_{IJ} u^J}{2} - s^I u_I \nonumber \\
\vec{k}_4 &\rightarrow \vec{k}_4 - \vec{v}^I q_I  \nonumber \\
\vec{k}_5 &\rightarrow \vec{k}_5 - \vec{v}^I s_I - u^I \vec{t}_I - u^I K_{IJ} \vec{v}^J \nonumber \\
k_7 &\rightarrow k_7 - m^I u_I 
\end{align}
Note that all coefficients except $k_1$ and $k_6$ can be transformed in principle by these relabellings.

We will now use this result to perform some specific computations. Consider an example with the $1/N$ Laughlin state (with $N$ even) and $U(1)\times p4 = U(1)\times[\Z^2 \rtimes \Z_4]$ symmetry. 

In this case we have $M=4$ and $K = N$. The integer $s$ is defined modulo $\gcd(4,N)$, which is either 2 or 4, since we consider $N$ to be even; the integers $q,m$ are defined modulo $N$. The equivalence class of the torsion vector $\vec{t}$ is given by the value of $t_x + t_y \mod 2$. The condition on the relabelling indices $u, \vec{v}$, Eq. \eqref{Eq:relabel_constraint}, now becomes
\begin{align}
N u &= 4 \Lambda_s \\
N \vec{v} &= (1-U(\frac{\pi}{2})) \vec{\Lambda}_t
\end{align}
The transformation of SPT coefficients, Eq. \eqref{relabel_SPT}, now reads
\begin{align}
k_2 &\rightarrow k_2 - q u \nonumber \\
k_3 &\rightarrow k_3 - \frac{N u^2}{2} - s u \nonumber \\
\vec{k}_4 &\rightarrow \vec{k}_4 - q\vec{v} \nonumber \\
\vec{k}_5 &\rightarrow \vec{k}_5 - s\vec{v} - u \vec{t} \nonumber \\
k_7 &\rightarrow k_7 - m u.
\end{align}
Note that the SPTs parametrized by $\vec{k}_4,\vec{k_5}$ are nontrivial only if $k_{4,x}+k_{4,y}$ (respectively $k_{5,x}+k_{5,y}$) is odd. Since $N$ must be even, the term $- N u \vec{v}$ in the transformation of $\vec{k}_5$ is trivial, and has been ignored. With $N$ a multiple of 4, we can without loss of generality take $u = 1$ to satisfy the constraint; however, we must choose $v_i$ to be even. Hence $\vec{v}$ will not be responsible for any nontrivial relabellings.

When $N$ is of the form $4N' + 2$ we must choose $u=2$, while $\vec{v}$ must be chosen so that $v_x + v_y$ is even. However, this means that $\vec{v}$ still gives a trivial contribution to the relabellings of $\vec{k}_4,\vec{k}_5$, and hence we only need to consider transformations due to $u$.

The SET equivalences for this example are summarized in Table \ref{table:Z4Relabellings}. Note that in our examples, it is crucial that $C,\vec{\cancel{R}}$ be discrete, so that we can add trivial terms such as $\frac{M s'_I}{2\pi}a^I \cup dC$. This is not possible for continuous symmetry gauge fields: a term $\frac{q_I}{2\pi}a^I \cup dA$ cannot be trivial on its own for any nonzero integer value of $q_I$. This means that there is no chain of equivalences relating different elements of $\H^3(U(1),U(1))$ while keeping the charge vector $\vec{q}$ fixed. This is consistent with the fact that the different $U(1)$ SETs with the same charge vector all have different Hall conductivities, and are thus physically distinct states of matter. 

In our final example below, we will count the number of distinct SETs associated to the $1/2$ Laughlin state with $U(1)\times G_{\text{space}}$ symmetry. We will only present the results, which can be derived using the arguments above. In this case, the parameters $q,s,t_x,t_y,m$ can correspond to the identity particle $I$ or to the semion $S$. The parameters $k_1$ and $k_6$ will not be affected by relabellings and will always contribute a factor of $\Z\times \Z$ to the overall SET classification; we assume they are fixed. The remaining SPT parameters $k_2,k_3,\vec{k}_4,\vec{k}_5,k_7$ are classified by the group $\Z_M\times\Z_M\times K_M\times K_M\times \Z_M$. The relabelling equation is now
\begin{align}
k_2 &\rightarrow k_2 - q u \nonumber \\
k_3 &\rightarrow k_3 - u^2 - s u \nonumber \\
\vec{k}_4 &\rightarrow \vec{k}_4 - q\vec{v}  \nonumber \\
\vec{k}_5 &\rightarrow \vec{k}_5 - s \vec{v} - u \vec{t} - 2 u \vec{v} \nonumber \\
k_7 &\rightarrow k_7 - m u 
\end{align}
where 
\begin{align}
2 u &= M \Lambda_s \nonumber \\
2 \vec{v} &= (1-U\left(\frac{2\pi}{M}\right)) \vec{\Lambda}_t.
\end{align}
We analyze $M=2,3,4,6$ separately below:
\begin{enumerate}
	\item For $M=2$, there are $2$ ways to choose each of  $q,s,t_x,t_y,m$. We also have $|\Z_2^3\times K_2^2| = 2^7$. We can choose $u$ and $v$ arbitrarily. There are now different cases. If $q=s=I$ there is a factor 2 reduction due to $u$. If exactly one of $q,s$ equals $S$, there is a factor $2^3=8$ reduction due to both $u$ and $\vec{v}$. By varying $\vec{v}$, we see that if $q=S$, all values of $\vec{k}_4$ are trivial, while if $s=S$, all values of $\vec{k}_5$ are trivial. Finally, if $q=s=S$, we have a factor $2^5=32$ reduction, and both $\vec{k}_4$ and $\vec{k}_5$ will be trivial. 
	
	Therefore we obtain
	\begin{equation}
	2^7\left(\frac{8}{2}+\frac{16}{8}+\frac{8}{32}\right) = 800
	\end{equation}
	SETs. \\  
	\item For $M=3$ there are 2 choices each for $q$ and $m$, but all possible choices for $s$ and $\vec{t}$ are trivial. Thus there are 4 symmetry fractionalization classes. We also have $|\Z_3^3\times K_3^2| = 3^5$. There are no relabellings involving either $u$ or $\vec{v}$, since we must choose both $u$ and $v_x - v_y$ to be multiples of $3$. Hence we get $3^5 = 243$ distinct SETs for each symmetry fractionalization class, and $4 \times 243 = 972$ SETs in total. \\
	\item For $M=4$, there are 2 choices each for $q,s,m$ and $t_x+t_y$, giving 16 choices of charge vectors in total.  We also have $|\Z_4^3\times K_4^2| = 2^8$. First we note that we have to choose $v_x + v_y$ even, implying that $\vec{v}$ is not responsible for any nontrivial equivalences. Therefore we only consider equivalences due to $u=2$. 
	
	We have $2^2\times 4^3 = 256$ SETs whenever $q=s=m=I$ (there are no relabellings); otherwise we have $2\times 4^3 = 128$ SETs. This gives $2 \times 256 + 14 \times 128 = 2304$ SETs in total. \\
	\item For $M=6$, there are 2 choices each for $q,s,m$, while $\vec{t}$ is anyway trivial, and so we do not have to consider relabellings involving $\vec{v}$.  We also have $|\Z_6^3| = 6^3$. The only relabellings come from setting $u=1$. Now we have $6^3 = 216$ SETs whenever $q=M=I$ and $s=S$ (there are no relabellings in these cases); otherwise we have $ 6^3/2 = 108$ SETs. This gives $1 \times 216 + 7 \times 108 = 972$ SETs in total. 
\end{enumerate}
If we specialize to the $M=4$ case, we find that there are $2304$ distinct SETs, in contrast to the naive estimate of $|\H^2(G,\A)\times \H^3(G,U(1))| = 4096$. Note that much of the analysis was simplified by our choice of the 1/2 Laughlin topological order. If we consider more complicated topological orders, the analysis will become much more involved.

        	\section{Topological terms and group cohomology}
	\label{Appendix:GrpCoho}
        
	The correspondence between the topological effective action and the group cohomology formulation runs deeper than giving the same overall classification. There is a one-to-one correspondence
        between topological terms in the action involving flat background $G$ gauge fields and cocycles in group cohomology. In this section we will explain this relationship through concrete calculations.
        
        Let us first summarize the relationship between $\mathcal{H}^3(G,U(1))$ and the topological terms in $\mathcal{L}_{SPT}$, which correspond to topological effective actions for (2+1)D SPT states.
        See Ref. \onlinecite{Chen2013,Dijkgraaf1989pz} for a more detailed discussion. For an overview of simplicial calculus, see Ref. \onlinecite{Kapustin2014}. 
        
	\begin{enumerate}
          
        \item A topological Lagrangian for an SPT involving flat $G$ gauge fields (defined on 1-simplices) can be integrated over a 3-simplex of a triangulation, which gives an action $S$ associated to a single 3-simplex.
          The resulting $e^{i S}$, which depends on the values of the flat gauge field defined on the 1-simplices, is thus a 3-cochain of $G$ valued in $U(1)$, i.e. an element of $C^3(G,U(1))$. \\
          
	\item In fact $e^{i S}$ is a 3-cocycle of $G$ valued in $U(1)$, i.e. an element of $Z^3(G, U(1))$. The 3-cocycle condition arises by demanding that the theory be independent of the triangulation.
                 
	\item Gauge transformations applied to the $G$ gauge fields on a triangulation change the value of $e^{i S}$ by an amount $e^{i \int df}$, which corresponds to a 3-coboundary of $G$ valued in $U(1)$,
                  or an element of $B^3(G,U(1))$.               
                \end{enumerate}
                Therefore we see that gauge-inequivalent topological actions for flat $G$ gauge fields fall into equivalence classes determined by the quotient $\H^3(G,U(1)) := \frac{Z^3(G,U(1))}{B^3(G,U(1))}$. It
                has been shown that this fully characterizes topological gauge theories for gauge group $G$ \cite{Dijkgraaf1989pz}, and also believed to fully characterize (2+1)D SPTs \cite{Chen2013,Senthil2015SPT}. It is also known
                to classify the fusion and braiding properties of symmetry defects in (2+1)D SETs once the symmetry fractionalization class has been fixed \cite{Barkeshli2019}.

        Let us now summarize the relationship between $\mathcal{H}^2(G, \mathcal{A})$ and $\mathcal{L}_{\text{frac}}$. 
	Consider the coupling of flat $G$ gauge fields to flat internal gauge fields describing the Abelian topological order
        (we assume that the symmetry does not permute anyons):        
	\begin{enumerate}
          
        \item Consider a single internal $U(1)$ gauge field $a$. Consider a topological term which is an integer multiple of $\frac{1}{2\pi} a \cup B$, where $B \in 2\pi \Z$ is
          obtained in terms of the $G$ gauge field and is defined on 2-simplices. Note that $B \in 2\pi\Z$ in order for this term to
          be invariant under large gauge transformations of $a$. 
          
        \item This action can be thought of as an action for $U(1)\times G$ symmetry. Demanding retriangulation invariance implies that
          $\frac{1}{2\pi}a\cup B$ must be a 3-cocycle: $\frac{1}{2\pi} d (a\cup B) = \frac{1}{2\pi} (da \cup B + a \cup d B) \in 2\pi \Z$. Since $a$ is flat, $da \in 2\pi \Z$, so we find $dB = 0$.

          A $G$ gauge transformation which takes $B \rightarrow B + d \Gamma$, where $\frac{1}{2\pi}\Gamma \in \Z$, changes the Lagrangian by a $2$-coboundary of $G$ with $\Z$ coefficients. 
          Therefore the gauge inequivalent actions fall into equivalence classes determined by the
          quotient $\H^2(G,\Z) := \frac{Z^2(G,\Z)}{B^2(G,\Z)}$.

          When there are $D$ independent internal gauge fields, the coefficient changes from $\Z$ to $\Z^D$.\\
          
	\item The $K$-matrix coupling ensures that if $B$ is of the form $K \vec{\Lambda}$ where $\frac{1}{2\pi}\vec{\Lambda}$ is an integer vector, the theory is trivial. This is because the anyon associated to the
          ``symmetry flux'' $B$ is trivial. Therefore the correct coefficients which classify physically distinct couplings of the $K$-matrix theory to the background $G$ gauge field
          are $\Z^D/K\Z^D \cong \A$. This is in fact the definition of $\A$, the group of anyons. Therefore the classification of such actions is given by $\H^2(G,\A)$.
          
	\end{enumerate}

        In what follows we describe in more detail the precise relation between the topological terms in the effective action and the group cohomology cocycles for the symmetry group discussed in this paper, $G = U(1) \times G_{\text{space}}$. Note that the symbol $\H$ refers to cohomology with measurable cochains, also known as Borel cohomology. For $n>0$, the groups $\H^n(G,\Z)$ and $H^n(BG,\Z)$ coincide. Here $H(BG,\Z)$ refers to the cohomology of the classifying space $BG$ of $G$. 
        
	\subsection{Cocycle representatives for $G = U(1)$ and $G = \Z_M$}
        
	\underline{$G = U(1)$}: In this case we have 
	\begin{align}
	\mathcal{L}_{frac} &= \frac{1}{2\pi} q_I a^I \cup dA \\
	\mathcal{L}_{SPT} &= \frac{k}{2\pi} A \cup dA
	\end{align}
	where  $q_I, k \in \Z$. Here $\vec{q}$ is the charge vector. Define a flat, real-valued gauge field $A$ such that $A_{12} = a$ and $A_{23} = b$.
        Formally $A$ is the lift of a $U(1)$ gauge field to $\R$. A corresponding element of $U(1) = \R/2\pi\Z$ is written as $[a] = a \mod 2\pi$; therefore
        $a = [a] + 2\pi n_a$ for some $n_a \in \Z$. Now for the 3-simplex $[0123]$, $\L_{frac}$ becomes 
	\begin{equation}
	\frac{q_I}{2\pi} a^I_{01} dA_{123} = \frac{q_I}{2\pi} a^I_{01} ([a] + [b] - [a+b] + dn(a,b)) 
	\end{equation} 
	where $dn(a,b) = n_a + n_b - n_{a+b}$. The quantity $\frac{q_I}{2\pi}([a] + [b] - [a+b] + dn(a,b))$ defines an anyon, i.e. an element in $\mathcal{A}$, and thus can be
        viewed as an $\mathcal{A}$-valued 2-cocycle, i.e. an element of $Z^2(U(1),\A)$. The quantity $\frac{q_I}{2\pi}dn(a,b)$ is an $\mathcal{A}$-valued 2-coboundary, i.e. an element of
        $B^2(U(1),\A)$. In general, coboundaries correspond to changes of lift. Inequivalent choices of $\vec{q}$ determine inequivalent classes in the cohomology group
        $\H^2(U(1),\A) \cong \A$.
	
	A similar analysis can be made for the SPT term $\frac{k}{2\pi} A \cup dA$. In this case, choose a 3-simplex $[0123]$ with
        $A_{01} = a, A_{12} = b, A_{23} = c$ (the other values are fixed by the flatness of $A$). Again, $A$ is formally a lift from $U(1)$ to $\R$. Then we have 
	\begin{align}
          &\frac{k}{2\pi} A \cup dA [0123] = \frac{k}{2\pi} A_{01} \times dA_{123}
          \nonumber \\
          = &\frac{k}{2\pi} ([a] + n_a)([b]+[c]-[b+c] + dn(b,c)) 
	\end{align}
	When evaluated modulo $2\pi$, the rhs is a 3-cocycle which represents a cohomology class in $\H^3(U(1),U(1)) \cong \Z$ identified by $k$; the terms which
        explicitly depend on $n$ arise by choosing alternative lifts. For each choice of charge vector $\vec{q}$, it is possible to add a $\Z$ worth of SPT states.
        This means that for each symmetry fractionalization class, one can obtain a set of topological phases related to each other by stacking $G$-SPT states,
        given by elements of $\H^3(G,U(1))$.
	
	\underline{$G = \Z_M$}: Effective SPT actions for $G=\Z_M$ have been related to $\Z_M$ group cocycles in previous work \cite{Tiwari2018}. The action for $G = \Z_M$ is
	\begin{align}
	\mathcal{L}_{frac} &= \frac{s^I}{2\pi} a^I \cup dC \\
	\L_{SPT} &= \frac{k}{2\pi} C \cup dC
	\end{align}
	
	Define a flat gauge field $C \in \frac{2\pi}{M}\Z$ such that $C_{12} = 2\pi a/M$ and $C_{23} = 2\pi b/M$ where $a,b$ are integers. Formally $C$ is a lift from $\Z_M$ to $\frac{2\pi}{M}\Z$. A
        corresponding element of
        $\Z_M$ is written as $\frac{2\pi[a]_M}{M} = \frac{2\pi a}{M} \mod 2\pi$, where we define $[a]_M = a \mod M$; therefore $a = [a]_M + M n_a$ for some $n_a \in \Z$. 
        Now $\L_{frac}$ becomes 
	\begin{align}
	& \frac{s_I}{2\pi} a^I_{01}dC_{123} \nonumber \\
	&= \frac{s_I}{M} a^I_{01} ([a]_M + [b]_M - [a+b]_M + M dn(a,b)) 
	\end{align} 
	The quantity $\frac{s_I}{M}([a]_M + [b]_M - [a+b]_M + M dn(a,b))$ is a 2-cocycle in the group $Z^2(\Z_M,\A)$. The quantity $s_I dn(a,b)$, which is the difference between two different choices of lifts, is a 2-coboundary in the group $B^2(\Z_M,\A)$. Note that the most general coboundary relation implies that shifting $s_I$ by a multiple of $M$ corresponds to changing the lift; therefore $s_I + M \Lambda_I$ for $\Lambda_I \in \Z$ is equivalent to $s_I$. 
        With these conditions we see that the equivalence classes of $\vec{s}$ are in bijection with
        cohomology classes $[\mathfrak{w}] \in \H^2(\Z_M,\A)$. When $\A = \Z_{n_1}\times \dots\times\Z_{n_r}$, we simply have
        $\H^2(\Z_M,\A) = \Z_{(M,n_1)}\times \dots\times\Z_{(M,n_r)} = \A/M\A$ ($M\A$ is defined as $\{M a| a \in \A\}$).
	
	Next we analyze $\L_{SPT}$. The Lagrangian integrated on a 3-simplex with $C_{01} = a, C_{12} = b, C_{23} = c$ gives 
	\begin{align}
	&\frac{k}{2\pi} C \cup dC [0123]\nonumber \\ &= \frac{2\pi k}{M^2}([a]_M + M n_a)([b]_M+[c]_M-[b+c]_M + M dn(b,c)) 
	\end{align}
        Taken modulo $2\pi$, this function is a 3-cocycle in $Z^3(\Z_M,U(1))$. Choosing $k$ to be a multiple of $M$ results in a 3-coboundary; therefore the classification is $\H^3(\Z_M,U(1)) \cong \Z_M$. Since the 3-cocycles of $\Z_M$ and $U(1)$ have a similar form, the resulting SPT terms, which are of the form $A \cup dA$ and $C \cup dC$,
        also have the same Chern-Simons structure. 
	
	\subsection{Calculation of $\H^2(G_{\text{space}},\Z)$ and $\H^3(G_{\text{space}},U(1))$}
	The part of the action with terms from the group $G_{\text{space}}$ is
	\begin{align}
	\mathcal{L}_{frac} &= \frac{s_I}{2\pi} a^I \cup dC +\frac{\vec{t}_I}{2\pi} a^I \cup d\vec{\cancel{R}}+\frac{m_I}{2\pi} a^I \cup A_{XY}\\
	\L_{SPT} &= \frac{k_3}{2\pi} C \cup dC +\frac{\vec{k}_5}{2\pi} C \cup d\vec{\cancel{R}}+\frac{k_7}{2\pi} C \cup A_{XY}
	\end{align}
	Since the group cocycles for $G_{\text{space}}$ are less common than those of $U(1)$ or $\Z_M$, we will first derive them abstractly and then discuss their relationship to the gauge fields $\vec{R}$ and $C$. A space group $G_{\text{space}}$ can always be written as a group extension of a point group $H$ by the group of translations $\Z^2$, with some action $\theta: H \rightarrow \text{Aut}(\Z^2)$, as summarized by the short exact sequence
	\begin{equation}
	1 \rightarrow \Z^2 \rightarrow G_{\text{space}} \rightarrow H \rightarrow 1.
	\end{equation}
	When a group $G$ can be expressed in terms of a direct product extension, we can use the K\"unneth formula and its associated decomposition to determine its cohomology groups. When $H$ is a rotation point group, the above extension is on the other hand always a semidirect product. If it were possible to apply the K\"unneth decomposition to the above semidirect product extension for the cohomology of $G_{\text{space}}$ with $\Z$ coefficients, we would obtain
	\begin{equation}\label{Eq:Kunneth_Gspace}
	\H^n(G_{\text{space}},\Z) = \prod\limits_{k=0}^n \H^k_{\theta_k}(H,\H^{n-k}(\Z^2,\Z)).
	\end{equation} 
	This equation will be further explained below; however we first note the following caveats. For a general semidirect product extension, it is not possible to use the K\"unneth decomposition. The more general technique that is applicable in this case involves what are referred to as \textit{spectral sequences} (see eg. Refs. \cite{Mccleary2000,ramos2017spectral} for an introduction). When $G=G_{\text{space}}$, however, the cohomology groups $\H^n(G_{\text{space}},\Z)$ can be numerically computed using a program such as GAP, as was done in Ref. \cite{Thorngren2018}. Although we do not show the calculations here, we can apply spectral sequence techniques (specifically, we use the Lyndon-Hochschild-Serre spectral sequence) and compare them to the known numerical results. From this, we can infer that the correct expansion for $\H^n(G_{\text{space}},\Z)$ is indeed given by the K\"unneth decomposition. Knowing this result, we can finally use the K\"unneth decomposition again to obtain the cohomology of the group $U(1)\times G_{\text{space}}$.
	
	In \eqref{Eq:Kunneth_Gspace}, the action $\theta_k$ is not on $\Z^2$ itself, but on the cohomology group $\H^{n-k}(\Z^2,A)$; it is induced by the action $\theta$ of $H$ on $\Z^2$, and will be discussed further below. 
	
	Let us first study the classification of symmetry fractionalization. It is easiest to first compute $\H^2(G_{\text{space}},\Z)$ and then shift to $\A$ coefficients. Eq. \eqref{Eq:Kunneth_Gspace} gives 
	\begin{align}
	&\H^2(G_{\text{space}},\Z) \cong \H^2_{\theta_2}(\Z_M,\H^0(\Z^2,\Z)) \nonumber \\
	& \times\H^1_{\theta_1}(\Z_M,\H^1(\Z^2,\Z))\times\H^0_{\theta_0}(\Z_M,\H^2(\Z^2,\Z)) \label{Kunneth1}\\
	&= \H^2_{\theta_2}(\Z_M,\Z)\times \H^1_{\theta_1}(\Z_M,\Z\times\Z) \times \H^0_{\theta_0}(\Z_M,\Z) \label{Kunneth 2} \\
	&= \Z_M\times K_M \times \Z.
	\end{align}
	The first line is the K\"unneth decomposition. In the second line, we substituted the known cohomology groups $\H^k(\Z^2,\Z) = \Z^{{2\choose k}}$. The result of evaluating these cohomology groups is shown on the last line; these calculations will be discussed further below. 
	
	Using Eq.\eqref{Eq:Kunneth_Gspace}, we can next compute 
	\begin{align}
	&\H^4(G_{\text{space}},\Z) \cong \H^4_{\theta_4}(\Z_M,\H^0(\Z^2,\Z)) \nonumber \\
	& \times\H^3_{\theta_3}(\Z_M,\H^1(\Z^2,\Z))\times\H^2_{\theta_2}(\Z_M,\H^2(\Z^2,\Z)) \\
	&= \H^4_{\theta_4}(\Z_M,\Z)\times \H^3_{\theta_3}(\Z_M,\Z\times\Z) \times \H^2_{\theta_2}(\Z_M,\Z) \\
	&= \Z_M\times K_M \times \Z_M.
	\end{align}
	
	For a finite group or a compact Lie group $G$, it is a general result that $\H^n(G,U(1)) \cong \H^{n+1}(G,\Z)$ when $n>0$. Thus we can compute the cohomology groups of $U(1)$ with both $\Z$ and $U(1)$ coefficients. However, this statement does not hold for arbitrary $G$. Verifying this requires additional spectral sequence computations, which we will not show here. However, these computations indeed reveal that
	\begin{equation}
	\H^3(G_{\text{space}},U(1)) \cong \H^4(G_{\text{space}},\Z) 
	\end{equation}
	for $G_{\text{space}} = \Z^2\rtimes \Z_M$ (we in fact expect this relation to hold for a general 2D space group, although we have not done the more general computation). Finally, we use the K\"unneth decomposition for the direct product $U(1)\times G_{\text{space}}$ and use the above results to obtain \footnote{This version of the Kunneth decomposition assumes that the coefficients have the discrete topology; we have assumed here that we can also use $U(1)$ coefficients in the formula.}
	\begin{align}
	&\H^3(U(1)\times G_{\text{space}},U(1)) \cong \H^3(U(1),U(1)) \nonumber \\
	& \times \H^2(G_{\text{space}},\Z) \times \H^3(G_{\text{space}},U(1)) \\
	&= \Z^2\times Z_M^3\times K_M^2.
	\end{align}
	
	Note that since $U(1)$ is a continuous group, and we are working with measurable (Borel) cohomology, it is difficult to compute its cohomology groups directly, and in doing so we must rely on technical mathematical results. To provide some additional intuition about the cohomology of $U(1)$ using results on finite groups, Ref. \cite{Chen2013} computed the cohomology groups of $\Z_n$ and showed how they were related to those of $U(1)$ upon taking an appropriate limit where $n \rightarrow \infty$. In a similar spirit, we can compute the cohomology of $G=\Z_n\times G_{\text{space}}$ for an arbitrary integer $n$ using the K\"unneth decomposition, and obtain
	\begin{align}
	&\H^3(\Z_n\times G_{\text{space}},U(1)) \cong \H^3(\Z_n,U(1)) \nonumber \\
	& \times \H^2(G_{\text{space}},\Z_n) \times \H^3(G_{\text{space}},U(1)) \\
	&= \Z_n^2\times Z_M^2\times K_M \times \Z_{(M,n)} \times (K_M \otimes \Z_n).
	\end{align}
	Thus we obtain a result which bears a significant resemblance to the claimed result for $U(1)\times G_{\text{space}}$: the difference is that some groups in the above classification depend on the commensuration between $n$ and $M$. If we choose $n$ to be a multiple of $M$, and take $n \rightarrow \infty$, so that the initial factors of $\Z_n$ are replaced by $\Z$, we recover the result for the group $U(1)\times G_{\text{space}}$.
	\subsection{Cocycle representatives for $\H^2(G_{\text{space}},\Z)$}
	Note that all 2-cocycles must satisfy the following condition:
	\begin{align}
	f_2(g_1,g_2) + f_2(g_1g_2,g_3) = f_2(g_2,g_3) + f_2(g_1,g_2g_3)
	\end{align}
	where, if $g_i = (\vec{r}_i,h_i)$, then $g_1g_2 = (\vec{r}_1+U(h_1)\vec{r}_2,h_1+h_2)$. In what follows, we assume that the translation gauge field $\vec{R}$ is valued in $2\pi\Z^2$, while the $\Z^2$ group elements $\vec{r}_i$ are assumed to be integer-valued. Cocycles in $Z^d(G_{\text{space}},\Z)$ are denoted as $f_d$.

          \subsubsection{$ \H^2_{\theta_2}(\Z_M,\H^0(\Z^2,\Z))$}
          The cocycles in the coefficient group $\H^0(\Z^2,\Z)$ in the first term of Eq. \eqref{Kunneth1} are constant functions valued in $\Z$. The $\Z_M$ rotations, which act on $\Z^2$, therefore do not change the value of these functions, so that $\theta_2$ is the trivial action. The first term is thus isomorphic to $\H^2(\Z_M,\Z)$, and the associated cocycle representatives of $G_{\text{space}}$ are $f_2(g_1,g_2) = \frac{s}{M}([h_1]_M+[h_2]_M-[h_1+h_2]_M)$ with $s \in \Z_M$ as discussed previously. The corresponding field-theoretic element is $\frac{s}{2\pi} dC$. \\

          \subsubsection{$\H^1_{\theta_1}(\Z_M,\H^1(\Z^2,\Z))$}

          Now we consider the second term of \eqref{Kunneth1}. The coefficient module $\H^1(\Z^2,\Z)$ has cocycle representatives of the form $p_{\vec{t}}$, where $p_{\vec{t}}(\vec{r}) = \vec{t} \cdot \vec{r}$ for some $\vec{t} \in \Z^2$. Under a rotation $U(h)$, $p_{\vec{t}}$ gets transformed as $\vec{t} \cdot {U(h)}\vec{r} = ({{U^T(h)}}\vec{t}) \cdot \vec{r} = p_{U^T(h)\vec{t}}(\vec{r})$. This means that the induced action on the coefficients is equivalent to the rotation action $\theta_1 = \theta$ on vectors in $\Z^2$. 
          
          The first observation is that the group $\H^1_{\theta_1}(\Z_M,\Z^2)$ classifies functions $f_1$ taking elements $h$ of $\Z_M$ to vectors in $ \Z^2$.
          We have  $\H^1_{\theta}(\Z_M,\Z^2) \cong\frac{\Z^2}{(I-U\left(\frac{2\pi}{M}\right))\Z^2} \cong K_M$, using standard results on the cohomology of cyclic groups (see for eg. Ref \cite{Chen2013}). A representative cocycle $f_1$ of this group has the form
          \begin{align}
          f_1(h)
          & = \frac{1-U(h)}{1-U(\frac{2\pi}{M})} f_1(2\pi/M),
          \end{align}
          where  $f_1(2\pi/M) = \vec{t}$ for some $\vec{t} \in \Z^2$, and the $U$ matrices act on $\vec{t}$ by rotation.
          
          Next we consider the more detailed decomposition $\H^1_{\theta_1}(\Z_M,\H^1(\Z^2,\Z))$. A cocycle of this group maps an element $h \in \Z_M$ to a \textit{cohomology class} $[p_{f_1(h)}] \in \H^1(\Z^2,\Z)$ whose representatives are functions $p_{f_1(h)}$.
          
           The desired 2-cocycle of $G_{\text{space}}$ is completely determined in terms of $p_{f_1(h)}$ as follows:
		\begin{align}
		f_2(g_1,g_2) &= p_{f_1(h_1)}(\vec{r_2}) \\
		&=   \left(\frac{1-U(h_1)}{1-U(\frac{2\pi}{M})}\vec{t}\right) \cdot \vec{r}_2
		\end{align} 
		This function, with parameter $\vec{t}$, satisfies the 2-cocycle condition for $G_{\text{space}}$. Values of $\vec{t}$ which are of the form $\vec{t} = (1-U(2\pi/M)) \vec{t'}$ are trivial, as the resulting cocycles are actually 2-coboundaries $db$, where $b(g) = \vec{t}' \cdot \vec{r}$.
		It is easy to motivate this function by looking at a 2-simplex [012]. If $\frac{1}{2\pi}\vec{R}_{01} = \vec{r}_1$, $C_{01} = h_1$ and $\frac{1}{2\pi}\vec{R}_{12} = \vec{r}_2$, then from flatness of $(\vec{R},C)$ we have $\frac{1}{2\pi}\vec{R}_{02} = \vec{r}_1+U(h_1)\vec{r}_2$. Therefore $\frac{1}{2\pi}d\vec{R}[012] = \vec{r}_1+\vec{r}_2-(\vec{r}_1+U(h_1)\vec{r}_2) = (1-U(h_1)) \vec{r}_2$. Since $U(h_1)$ is a power of $U\left(\frac{2\pi}{M}\right)$, this function is always a multiple of $(1-U\left(\frac{2\pi}{M}\right))$. Therefore, $f_2(g_1,g_2) = \vec{t}\cdot \frac{1-U(h_1)}{1-U\left(\frac{2\pi}{M}\right)} \vec{r}_2$ is integer valued for all $\vec{t} \in \Z^2$. However, it cannot be generated on a 2-simplex by a 2-coboundary $df(g_1,g_2)$ (the only function that would give $df$ is $f(g) = \vec{t}\cdot(1-U\left(\frac{2\pi}{M}\right))^{-1} \vec{r}$, which is not integer-valued, unless $\vec{t}$ has the trivial form). The field theory element giving this value is $ \frac{\vec{t}}{2\pi}\cdot d\vec{\cancel{R}} = \frac{\vec{t}}{2\pi} \cdot (1-U\left(\frac{2\pi}{M}\right))^{-1} d\vec{R}$.

\subsubsection{$ \H^0_{\theta_0}(\Z_M,\H^2(\Z^2,\Z)) $}

Finally, we study the third term of \eqref{Kunneth1}. The coefficient module $\H^2(\Z^2,\Z)$  has representatives $w_m$ for $m \in \Z$, satisfying $w_m(\vec{r}_1,\vec{r}_2) - w_m(\vec{r}_2,\vec{r}_1) = m \vec{r}_1\times \vec{r}_2$. Although the rotation action changes the form of $w_m$, the above cross product (and hence the value of $m$) is rotationally invariant, and in this sense the action $\theta_0$ is trivial. Now the group $ \H^0_{\theta_0}(\Z_M,\Z)$ classifies functions $f_0$ taking each $h \in \Z_M$ to some fixed integer $f_0(h) = m \in \Z$. Therefore a cocycle in the group $ \H^0_{\theta_0}(\Z_M,\H^2(\Z^2,\Z)) $ should take $h$ to the cohomology class $[w_{f_0(h)}]$ whose representatives $w_{f_0(h)}$ are such that $w_{f_0(h)}(\vec{r}_1,\vec{r}_2) - w_{f_0(h)}(\vec{r}_2,\vec{r}_1)$ is rotationally invariant. 

It can be verified that the following function is a 2-cocycle of $G_{\text{space}}$ with these properties: 
\begin{equation}
f_2(g_1,g_2) = w_{f_0(h_1)}(\vec{r}_1,U(h_1)\vec{r}_2) = m r_{1,x} (U(h_1)\vec{r}_2)_y
\end{equation}
In this case we have $w_{f_0(h_1)}(\vec{r}_1,U(h_1)\vec{r}_2) - w_{f_0(h_1)}(U(h_1)\vec{r}_2,\vec{r}_1) = m \vec{r}_1\times U(h_1)\vec{r}_2$. The cross product is invariant under rotations and is a measure of area. If operations 2 and 1 are performed successively, the rotation $h_1$ changes the relative orientation of axes used to measure the two translations. The vector $\vec{r}_2$ is therefore rotated by $U(h_1)$ so as to meaningfully take a cross product with $\vec{r}_1$. 
		
 Consider the quantity $w'_{f_0(h_1)}(\vec{r}_1,U(h_1)\vec{r}_2) = \frac{m}{2} \vec{r}_{1}\times {U(h_1)}\vec{r}_{2}$. Although it is not an integer-valued cocycle, it satisfies the 2-cocycle condition with $\frac{1}{2}\Z$ coefficients (hence it can be used to obtain a topologically invariant action on 3-simplices). This function satisfies $w'_{f_0(h_1)}(\vec{r}_1,U(h_1)\vec{r}_2) - w'_{f_0(h_1)}(U(h_1)\vec{r}_2,\vec{r}_1) = m \vec{r}_1\times U(h_1)\vec{r}_2$, i.e. it has the same gauge-invariant property as $f_2(g_1,g_2)$; moreover, it is already rotationally invariant. We use this $\frac{1}{2}\Z$-valued cocycle in the field theory because it is closely related to the integer-valued space group cocycles, and furthermore the cross product is an intuitive measure of area. The corresponding field theory object is $\frac{m}{2\pi} A_{XY}$, where $A_{XY}[012] = \frac{1}{4\pi}(\vec{R}_{01})\times U(C_{01})\vec{R}_{12}$. The gauge transformation behaviour of $A_{XY}$ and its physical relationship to the area element were discussed in Section \ref{Sec:Area}.

                \subsubsection{Classification}
                
	   The classification $\H^2(G_{\text{space}},\Z)$ is seen from the discussion above to be $\Z_M\times K_M \times \Z$. To obtain the classification of symmetry fractionalization, we use the Universal Coefficient Theorem \cite{Sato1996} to write 
	   \begin{align}
	   &\H^2(G_{\text{space}},\A) \nonumber \\&= \H^2(G_{\text{space}},\Z)\otimes \A \times \text{Tor}(\H^3(G_{\text{space}},\Z),\A) \\
	   &= \H^2(G_{\text{space}},\Z)\otimes \A
	   \end{align}
	    (we can check that the group $\H^3(G_{\text{space}},\Z)$ vanishes when $G_{\text{space}}$ is orientation-preserving). The $\otimes$ (tensor product) symbol defines the tensor product $G \otimes H$ of abelian groups $G$ and $H$. The group $G \otimes H$ is defined as the set of pairs $g \otimes h$ where $g \in G, h \in H$, where $\otimes$ is a bilinear operation such that $g \otimes h$ is trivial if either $g$ or $h$ is trivial. For example if $ng = 1_G$ (the identity element of $G$), $n (g \otimes h) = (n g) \otimes h = 1_{G \otimes H}$; and this argument runs similarly for $h$. The group $G \otimes H$ is completely defined by the following properties:
	   \begin{align}
	   G \otimes H & \cong  H\otimes G\\
	   (\prod_{i}G_{i})\otimes(\prod_{j}H_{j}) & \cong\prod_{i,j}(G_{i}\otimes H_{j})\\
	   G \otimes \Z & \cong G\\
	   \Z_m \otimes \Z_n & \cong\mathbb{Z}_{d},d=\gcd(m,n)
	   \end{align}
	   The topological terms classified by $\H^2(G_{\text{space}},\Z) \otimes \A$ are thus consistent with the group structure of symmetry fluxes (classified by $\H^2(G_{\text{space}},\Z)$) as well as that of anyons (classified by $\A$). The formal effect of the $\otimes$ symbol is to replace the $\Z$ coefficients by $\A$ coefficients. This means that the coefficients $s,\vec{t},m$ are replaced by vectors $s_I,\vec{t}_I,m_I$ in $\Z^D$. Moreover, if these parameters take the form $K \vec{\Lambda}$, they are trivial. With this change, the above cocycles all become cocycle respresentatives for $\H^2(G_{\text{space}},\A) \cong (\A/M\A) \times (K_M \otimes \A) \times \A$. (Note that $\Z_M\otimes \A \cong \A/M\A$.) Effective actions corresponding to these cocycles are recovered by taking a cup product of the vector $a_I$ of internal gauge fields with the field theory term corresponding to a representative of $\H^2(G,\A)$.
   
	\subsection{Cocycle representatives for $\H^3(G_{\text{space}},U(1))$}
	With our knowledge of $\H^2(G_{\text{space}},\Z)$, it is easy to understand the group $\H^3(G_{\text{space}},U(1))$. We can derive its cocycle representatives in the following direct way. The $G_{\text{space}}$ charges are classified by $\H^1(G_{\text{space}},U(1)) \cong \Z_M$ (corresponding to the charges of $C$), whose generator is represented by the cocycle $f_1(h) = 2\pi [h]_M/M \mod 2\pi$. The associated field theory element is just $C$. The fluxes are classified by the group $\H^2(G_{\text{space}},\Z)$. Therefore SPT cocycles, which associate symmetry flux to an elementary symmetry charge, are all of the form $\nu(g_1,g_2,g_3) = \frac{2\pi [h_1]}{M} \beta (g_2,g_3) \mod 2\pi$, where $[\beta] \in \H^2(G_{\text{space}},\Z)$. These functions satisfy the 3-cocycle condition for $\H^3(G_{\text{space}},U(1))$, and correspond to taking the cup product of a cocycle in $\H^1(\Z_M,U(1))$ with another from $\H^2(G_{\text{space}},\Z)$. 
	
	To obtain the relevant SPT cocycle representatives, consider the three subgroups $S_k, k=0,1,2$, of $\H^2(G_{\text{space}},\Z)$, defined as 
	\begin{align}
	S_k := \H^k(\Z_M,\H^{2-k}(\Z^2,\Z)).
	\end{align}
	From the definition of the tensor product, the classification of SPT terms obtained by associating an elementary $\Z_M$ charge to a flux represented by a cocycle of $S_i$ is $\Z_M \otimes S_i$. The full SPT classification is therefore $\prod_{i=0}^2 \Z_M \otimes S_i = (\Z_M \otimes \Z_M) \times (\Z_M \otimes K_M) \times(\Z_M \otimes \Z) = \Z_M \times K_M \times \Z_M$. This is the same as the K\"unneth decomposition result: $\H^3(G_{\text{space}},U(1)) \cong \H^3(\Z_M,U(1))\times \H^2_{\theta}(\Z_M,U(1)\times U(1))\times \H^1(\Z_M,\H^2(\Z^2,U(1))) \cong \Z_M^2\times K_M$. Therefore the flux-charge construction accounts for all the group cohomology SPTs. The cocycles so obtained are moreover in one-to-one correspondence with cocycle representatives of $\H^4(G_{\text{space}},\Z)$.
	
	The cocycles for mixed SPTs of $U(1)$ and $G_{\text{space}}$ symmetry are obtained by a cup product of a 1-cocycle representative of $\H^1(U(1),U(1))$ (generated by $f_1(a) = [a] \mod 2\pi$) and a 2-cocycle representative of $\H^2(G_{\text{space}},\Z)$. Finally, the full $\H^3(G_{\text{space}}\times U(1),U(1))$ classification can also be obtained from the K\"unneth decomposition: it equals $\Z^2\times \Z_M^3\times K_M^2$. In this case, the possible charges are classified by the group $\Z\times\Z_M$, corresponding to charge of $A$ and $C$ respectively. These charges couple to fluxes, i.e. representatives of the group $\H^2(G_{\text{space}},\Z)$, to give the full SPT action for the group $G_{\text{space}} \times U(1)$.

        \bibliography{library_prl} 
      
\end{document}